\documentclass[twocolumn]{aastex62}

\graphicspath{{./}{figures/}}
\revised{\today}
\usepackage{multirow} 
\usepackage{longtable}
\usepackage{xcolor}

\shorttitle{The Swansong of the Galactic Center Source X7}
\shortauthors{A. Ciurlo, R. Campbell, M. Morris et al.}

\begin{document}

\title{The Swansong of the Galactic Center Source X7: An Extreme Example of Tidal Evolution near the Supermassive Black Hole}

\correspondingauthor{Anna Ciurlo}
\email{ciurlo@astro.ucla.edu}

\author[0000-0001-5800-3093]{Anna Ciurlo}
\affil{Department of Physics and Astronomy, University of California, Los Angeles, CA 90095, USA}

\author[0000-0002-3289-5203]{Randall D. Campbell}
\affiliation{W. M. Keck Observatory, Waimea, HI 96743, US}

\author[0000-0002-6753-2066]{Mark R. Morris}  
\affiliation{Department of Physics and Astronomy, University of California, Los Angeles, CA 90095, USA}

\author[0000-0001-9554-6062]{Tuan Do}
\affiliation{Department of Physics and Astronomy, University of California, Los Angeles, CA 90095, USA}

\author[0000-0003-3230-5055]{Andrea M. Ghez}
\affiliation{Department of Physics and Astronomy, University of California, Los Angeles, CA 90095, USA} 

\author{Eric E. Becklin}
\affiliation{Department of Physics and Astronomy, University of California, Los Angeles, CA 90095, USA}

\author[0000-0001-7017-8582]{Rory O. Bentley}
\affiliation{Department of Physics and Astronomy, University of California, Los Angeles, CA 90095, USA}

\author[0000-0003-3765-8001]{Devin S. Chu}
\affiliation{Department of Physics and Astronomy, University of California, Los Angeles, CA 90095, USA}

\author[0000-0002-2836-117X]{Abhimat K. Gautam}
\affiliation{Department of Physics and Astronomy, University of California, Los Angeles, CA 90095, USA}

\author[0000-0003-1754-2570]{Yash A. Gursahani}
\affiliation{Department of Physics and Astronomy, University of California, Los Angeles, CA 90095, USA}

\author[0000-0001-9554-6062]{Aur\'elien Hees}  
\affiliation{Sytemes de Reference Temps Espace, Observatoire de Paris, Universit\'e Paris-Sciences-et-Lettres, \\ Centre National de la Recherche Scientifique, Sorbonne Universite, \\ Laboratoire National de m\'etrologie et d'Essais, 61 avenue de l'Observatoire, 75014 Paris, France}

\author[0000-0003-2400-7322]{Kelly Kosmo O'Neil}
\affiliation{Department of Physics and Astronomy, University of California, Los Angeles, CA 90095, USA}

\author[0000-0001-9611-0009]{Jessica R. Lu}
\affiliation{Astronomy Department, University of California Berkeley, Berkeley, CA 94720, USA}

\author{Gregory D. Martinez}
\affiliation{Department of Physics and Astronomy, University of California, Los Angeles, CA 90095, USA}

\author[0000-0002-9802-9279]{Smadar Naoz}
\affiliation{Department of Physics and Astronomy, University of California, Los Angeles, CA 90095, USA}
\affiliation{Mani L. Bhaumik Institute for Theoretical Physics, Department of Physics and Astronomy, UCLA, Los Angeles, CA 90095, USA}

\author{Shoko Sakai}
\affiliation{Department of Physics and Astronomy, University of California, Los Angeles, CA 90095, USA}

\author{Rainer Sch\"odel}  
\affiliation{Instituto de Astrof\'isica de Andaluc\'ia, Consejo Superior de Investigaciones Cientificas, \\ Glorieta de la Astronomia S/N, 18008 Granada, Spain}


\begin{abstract}
We present two decades of new high-angular-resolution near-infrared data from the W. M. Keck Observatory that reveal extreme evolution in X7, an elongated dust and gas feature, presently located half an arcsecond from the Galactic Center supermassive black hole. 
With both spectro-imaging observations of Br-$\gamma$ line-emission and Lp (3.8 $\mu$m) imaging data, we provide the first estimate of its orbital parameters and quantitative characterization of the evolution of its morphology and mass.  We find that the leading edge of X7 appears to be on a mildly eccentric (e$\sim$0.3), relatively short-period (170 years) orbit and is headed towards periapse passage, estimated to occur in $\sim$2036. Furthermore, our kinematic measurements rule out the earlier suggestion that X7 is associated with the stellar source S0-73 or with any other point source that has overlapped with X7 during our monitoring period. Over the course of our observations, X7 has (1)  become more elongated, with a current length-to-width ratio of 9, (2) maintained a very consistent long-axis orientation (position angle of 50$^{\circ}$), (3) inverted its radial velocity differential from tip to tail from -50 to +80~km/sec, and (4) sustained its total brightness (12.8 Lp magnitudes at the leading edge) and color temperature (425~K), which suggest a constant mass of $\sim$50~M$_{Earth}$.  We present a simple model showing that these results are compatible with the expected effect of tidal forces exerted on it by the central black hole and we propose that X7 is the gas and dust recently ejected from a grazing collision in a binary system. 
\end{abstract}



\section{Introduction} \label{sec:intro}
\begin{figure}[ht]
    \centering
    \includegraphics[width=8cm]{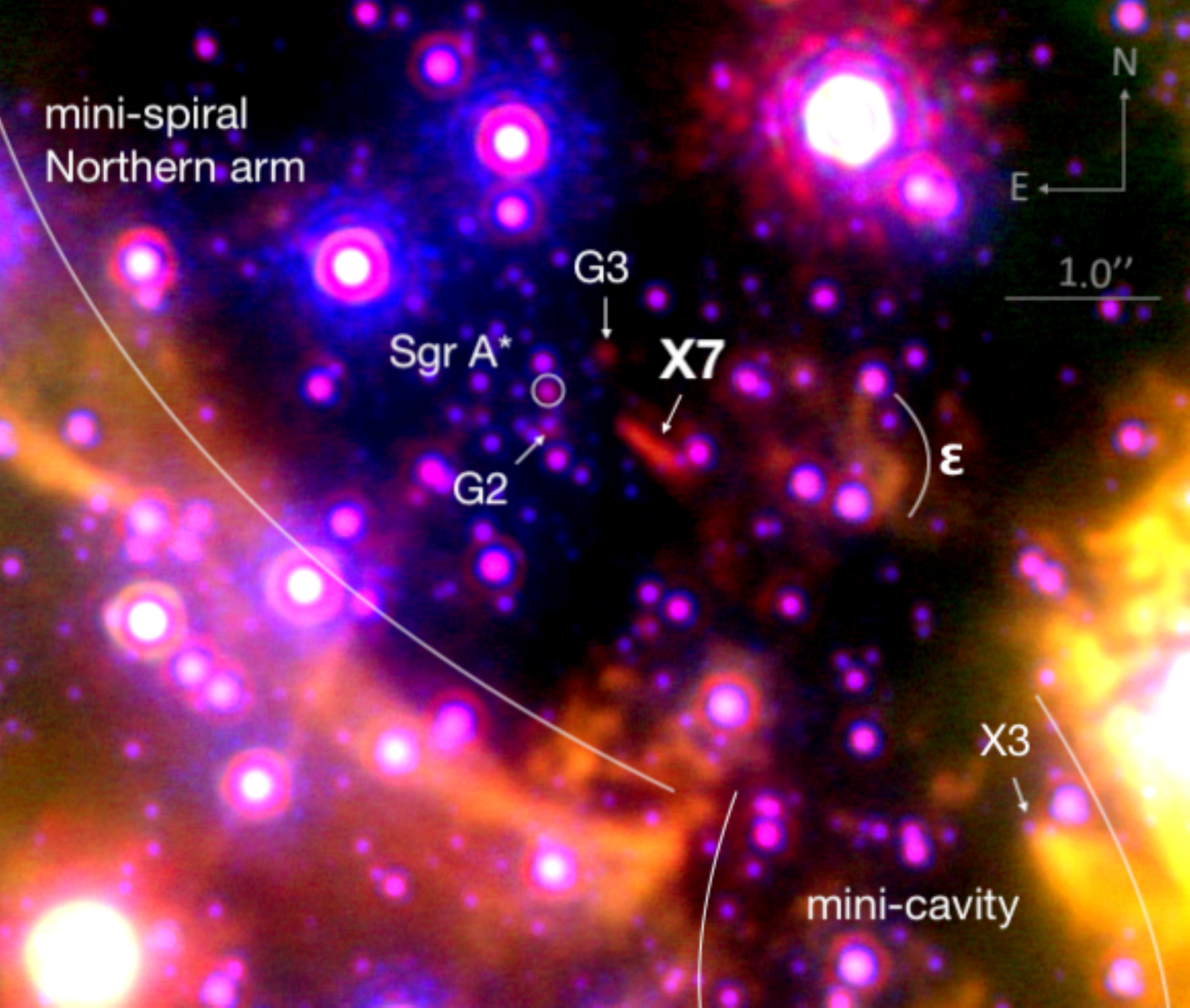}
    \caption{X7 in context: gas and dust structures in the Galactic Center. Three-color image of the central $0.27\times0.27$ parsecs of the Galactic Center obtained with the NIRC2 imager at the W. M. Keck Observatory. 
    Lp (3.8~$\mu$m) is shown in red, Ms (4.7~$\mu$m) in green, Kp (2.1~$\mu$m) in blue.  
    All three bands were observed in summer 2021 using adaptive optics (Section~\ref{sec:data} provides details on these observations).
    Thermal emission from warm dust in X7 is well detected at Lp (3.8 $\mu$m).
    The source X3, originally thought to be of the same nature as X7 \citep{Muzic+07}, is also visible in the Southwest corner.
    Other prominent, larger-scale features include the tip of the Northern Arm of the mini-spiral  \citep{Lo83} located to the south of X7, the top part of the mini-cavity \citep{MYZ87,Yusef-Zadeh90} in the southwest, and the Epsilon source \citep{Yusef-Zadeh90}, immediately East of X7.}
    \label{fig:rgb}
\end{figure}

The immediate entourage of the supermassive black hole (SMBH) at the center of the Milky Way Galaxy includes dense, co-spatial clusters of both young and old stars \citep{Paumard06, Do13, Lu13}, as well as orbiting streams of gas and dust on scales from 0.1 to 1~pc (see reviews by \citealt{Morris96, Genzel10}). 
At smaller scales ($\sim$0.02 pc), a collection of so-called G objects has been found \citep{Gillessen12, Phifer13, Pfuhl15, Witzel17, Ciurlo20}: compact gas/dust features that have been interpreted as stellar objects enshrouded by extended dust photospheres, possibly as a result of relatively recent binary mergers \citep{Witzel+14, Stephan16, Ciurlo20}. 
In at least two cases, the sizes of the G objects have apparently exceeded their tidal radii as they passed through their orbital periapse near the SMBH, causing them to shed an observable amount of their gas/dust envelopes \citep{Witzel17,Gillessen+19}, but thereafter, they reverted to their compact form.
One of the most intriguing objects found within $\sim$0.02~pc of the SMBH is X7, the subject of this paper. The position of X7 relative to other gas and dust constituents of the central light-year of the Galaxy, including the electromagnetic counterpart of the SMBH, Sgr~A* \citep{Ghez05, Genzel03}, is shown in Figure~\ref{fig:rgb}. 

First noted by \cite{Clenet04}, then named by \citet{Muzic+07}, X7 is a filamentary dust/gas feature that, like the G objects, is observable in both near-infrared thermal dust emission and in emission lines from ionized gas.  
As it orbits the SMBH, X7 has been undergoing dramatic evolution over the past 20 years.
X7's early (mid-2000s) appearance resembled a cometary shape, which led \cite{Muzic+07} to suggest that X7 results from a bow shock caused by winds (either from nearby massive stars or from the supermassive black hole itself). 
However, the bow shock appearance has not persisted in subsequent years. More recently, \cite{Peissker21} proposed instead that X7 is the product of the interaction between the circumstellar envelope of an S-star (a member of a group of early-type stars on tight, eccentric orbits around the SMBH) and a wind emanating from the neighborhood of Sgr~A*. 

In this study we use the GCOI\footnote{Galactic Center Orbits Initiative (GCOI, PI: Ghez), a growing 27-year database composed of data acquired at the W.M. Keck Observatory.} database from the long-term monitoring of this region with the W. M. Keck Observatory, to characterize the morphological and dynamical evolution of X7. We find that X7 is a complex feature with a rapidly evolving spatial-velocity structure most likely due to tidal interactions with the SMBH. Preliminary results of this analysis have previously been presented in \cite{Campbell21}.
With the GCOI database we have access to a set of imaging data that is similar, but completely independent of, the one presented by \cite{Peissker21}. 
Additionally, we present and analyze a much more extensive set of spectroscopic data than what has previously been published, which allows us to characterize the dynamical evolution of X7 over more than 15 years for the first time. 
These data, in combination with the imaging data, provide comprehensive new insights into the characteristics and behavior of this object. 
For the first time we compute the orbit of the leading edge of X7 and use it to model X7's response to the gravity of the SMBH. 
We find here that the evolution of X7's shape can be well explained  by the effect of the gravity of the SMBH alone.
We show that the initial bow-shock shape of X7 has evolved to a more linear morphology and that it is starting to undergo fragmentation. The orientation of X7's elongation is inconsistent with its direction of motion, with models of the collective stellar winds in the local region and with a spherical outflow from the SMBH. 
The long-term monitoring with integral field spectroscopy data reveal a clear positional and dynamical separation of X7 from stellar sources, including the one proposed by \citet{Peissker21} and from nearby G object sources (G4 and G5, \cite{Ciurlo20}) due to distinct differences in both radial velocity and proper motion.
We do note and discuss an intriguing similarity in orbital motion with G3.

We describe the observations in Section~\ref{sec:data} and in Section~\ref{sec:methods} the methodology we employed to parameterize X7's properties.
In Section~\ref{sec:results} we present our results on X7's morphology, orbit, length, brightness and mass. 
In Section~\ref{sec:toymod} we model X7's tidal evolution. 
Several scenarios for the X7's evolution and origin are discussed in Section~\ref{sec:discussion}.
The summary and conclusions of our study are reported in Section~\ref{sec:concl}.


\section{Datasets} 
\label{sec:data}

This study uses imaging (Section~\ref{subsec:imaging}) and spectroscopic (Section~\ref{subsec:spectro}) data  consisting of new data and existing data, taken as part of the GCOI. 
While the majority of this dataset was acquired using the laser guide star adaptive optics (AO) systems that operate with optical tip-tilt systems \citep{Wizinowich06}, using a guide star located $\sim$20~arcsec away from Sgr~A*, some of the latest datasets were taken with the newest AO systems. 
On Keck I, which hosts the integral field unit OSIRIS \citep{Larkin06} used in this study, the AO system has been upgraded to include an infrared tip-tilt system, TRICK, enabling the use of IRS~7, a brighter and closer star for the tip-tilt corrections for half of our 2020 and 2021 spectroscopic observations.    
On Keck II, which hosts the imager NIRC2 (PI: K. Matthews), a near-infrared natural-guide-star AO system that includes a pyramid wavefront sensor \citep{Bond20}, introduced the opportunity to use IRS~7 for both high- and low-order wavefront corrections for all of our 2021 imaging observations.
All three AO systems deliver very near diffraction-limited performance. 

\subsection{Imaging data}
\label{subsec:imaging}

\begin{table}[ht]
    \centering
    \begin{singlespace}
    \begin{tabular}{|c|cccc|}
    \hline
    \multirow{2}{*}{Date}  & \multirow{2}{*}{filter} & {itime [sec]}  & frames & FWHM  \\ 
 &  & $\times$coadds & number  & [mas]\\ 
\hline
2017-07-16 & Lp &  0.50$\times$30   & 368 & 96         \\
2019-08-14 & Lp &  0.50$\times$60   & 294 & 90         \\
2020-07-31 & Lp &  0.20$\times$30   & 29  & 98         \\
2021-07-13 & Lp &  0.50$\times$60   & 40  & 102        \\
2021-08-15 & Lp &  0.35$\times$85   & 43  & 93$^{*}$   \\
\hline 
2015-04-02 & Kp &  2.8$\times$10    & 18  & 72        \\
2021-08-15 & Kp &  2.1$\times$14    & 36  & 49$^{*}$  \\
\hline
2015-04-02 & Ms &  0.2$\times$600   & 32  & 122       \\
2021-08-21 & Ms &  0.2$\times$200   & 34  & 103$^{*}$ \\ 
\hline
    \end{tabular}
    \end{singlespace}
    \caption{
    Newly reported GCOI NIRC2 observations. All observations used a 10~mas pixel scale. 
    The epochs marked with a $^*$ have been observed in natural guide star mode using an infrared wavefront sensor instead of the usual laser-guide-star configuration.
    }
    \label{tab:newnobs}
\end{table}
A subset of the GCOI imaging data set used in this study, collected between 2002 and 2021, is primarily composed of NIRC2 observations taken through the Lp band-pass filter (3.776~$\mu$m central wavelengths) during 22 nights, of which the most recent 5 observations (post-2015) are newly reported.  
The observations cover a field of view of roughly 10" x 10" around Sgr~A*, with a 10~mas/pixel scale. For each night, the individual  data frames are calibrated and combined following the same procedure used for previously published data \citep{Ghez04, Ghez05b, Phifer13, Witzel+14}.  
The final images, which have an average resolution of 93~mas (FWHM), are placed in an absolute coordinate system with Sgr~A* at the center, using the known positions and proper motions of 5 stars (IRS16~SW, IRS16~C, IRS16~NW, IRS29~N and S1-23). The reference stars' positions are measured by the GCOI pipeline (see \citealt{Sakai19, Jia19}) and the alignment procedure is the same as employed for the OSIRIS data in \cite{Ciurlo20}. 
Table~\ref{tab:newnobs} provides details on the newly reported observations at Lp as well as two supplemental epochs at Kp and Ms; Figure~\ref{fig:rgb} shows a three-color image constructed from the Kp, Lp, and Ms data taken in 2021.

\subsection{Spectroscopic data}
\label{subsec:spectro}

\begin{table}[ht]
\centering
\begin{singlespace}
    \begin{tabular}{|c|cccc|}
    \hline
    \multirow{2}{*}{Date} & scale & itime  & frames & FWHM  \\ 
 & [mas] & [sec] & number  & [mas] \\ 
\hline
2019-05-11 & 35 & 900 &   7   & 88       \\
2020-05-25 & 35 & 300 &   5   & 63$^{*}$ \\
2020-07-23 & 35 & 900 &   8   & 63       \\
2020-07-30 & 35 & 900 &   9   & 59       \\
2020-08-03 & 35 & 900 &   10  & 63       \\
2021-05-07 & 35 & 900 &   7   & 60$^{*}$ \\
\hline
2020-08-13 & 20 & 900 &   7   & 46$^{*}$ \\ 
\hline
\end{tabular}
\caption{
    Newly reported GCOI OSIRIS observations. Epochs marked with a $^*$ have been observed with TRICK, an infrared tip-tilt sensor used in conjunction with the usual optical one.
}
\end{singlespace}
\label{tab:newoobs}
\end{table}

Our spectroscopic data set, gathered between 2006 and 2021, consists of 33 nights of OSIRIS observations, of which the most recent 7 (post-2018) are newly reported.   
We use 32 observations taken with a pixel scale of 35~mas/pixel. 
Additionally, we use one 20~mas/pixel observation to illustrate the structure of X7 with slightly higher spatial resolution.
The 35~mas/pixel observations selected from the GCOI archive are those lacking substantial residuals from telluric OH line subtraction (visible as extremely strong absorption lines).
All observations were taken at a position angle of 285~degrees with a spectral resolution of R$\sim$3800 and were obtained through the Kn3 bandpass filter (2.121-2.220$\mu$m),  which covers the hydrogen recombination line Br-$\gamma$ (rest wavelength 2.1661~$\mu$m), forbidden lines of iron [FeIII] (rest wavelengths 2.1457 and 2.2184~$\mu$m) and (near the edge of the filter) the molecular hydrogen H$_2$~1-0~S(1) line (rest wavelength 2.1218~$\mu$m).  
The new data were calibrated in the same manner as the existing GCOI OSIRIS data used for this study, which have been reported in earlier GCOI publications \citep{Ghez08, Boehle16, Chen18, Do19GR}.
The selected observations consist of calibrated and mosaicked data cubes from the GCOI Archive. 
For the 35~mas/pix observations, this resulted in a final field of view of roughly to 3''$\times$2.5'' centered on Sgr~A* and a typical angular resolution of 78~mas (evaluated on the star S0-2) while the 20~mas/pix observation has resolution of 46~mas. 
The GCOI experimental design prioritized the observation of short period stars close to Sgr~A* rather than X7, resulting in X7 being only partially within this smaller field of view.
The GCOI data-cubes are furthermore photometrically and astrometrically calibrated and stellar-continuum-subtracted by the procedures described by \citet{Ciurlo20}. 
Table~\ref{tab:newoobs} summarizes the newly reported spectroscopic observations.


\section{Methodology} 
\label{sec:methods}

\begin{figure*}[ht]
    \centering
    \includegraphics[width=8cm]{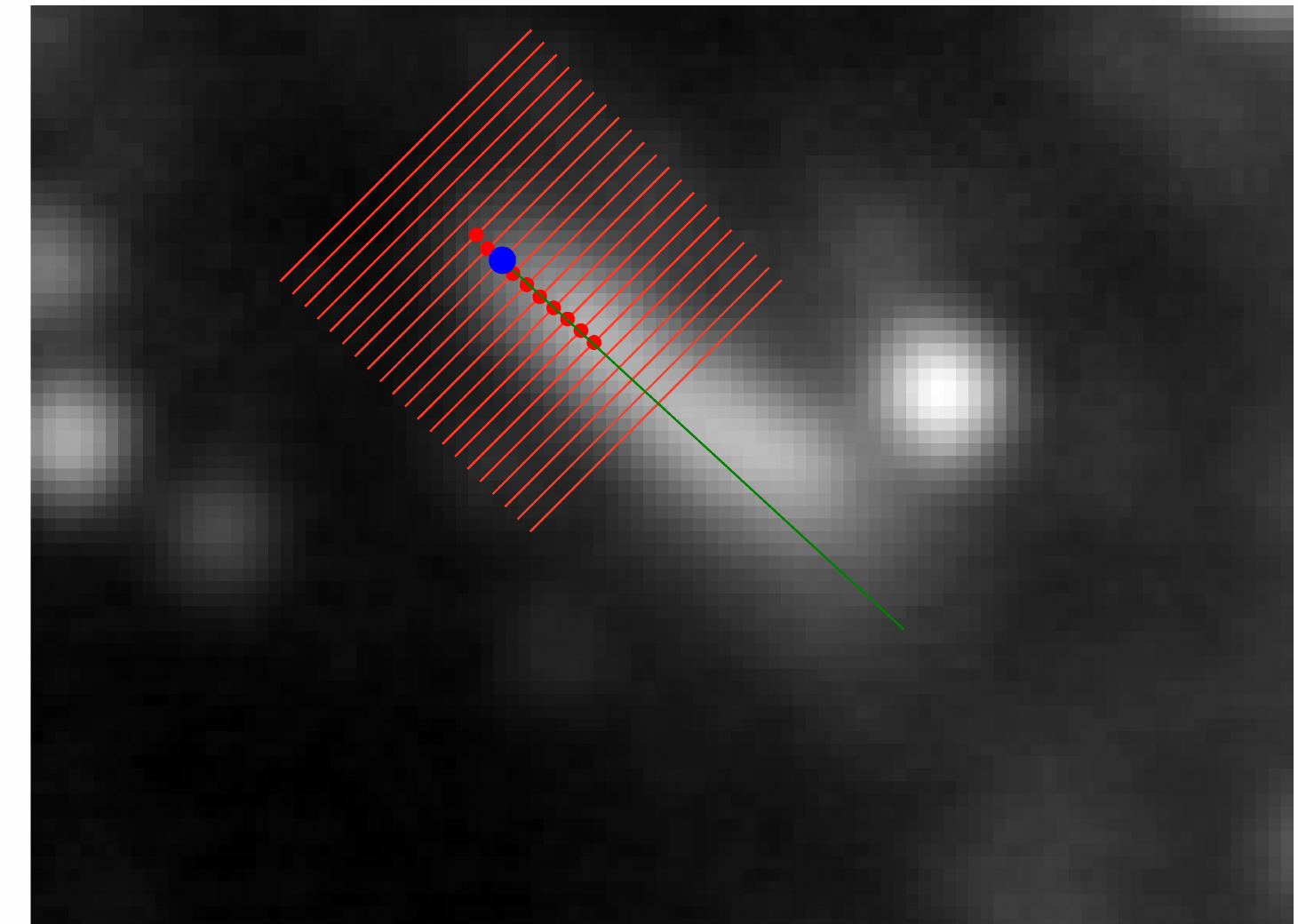}
    \includegraphics[width=8cm]{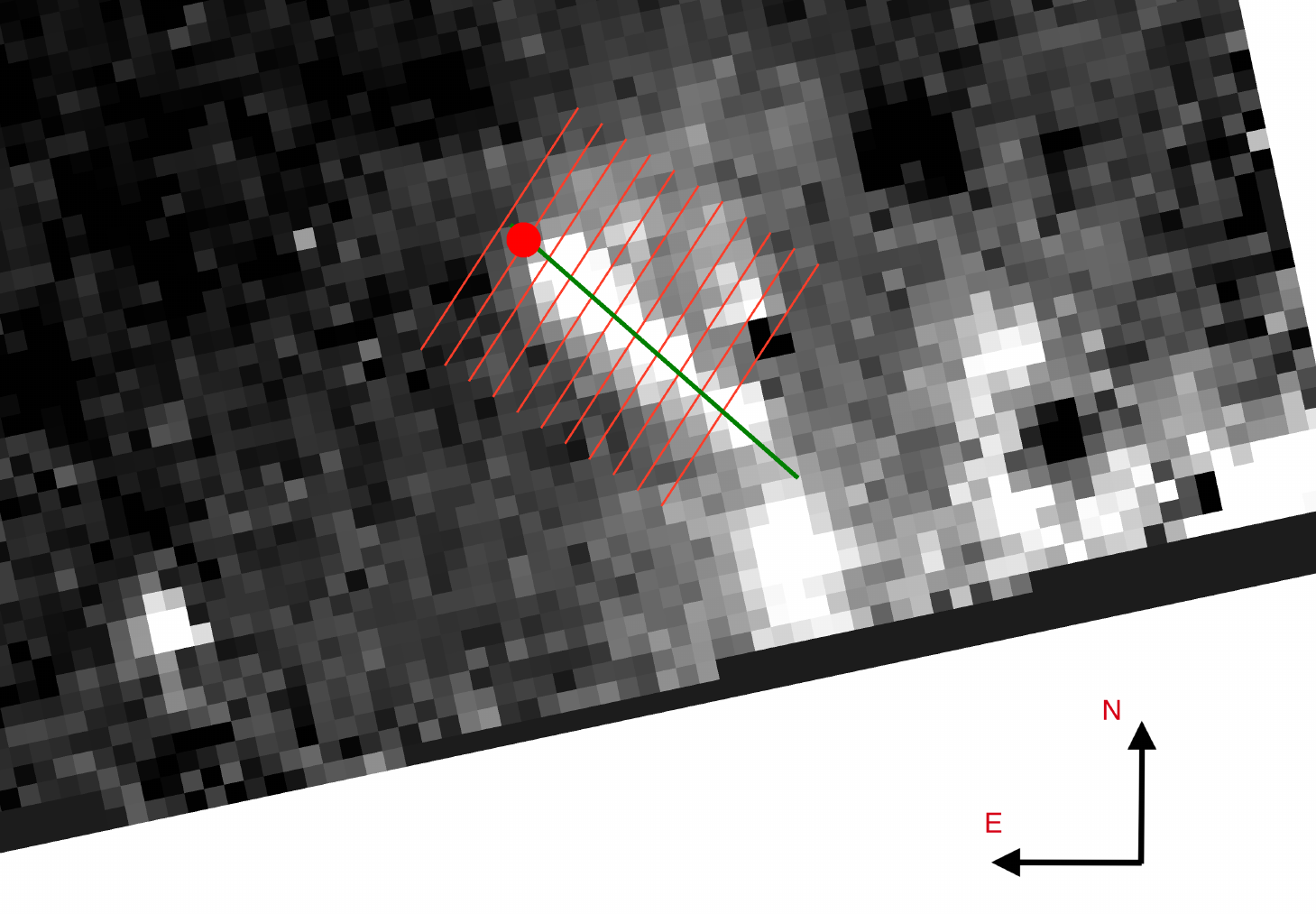}
    \includegraphics[width=8cm]{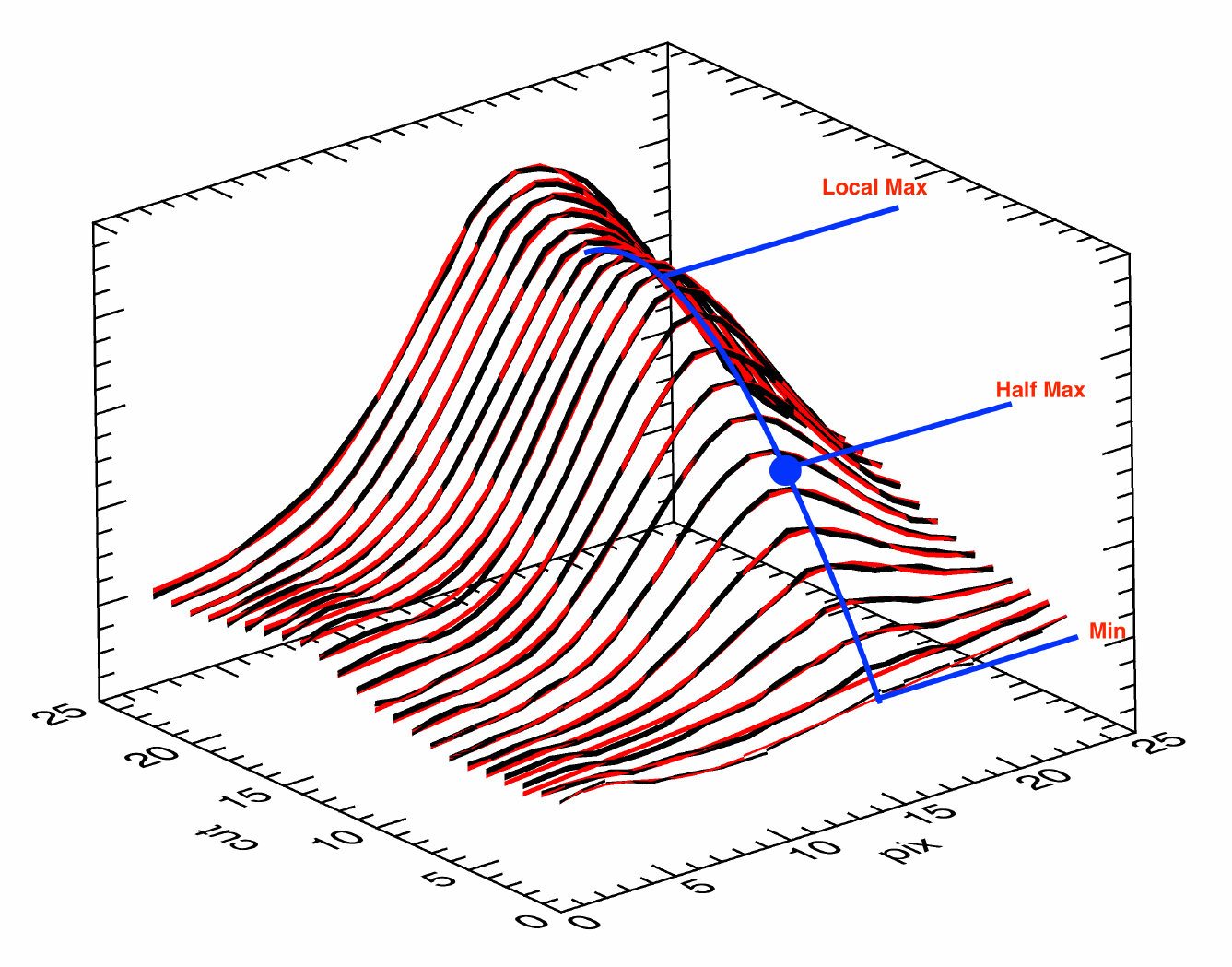}
    \includegraphics[width=8cm]{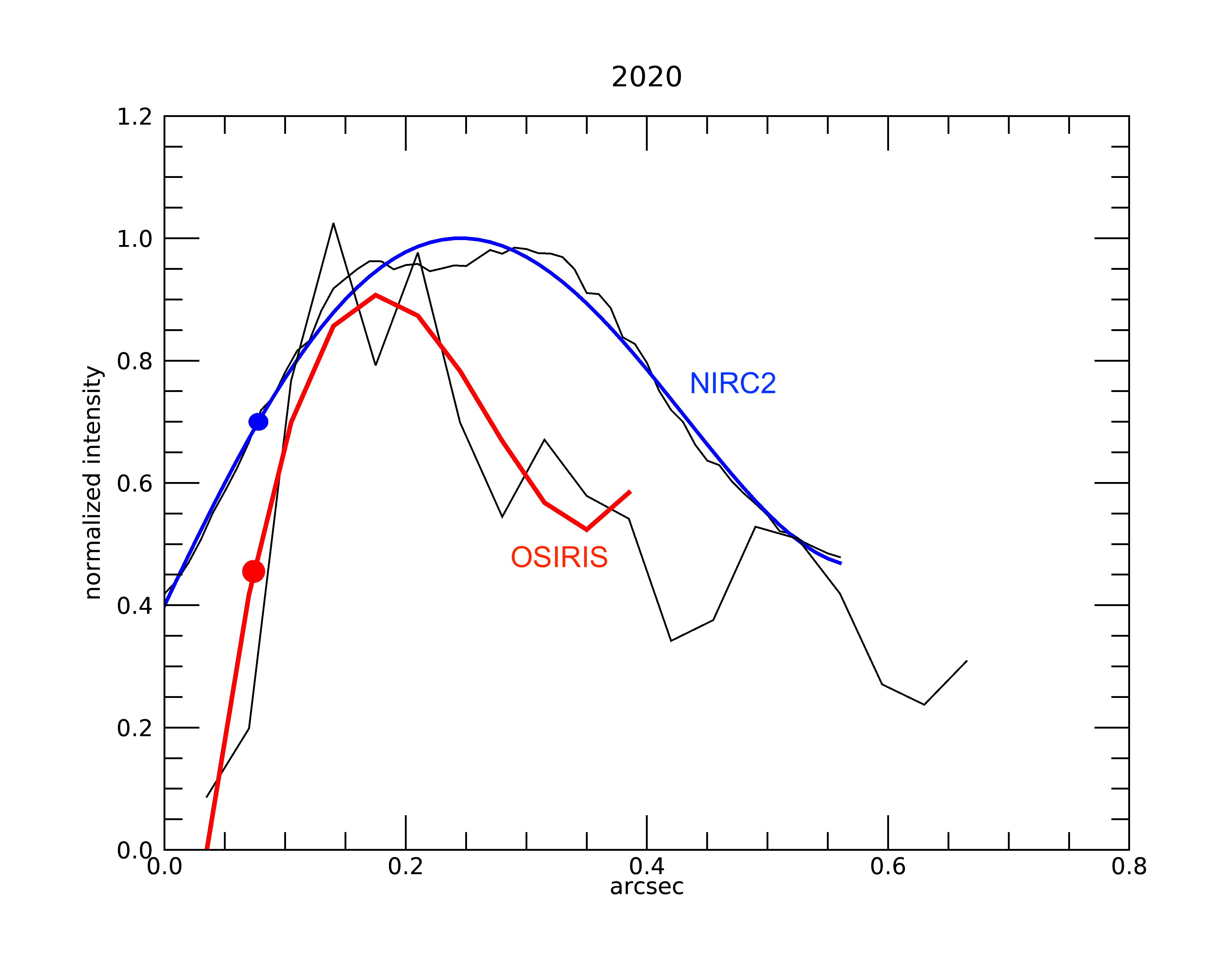}
    \caption{
    Defining the position, orientation, and profile of X7.
    {\it Top left:} NIRC2 2020 Lp image showing a series of cuts along X7 ridge. The central peak of each cut (shown as a red dot) is determined with a Gaussian. A linear fit to these peaks (green line) defines the ridge of X7 while the half-maximum along ridge profile defines the location of the tip (shown as a blue dot).
    {\it Bottom left:} 3D rendering of the NIRC2 cuts (the vertical axis represents the intensity) in black, together with the corresponding the Gaussian fit over-plotted in red and the tip measurement in blue.
    {\it Top right:} OSIRIS 2020 Br-$\gamma$ narrow-band image of X7 that highlights the series of cuts used to measure the tip location. 
    {\it Bottom right:} Comparison of X7's ridge profile in dust (NIRC2-Lp) and gas (OSIRIS Br-$\gamma$) emission (extracted along the green lines shown in the top panels). 
    A polynomial fit to each of the profiles, along with the measured location of the tip, is shown in blue for NIRC2 and in red for OSIRIS. 
    The profiles are shifted along the direction of the ridge to match the half-max points. }
    \label{fig:cuts}
\end{figure*}

Precise proper motion measurements of an evolving resolved object cannot be determined with a classical centroid method. 
Hence, we adopt an alternative method and characterize the position, orientation, and length of X7 by taking a series of nearly perpendicular line cuts, as shown in Figure~\ref{fig:cuts} (top panels). 
Through these cuts we define the northeast edge of X7's ridge, closest to the SMBH, as the ''tip'' and the rest of the ridge as the ''tail''.

For the NIRC2 data, we take cuts every 14 mas and adapt the number of cuts to the apparent lengthening of X7 in the plane of the sky: 15 cuts prior to 2011 and 21 cuts in 2011-2021.  
These cuts  cover the leading northeast half of X7's ridge (Figure~\ref{fig:cuts}, top left panel). 
Small deviations of the inclination of the cuts with respect to the ridge do not have a significant impact on our measurements.   
At this stage, we avoid the southwest bottom half because the early flared appearance of the tail and overlapping surrounding features bias our measurements.
The profile of each cut is fit by a Gaussian (red curves in Figure~\ref{fig:cuts}, bottom left panel). 
In turn, the peaks obtained through the Gaussian fit (5 of them pre-2011, 7 onward, represented as red dots in Figure~\ref{fig:cuts}, top left panel) are fit with a straight line (green line in Figure~\ref{fig:cuts}, top left panel). 
Then, the intensity profile of X7's ridge is extracted along the line. 
This ridge profile is fit to a fourth-order polynomial and the tip position is defined as the point, along the leading edge of the ridge, where the intensity is half of the ridge maximum.
This definition is arbitrary but objective, as it defines the location where X7's intensity is rising rapidly, and is clearly distinguished from the background. 
We estimate the positional error of the tip as the variance of measurements obtained by changing the measurement parameters: the lengths of the cuts, the number of cuts, starting location of cuts, and order of the polynomial for the ridge profile fit.
The line fit to the peaks is also used to find the orientation of X7 with respect to Sgr~A* (Figure~\ref{fig:cuts}, left panel).
Additionally, we define the total length of X7 as the distance between the half power points of the polynomial fit (i.e between the tip and the the half max point closest to the southwest corner where the ridge intensity starts to decrease).

We measured the Lp surface brightness at the peak intensity (which is close to the tip but encloses less background and is therefore a better determination for the surface brightness) through aperture photometry. The flux was extracted over a 0.09'' aperture radius, subtracting the local background from the median flux in an annulus of inner and outer radii 0.21 and 0.29'' respectively. 
We compare this measured flux to the one of nearby stars of know magnitude.
The variations in Lp magnitudes from year to year is only about 0.1 However, the absolute photometric error is larger, on the order of 0.5 mag (Table~\ref{tab:res}) due to high background, uncertainties in the Lp magnitude of reference stars, variation in AO performances, and the fact that we're comparing an extended source to a point source. 

\begin{figure}[ht]
    \centering
 \includegraphics[width=8cm]{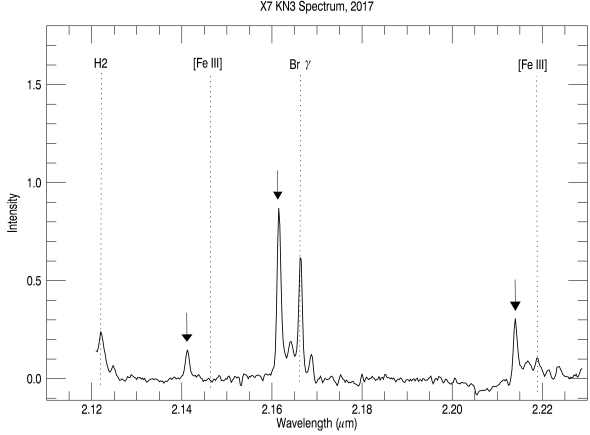}
    \caption{Spectrum of the X7 tip in 2017 with the Kn3 filter. The rest wavelengths of emission features of interest are highlighted with dotted lines whereas the blue-shifted emission associated with X7 is highlighted with arrows. 
   }
    \label{fig:fullsp}
\end{figure}

To measure the astrometry of the tip and the orientation of the tail in the OSIRIS data, we use a similar technique as that employed for NIRC2 data . However, in the case of OSIRIS we also need to construct an intensity map by selecting a slice of the cube in the wavelength dimension to isolate X7 from the rest of the emission. The 2017 OSIRIS Kn3 spectrum of the tip of X7 is shown in Figure~\ref{fig:fullsp}: we focus on the Br-$\gamma$ line which is the most prominent feature of X7’s spectrum (two emission lines of [Fe III] are also associated with X7 but are less intense).
We use an iterative process to determine both the tip location and the tip's radial velocity since they are correlated measurements: first we do a qualitative assessment of the spatial location of the peak emission, extract a spectrum over a 0.105” aperture diameter and measure the radial velocity with a Gaussian fit to the Br-$\gamma$ emission line (same procedure as described in \citealt{Ciurlo20}). Second we obtain a Br-$\gamma$ emission line map by collapsing the cube over 9 channels ($\sim$300 km/s) around the measured radial velocity. Third we use the line map to measure the tip location as explained above for the NIRC2 measurement, with the only difference being the number of cuts used to cover the same length of the ridge, due to the different spatial sampling in the two instruments (0.01” per pixel for NIRC2 versus 0.035” for OSIRIS). For OSIRIS, a total of 11 cuts were used for the orientation measurement. The position uncertainty is determined as for NIRC2, with the addition of varying the aperture size for the spectral extraction and the velocity width to use for the slice of the OSIRIS data cube. The new tip location is used to remeasure the radial velocity. 
The largest source of uncertainty to the radial velocity measurement is the location at which the spectra is extracted. Therefore, we characterize the uncertainties by varying the extraction position by 1-1.5 pixels. The standard deviation of these measurements it's added in quadrature to the statistical error which is quite small (1--2~km/s).
Additionally, we characterize the year-by-year evolution of the radial velocity gradient along the ridge (from tip to tail). To do so we pick one epoch per year and extract spectra (1.5 pixel aperture radius) on each pixel along the ridge (see Figure 6, third and sixth columns). The gradient, or slope, is determined from a linear fit to the radial velocity distribution along the ridge.

We measure the Br-$\gamma$ surface brightnesses evolution of X7 by measuring it in every epoch at the peak of intensity (as discussed for the Lp data) using a 0.0525'' aperture radius, and an annulus of inner and outer radii of 0.0875 and 0.21'', respectively, for background correction.
Additionally, the total surface brightness of the Br-$\gamma$ line can be used to estimate the density and the mass of X7 (see Section~\ref{subsec:mass}).  
We select the 2017 for this derivation since it has the smallest difference between tip and tail radial velocity. 
In order to maximize the signal-to-noise ratio, we combined all 2017 datasets (same procedure as in \citealt{Ciurlo20}).
This permits X7 emission to be easily isolated from superimposed sources when slicing the cube (we select the wavelength range 2.1594--2.1628~$\mu$m which corresponds to 14 spectral channels). 
We then create a mask to isolate X7 emission in the spatial dimensions and extract its overall spectrum. 
The total flux of X7 is then determined by fitting a Gaussian to the integrated Br-$\gamma$ line profile to measure the total flux of X7. 

All results of this analysis are reported in Table~\ref{tab:res}.

\LTcapwidth=\textwidth
\begin{longtable*}{|ccc|cc|ccccc|}
\hline
\multirow{2}{*}{Date} &  \multirow{2}{*}{Ref.} & 
                                data        & RA Dec$_{tip}$         & RV$_{tip}$  & PA    & length & RV slope  & flux$_{tip}(Br{\gamma})$ & flux$_{tip}(Lp)$\\ 
                      &       & type        & [mas]                  & [km/s]      & [deg] & [mas]  & [km/s/''] & [mJy]                & [mag]     \\ 
\hline
2002-05-31            & a     & Lp          & 425$\pm$33 -523$\pm$29 & -           & -132  & 225    & -         & -                    & 12.64$\pm$0.37  \\  
2003-06-10            & a     & Lp          & 419$\pm$34 -531$\pm$36 & -           & -127  & 243    & -         & -                    & 12.58$\pm$0.40\\
2004-06-28            & b     & Lp          & 417$\pm$46 -485$\pm$43 & -           & -129  & 242    & -         & -                    & 12.77$\pm$0.44\\
2004-07-26            & b     & Lp          & 417$\pm$36 -508$\pm$36 & -           & -129  & 201    & -         & -                    & 12.72 $\pm$0.49\\
2005-07-31            & c     & Lp          & 412$\pm$26 -484$\pm$26 & -           & -131  & 205    & -         & -                    & 12.91 $\pm$0.50\\
2006-05-21            & c     & Lp          & 422$\pm$24 -450$\pm$23 & -           & -132  & 247    & -         & -                    & 12.98$\pm$0.49 \\
2006-06-18            & d     & Br-$\gamma$ & 364$\pm$36 -385$\pm$22 & -705$\pm$27 &       & -      & +253      &  0.90$\pm$0.12       & -          \\ 
2006-06-30            & d     & Br-$\gamma$ & 418$\pm$24 -393$\pm$13 & -707$\pm$15 &       & -      & -         &  0.86$\pm$0.13       & -          \\ 
2006-07-01            & e     & Br-$\gamma$ & 391$\pm$45 -399$\pm$35 & -691$\pm$21 &       & -      & -         &  0.92$\pm$0.11       & -          \\
2008-07-25            & e     & Br-$\gamma$ & 372$\pm$21 -391$\pm$12 & -713$\pm$8  &       & -      & +244      &  0.82$\pm$0.10       & -          \\ 
2009-05-05            & e     & Br-$\gamma$ & 397$\pm$45 -391$\pm$20 & -710$\pm$5  &       & -      & +181      &  1.10$\pm$0.14       & -          \\ 
2009-07-22            & f     & Lp          & 421$\pm$27 -428$\pm$26 & -           & -131  & 273    & -         & -                    &  12.76$\pm$0.48\\
2009-05-06            & e     & Br-$\gamma$ & 393$\pm$55 -383$\pm$29 & -716$\pm$7  &       & -      & -         &  1.05$\pm$0.14       & -          \\ 
2010-05-05            & e     & Br-$\gamma$ & 389$\pm$52 -370$\pm$33 & -702$\pm$6  &       & -      & -         &  1.07$\pm$0.13       & -          \\ 
2010-05-08            & e     & Br-$\gamma$ & 404$\pm$64 -368$\pm$37 & -704$\pm$7  &       & -      & +165      &  1.07$\pm$0.14       & -          \\ 
2011-07-10            & e     & Br-$\gamma$ & 377$\pm$54 -344$\pm$39 & -694$\pm$6  &       & -      & +86       &  0.93$\pm$0.17       & -          \\ 
2012-05-16            & f     & Lp          & 428$\pm$23 -380$\pm$23 & -           & -133  & 313    & -         & -                    & 12.96$\pm$0.49\\
2012-05-17            & f     & Lp          & 426$\pm$23 -377$\pm$24 & -           & -134  & 285    & -         & -                    &  13.02$\pm$0.48\\
2012-07-22            & e     & Br-$\gamma$ & 381$\pm$55 -336$\pm$29 & -683$\pm$8  &       & -      & 48        &  0.71$\pm$0.19       & -          \\ 
2013-04-24            & f     & Lp          & 420$\pm$21 -367$\pm$21 & -           & -132  & 256    & -         & -                    &  12.83$\pm$0.47\\
2013-05-14            & e     & Br-$\gamma$ & 397$\pm$46 -322$\pm$24 & -683$\pm$5  &       & -      & -         &  1.01$\pm$0.12       & -          \\
2013-07-27            & e     & Br-$\gamma$ & 385$\pm$55 -297$\pm$29 & -682$\pm$6  &       & -      & +41       &  0.99$\pm$0.11       & -          \\
2014-03-19            & c     & Lp          & 421$\pm$24 -347$\pm$24 & -           & -128  & 240    & -         & -                    &  12.59$\pm$0.43\\
2014-03-20            & c     & Lp          & 418$\pm$19 -355$\pm$19 & -           & -132  & 243    & -         & -                    &  12.88$\pm$0.49\\
2014-05-11            & c     & Lp          & 419$\pm$21 -341$\pm$22 & -           & -131  & 284    & -         & -                    & 12.86$\pm$0.49 \\
2014-07-03            & g     & Br-$\gamma$ & 385$\pm$53 -288$\pm$36 & -670$\pm$4  &       & -      & +27       &  1.11$\pm$0.13       & -          \\
2014-07-03            & c     & Lp          & 422$\pm$23 -347$\pm$23 & -           & -128  & 246    & -         & -                    &  12.67$\pm$0.48\\
2014-08-04            & c     & Lp          & 421$\pm$21 -335$\pm$22 & -           & -131  & 260    & -         & -                    &  12.92$\pm$0.49\\
2015-03-31            & c     & Lp          & 414$\pm$22 -327$\pm$23 & -           & -131  & 276    & -         & -                    &  12.89$\pm$0.49\\
2015-05-04            & g     & Br-$\gamma$ & 418$\pm$69 -317$\pm$43 & -663$\pm$7  &       & -      & -         &  0.92$\pm$0.14       & -          \\
2015-07-21            & g     & Br-$\gamma$ & 362$\pm$55 -268$\pm$29 & -661$\pm$4  &       & -      & -11       &  1.29$\pm$0.15       & -          \\
2016-05-17            & c     & Lp          & 410$\pm$19 -293$\pm$20 & -           & -129  & 309    & -         & -                    & 12.95 $\pm$0.50\\
2017-05-17            & h     & Br-$\gamma$ & 374$\pm$52 -238$\pm$33 & -626$\pm$3  &       & -      & -         &  1.09$\pm$0.13       & -          \\ 
2017-05-18            & h     & Br-$\gamma$ & 371$\pm$55 -225$\pm$29 & -624$\pm$3  &       & -      & -         &  1.06$\pm$0.12       & -          \\
2017-05-19            & h     & Br-$\gamma$ & 345$\pm$69 -226$\pm$43 & -624$\pm$6  &       & -      & -         &  0.88$\pm$0.18       & -          \\ 
2017-07-16            & new   & Lp          & 401$\pm$19 -261$\pm$20 & -           & -132  & 363    & -         & -                    &  12.77$\pm$0.46\\
2017-07-19            & h     & Br-$\gamma$ & 368$\pm$68 -222$\pm$36 & -622$\pm$4  &       & -      & -55       &  1.02$\pm$0.12       & -          \\ 
2017-07-27            & h     & Br-$\gamma$ & 365$\pm$59 -204$\pm$31 & -623$\pm$5  &       & -      & -         &  0.65$\pm$0.21       & -          \\ 
2017-08-14            & h     & Br-$\gamma$ & 349$\pm$61 -219$\pm$38 & -621$\pm$3  &       & -      & -         &  1.45$\pm$0.11       & -          \\ 
2018-04-24            & h     & Br-$\gamma$ & 357$\pm$64 -193$\pm$37 & -611$\pm$7  &       & -      & -         &  0.82$\pm$0.11       & -          \\
2018-05-23            & h     & Br-$\gamma$ & 362$\pm$64 -211$\pm$37 & -609$\pm$4  &       & -      & -         &  0.89$\pm$0.10       & -          \\ 
2018-07-22            & h     & Br-$\gamma$ & 349$\pm$64 -200$\pm$37 & -607$\pm$5  &       & -      & -83       &  1.00$\pm$0.12       & -          \\ 
2018-07-31            & h     & Br-$\gamma$ & 361$\pm$64 -209$\pm$37 & -605$\pm$4  &       & -      & -         &  0.77$\pm$0.18       & -          \\
2018-08-11            & h     & Br-$\gamma$ & 355$\pm$64 -202$\pm$37 & -606$\pm$5  &       & -      & -         &  1.25$\pm$0.10       & -          \\ 
2019-05-11            & new   & Br-$\gamma$ & 354$\pm$43 -183$\pm$25 & -593$\pm$4  &       & -      & -109      &  1.42$\pm$0.09       & -          \\
2019-08-14            & new   & Lp          & 395$\pm$21 -229$\pm$13 & -           & -132  & 348    & -         & -                    &  12.93$\pm$0.49\\
2020-05-25            & new   & Br-$\gamma$ & 353$\pm$64 -178$\pm$37 & -556$\pm$8  &       & -      & -         &  1.44$\pm$0.10       & -          \\
2020-07-23            & new   & Br-$\gamma$ & 349$\pm$64 -169$\pm$37 & -560$\pm$3  &       & -      & -         &  1.76$\pm$0.14       & -          \\
2020-07-30            & new   & Br-$\gamma$ & 352$\pm$64 -167$\pm$37 & -555$\pm$13 &       & -      & -183      &  1.36$\pm$0.15       & -          \\
2020-07-31            & new   & Lp          & 365$\pm$20 -202$\pm$21 & -           & -131  & 419    & -         & -                    &  12.84$\pm$0.43\\
2020-08-03            & new   & Br-$\gamma$ & 357$\pm$64 -167$\pm$37 & -558$\pm$4  &       & -      & -         &  1.34$\pm$0.16       & -          \\
2021-05-07            & new   & Br-$\gamma$ & 356$\pm$57 -159$\pm$36 & -546$\pm$8  &       & -      & -174      &  0.84$\pm$0.12       & -          \\
2021-07-13            & new   & Lp          & 362$\pm$16 -153$\pm$16 & -           & -133  & 438    & -         & -                    & 12.78$\pm$0.42 \\
2021-08-15            & new   & Lp          & 358$\pm$21 -179$\pm$21 & -           & -133  & 422    & -         & -                    & 12.88$\pm$0.40 \\
\hline
                      &       & average     & 387        -306        & -641        & -131  & 288    & 30        & 1.05                 & 12.8 \\
\hline 
\caption{Summary of the measured X7 properties. RA and Dec are reported as offsets from Sgr ~A* (positive offsets Westward). Previously reported GCOI observations references: $a$) \citealt{Ghez04}, $b$) \citealt{Ghez05b}, $c$) \citealt{Witzel+14}, $d$) \citealt{Ghez08}, $e$) \citealt{Boehle16}, $f$) \citealt{Phifer13}, $g$) \citealt{Chu18}, $h$) \citealt{Do19GR}.}
\label{tab:res}
\end{longtable*}

\section{Results} \label{sec:results}

\subsection{Morphology and dynamical evolution}

\begin{figure*}[ht]
    \centering
    \includegraphics[width=17cm]{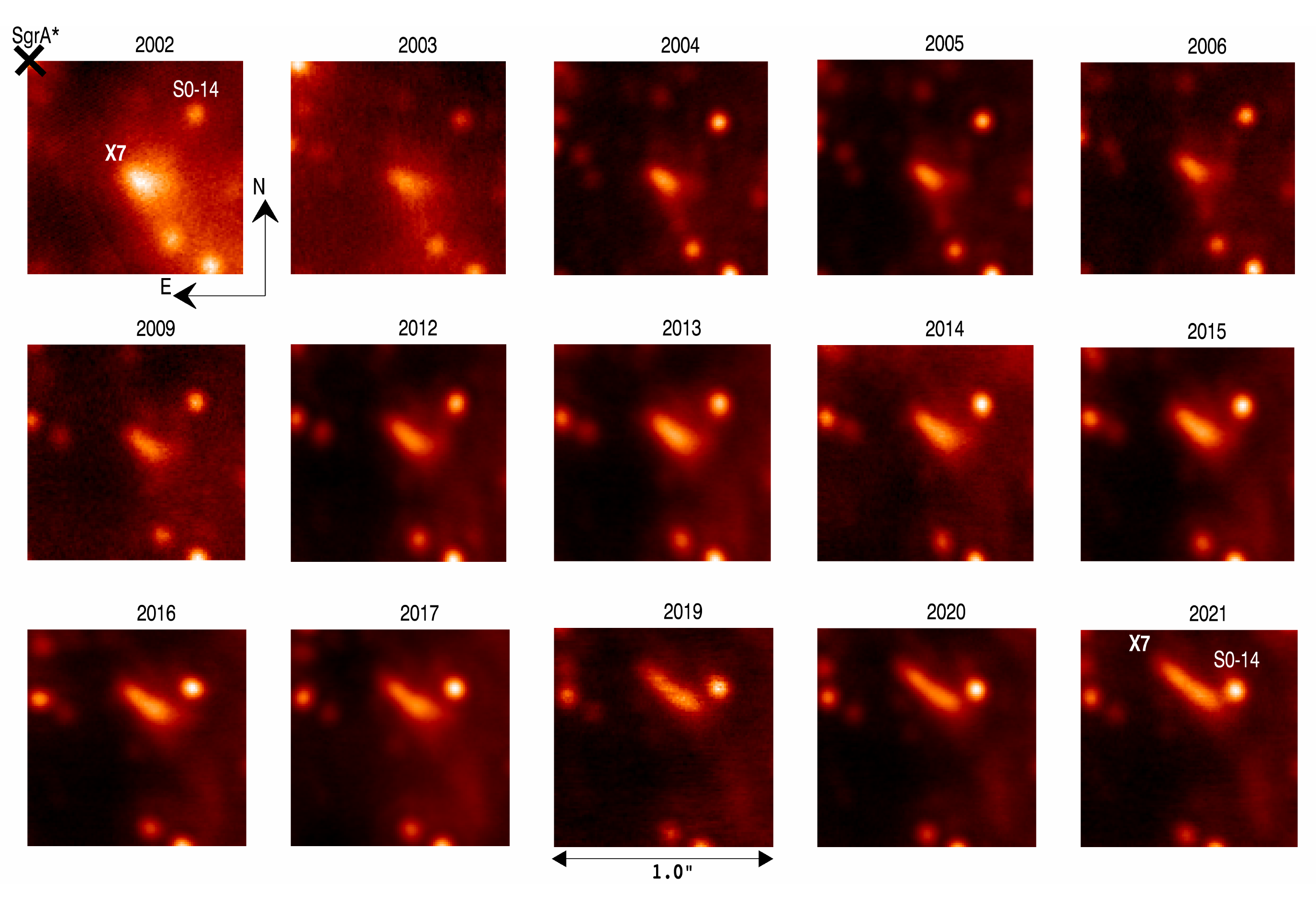}
    \caption{Morphology evolution of X7's thermal dust emission. NIRC2 Lp band (3.8 $\mu$m) images between 2002 and 2021. The images are oriented with equatorial north at the top, and with Sgr~A* positioned in the upper left corner of each panel. 
    }
    \label{fig:shapevol}
\end{figure*}
\begin{figure}[htb]
    \centering
    \includegraphics[width=8cm]{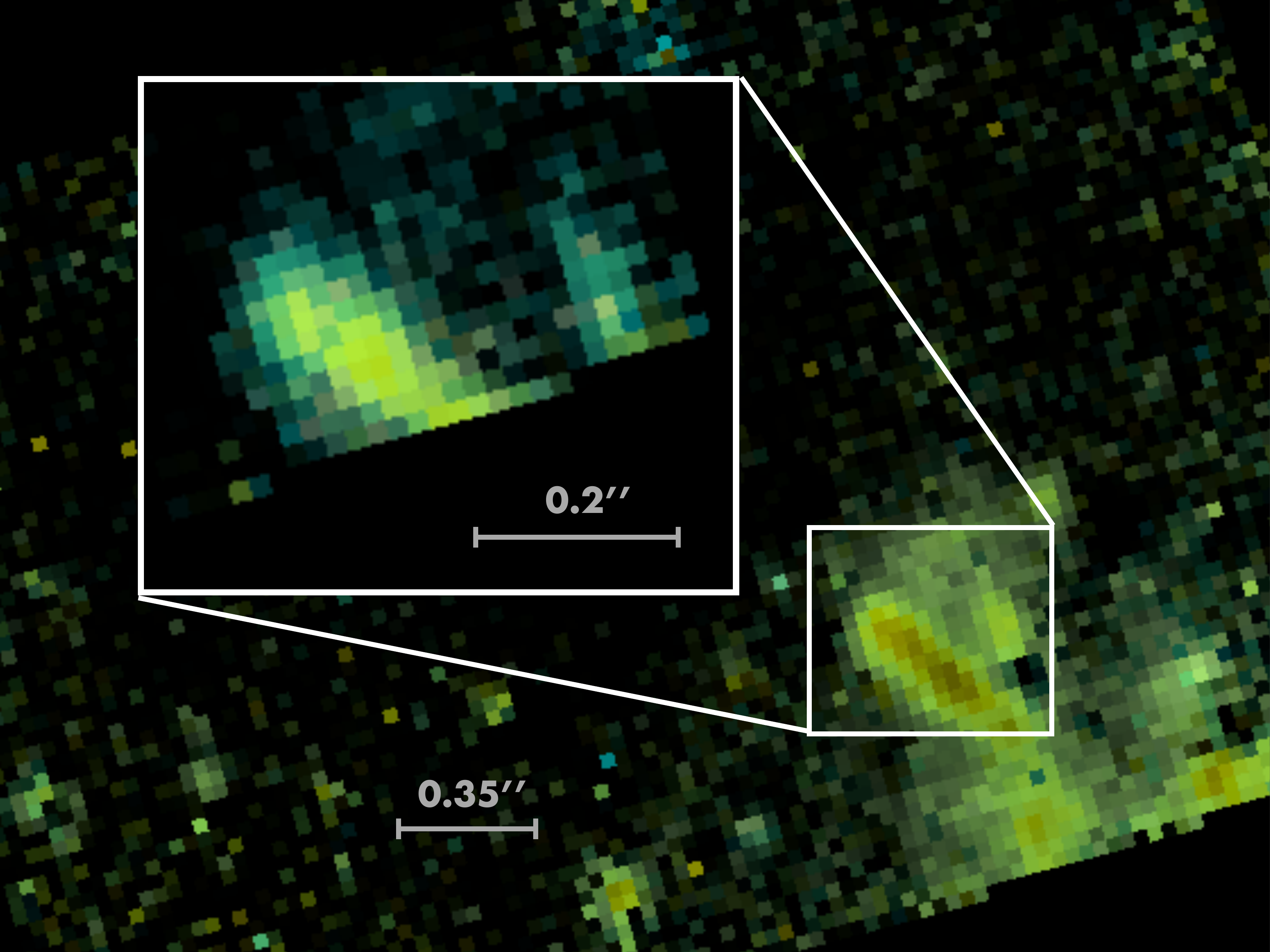}
    \caption{Zoom-in of the structure of X7 showing evidence for its fragmentation. The figure shows maps of Br-$\gamma$ emission in 2020 using 35 and 20~mas (inset at left) pixel scales, obtained by integrating over velocities in the data-cubes at which X7's blueshifted Br-$\gamma$ emission is present (corresponding to wavelengths of 2.1612 to 2.1628~$\mu$m). 
    The line maps are cut out from the data cube at velocities around that of X7. In the 20~mas map X7 is partially cropped owing to the smaller field of view.    }
    \label{fig:break}
\end{figure}

\begin{figure*}[ht]
    \centering
    \includegraphics[width=17cm]{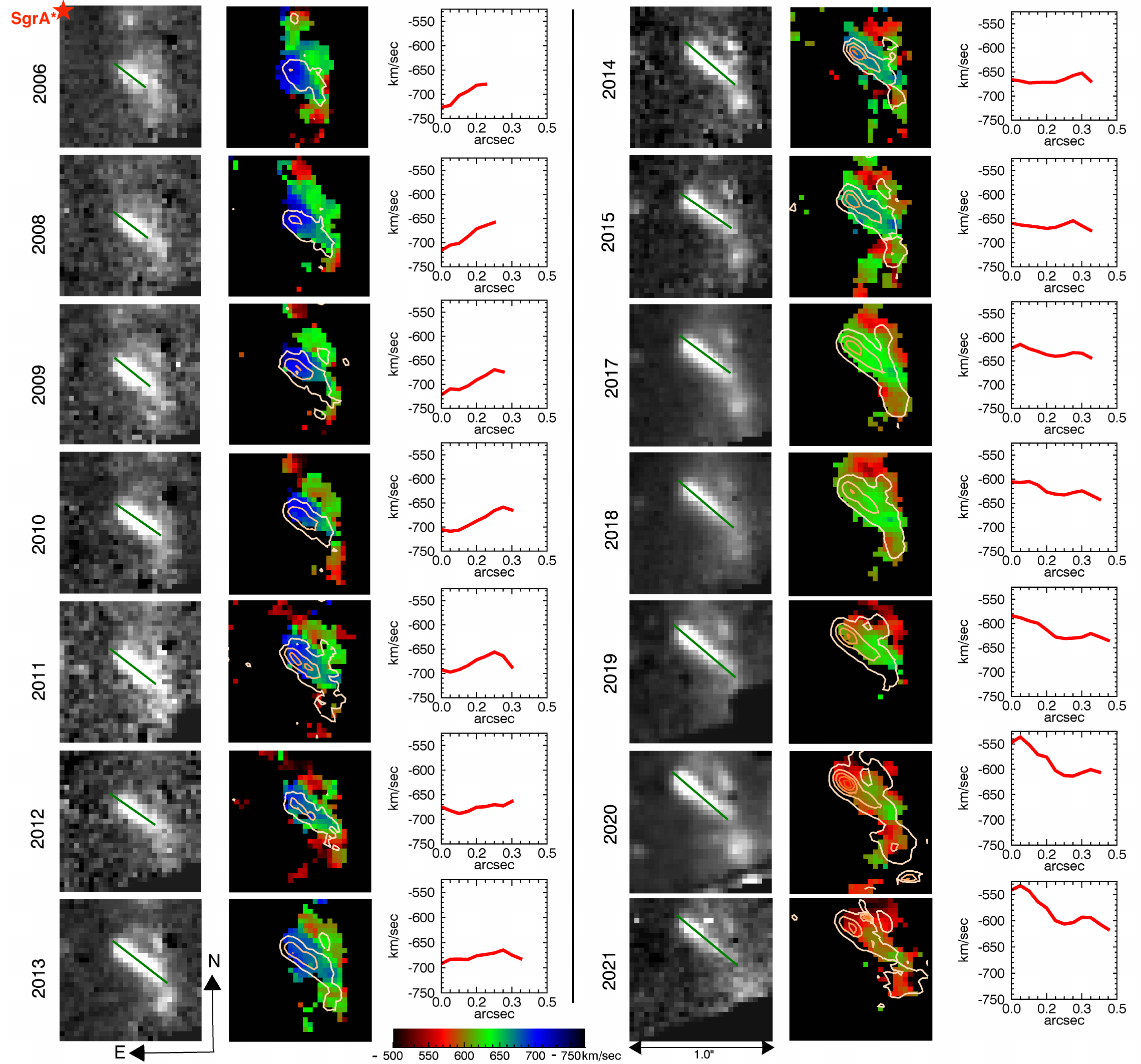}
    \caption{Morphological and dynamical evolution of X7's gas emission. 
    Br-$\gamma$ integrated intensity (left columns) and radial velocity for the peak of the local profile (center columns) obtained between 2006 and 2021.  Br-$\gamma$ total intensity contours are overlaid on the color-coded velocity maps. 
    Both maps are constructed from a narrow wavelength band centered around the blue-shifted Br-$\gamma$ emission extracted from the OSIRIS data-cubes with stellar continua removed. 
    The field shown is a 1.0$\times$1.0" area southwest of Sgr~A*, positioned so that Sgr~A* is located at the upper left corner of each panel.
    The intensity images illustrate proper motion and morphology evolution of the X7 gas emission, similarly to what Figure~\ref{fig:shapevol} shows for the dust emission. 
    The velocity maps highlight the changes in velocity structure, in particular the tip region as it becomes less blue-shifted going from blue (-725 ${\pm}$ 15 km/sec) to reddish-green (-540 ${\pm}$ 15 km/sec) in the color-coded maps over the 15 year time span. 
    The dramatic changes in the velocity structure of X7 are further demonstrated by the accompanying plots (right) of the radial velocity along the X7 ridge as a function of distance from the tip. 
    The position angle of the ridge is highlighted as a green line on the intensity maps and remains relatively constant at 52.4 $^{\circ}$ ${\pm}$ 6.7$^{\circ}$ . }
    \label{fig:rvevol}
\end{figure*}

Over our $\sim$20 years of observations, X7 has undergone significant changes in appearance, shape, velocity and position. 
\begin{itemize}
\item In the earlier epochs, X7 appeared to have a flared tail which, however,  does not persist after 2006, when the shape of X7 begins to become more linear (Figure~\ref{fig:shapevol}).  
More recently, the higher-resolution Br-$\gamma$ emission map, obtained with a smaller OSIRIS platescale (20~mas), seems to indicate that the ridge (or tail) of X7 is fragmenting (Figure~\ref{fig:break}).
\item The orientation of the ridge of X7 on the plane of the sky does not appear to have changed significantly during the period of our observations: we find that X7 maintains a constant position angle of -249$^{\circ}$ eastward from North with a root-mean-square of 1.8$^{\circ}$.
\item X7 has nearly doubled its length in the plane of the sky over 18 years, growing from 0.25" in 2003 to over 0.4" ($\sim$3300~AU) in 2021. 
\item As X7 moves across the plane of the sky and stretches in length, its internal radial velocity structure changes (see Figure~\ref{fig:rvevol}).
Over 15 years, the tip has decelerated by approximately 200 km/sec. On the other hand, the tail does not undergo such a dramatic change but still shows significant velocity evolution, decelerating by $\sim$50~km/s.
\end{itemize}
In Appendix~\ref{app:xtps} we discuss additional fainter structures apparent in the Br-$\gamma$ emission-line images. 

\subsection{Orbit of the tip}
\label{subsec:orbit}

\begin{figure*}[ht]
    \centering
    \includegraphics[width=7.7cm]{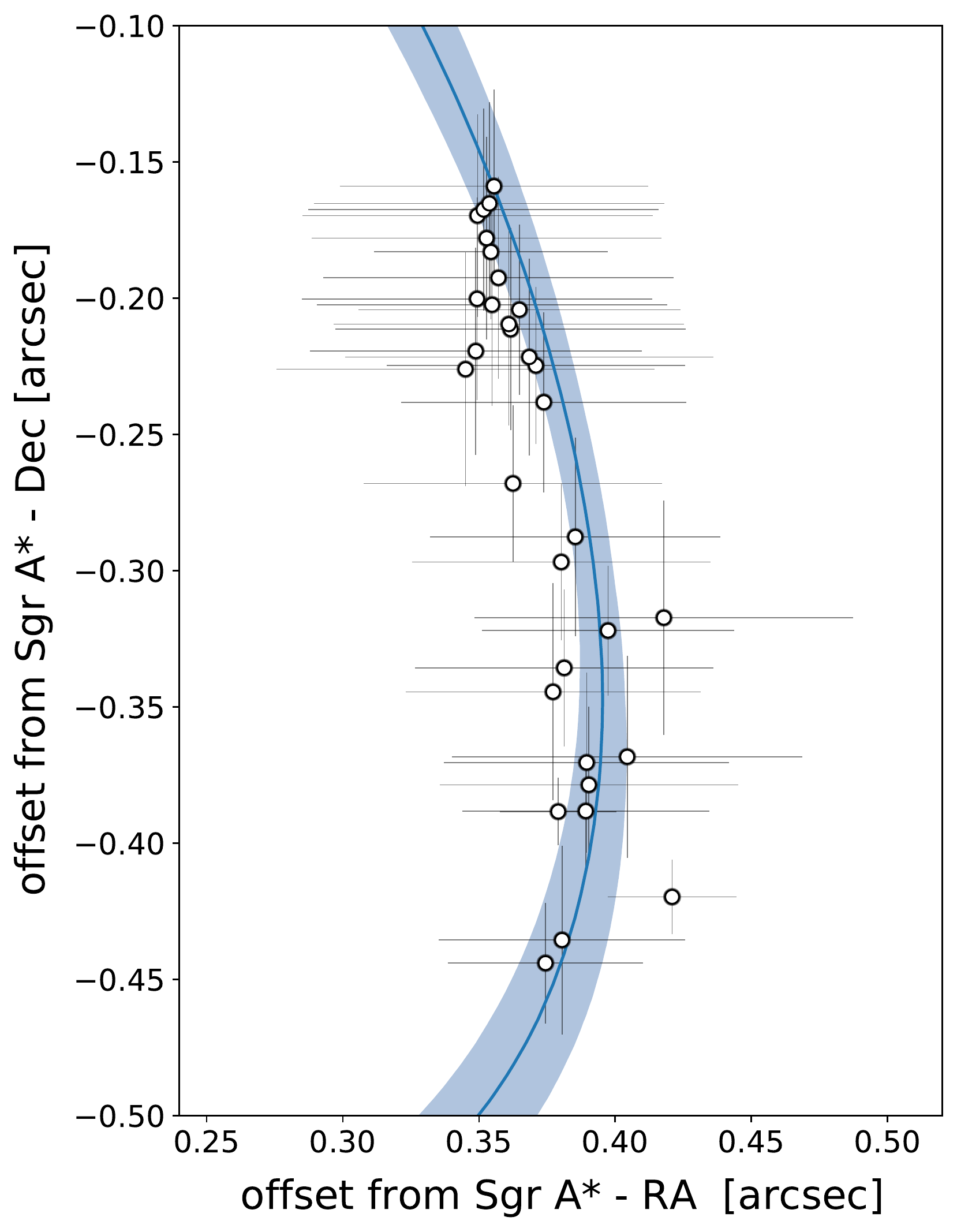}
    \includegraphics[width=7.8cm]{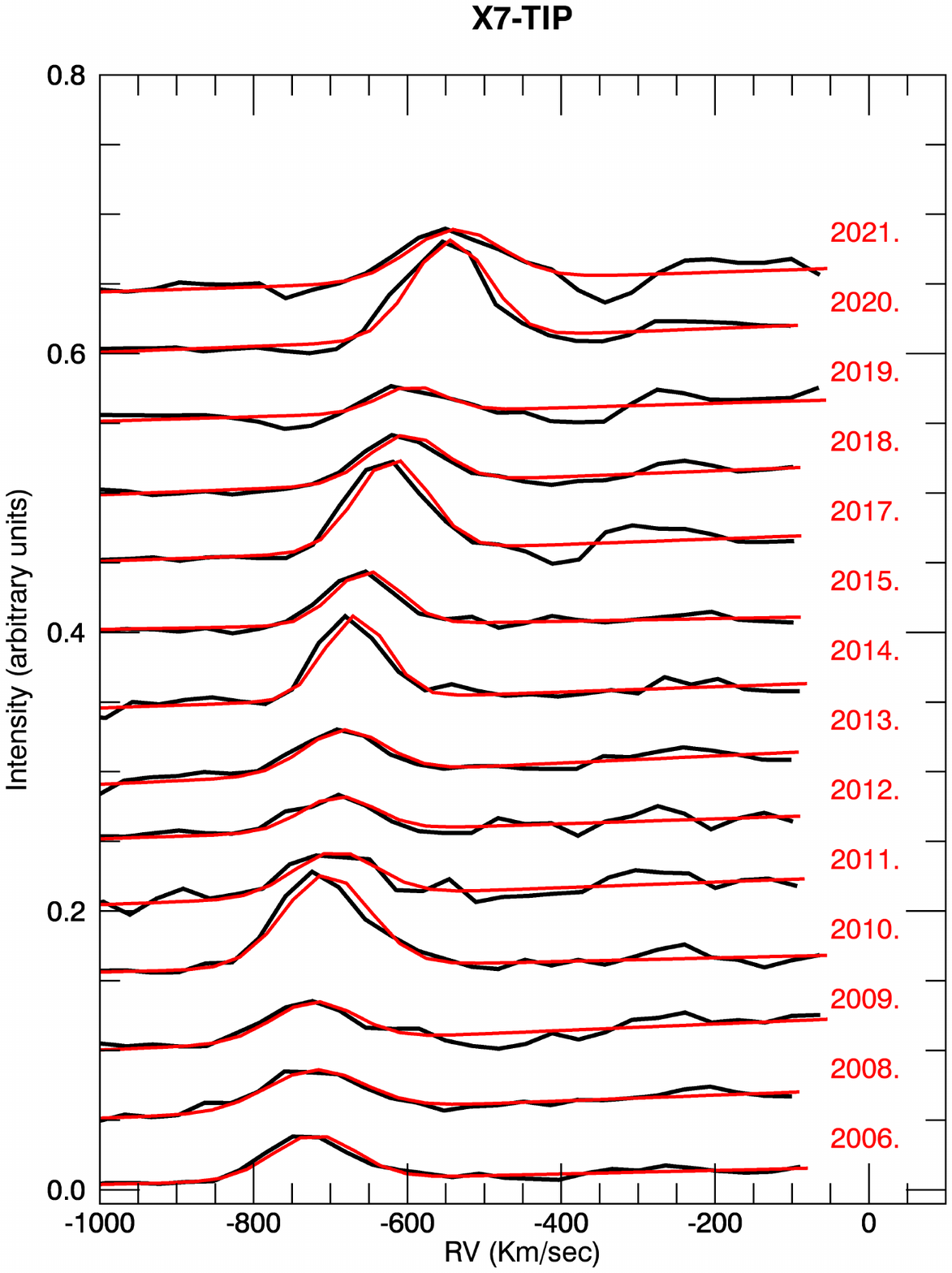}
   \includegraphics[width=5.9cm]{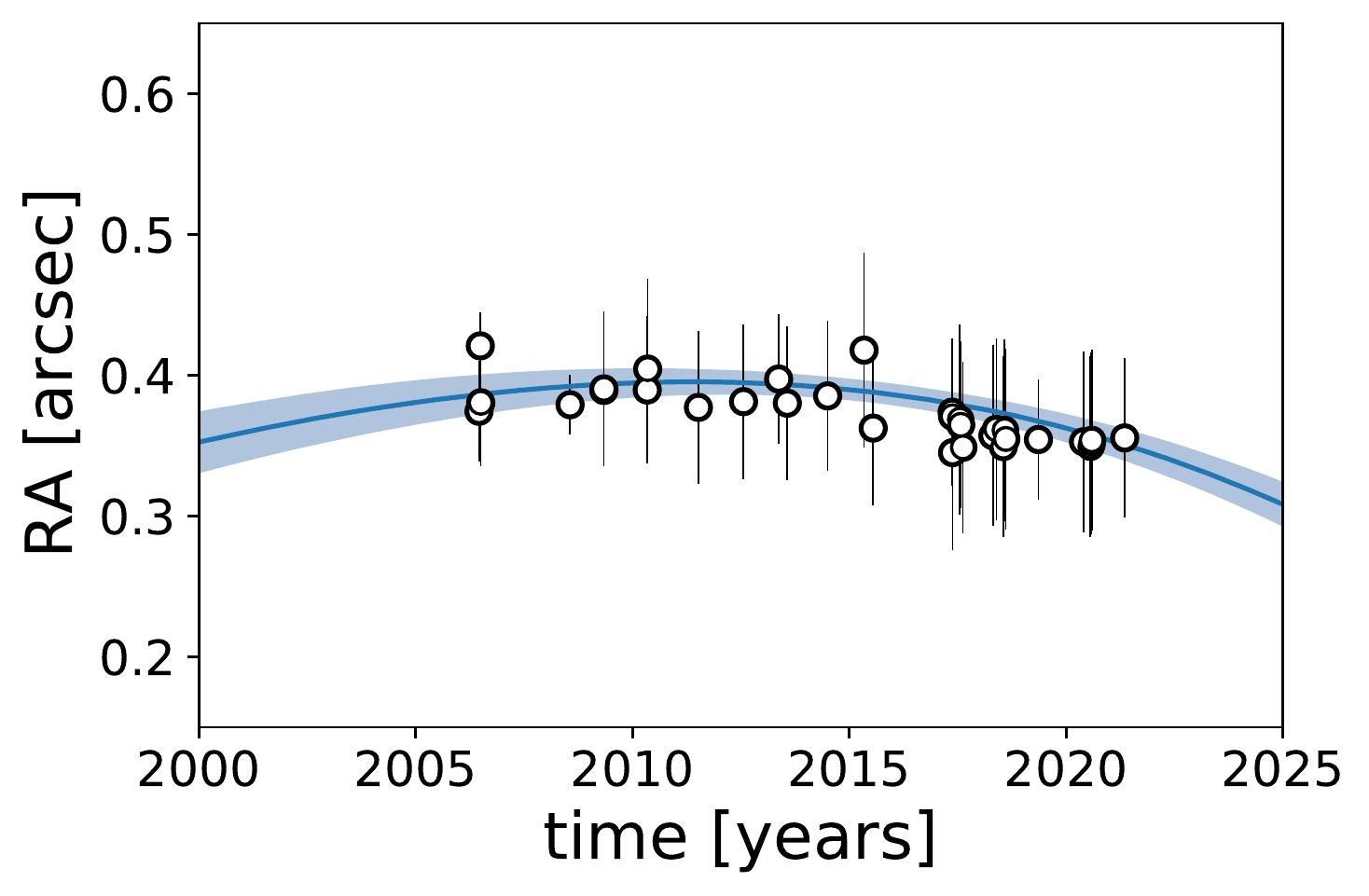}
    \includegraphics[width=5.9cm]{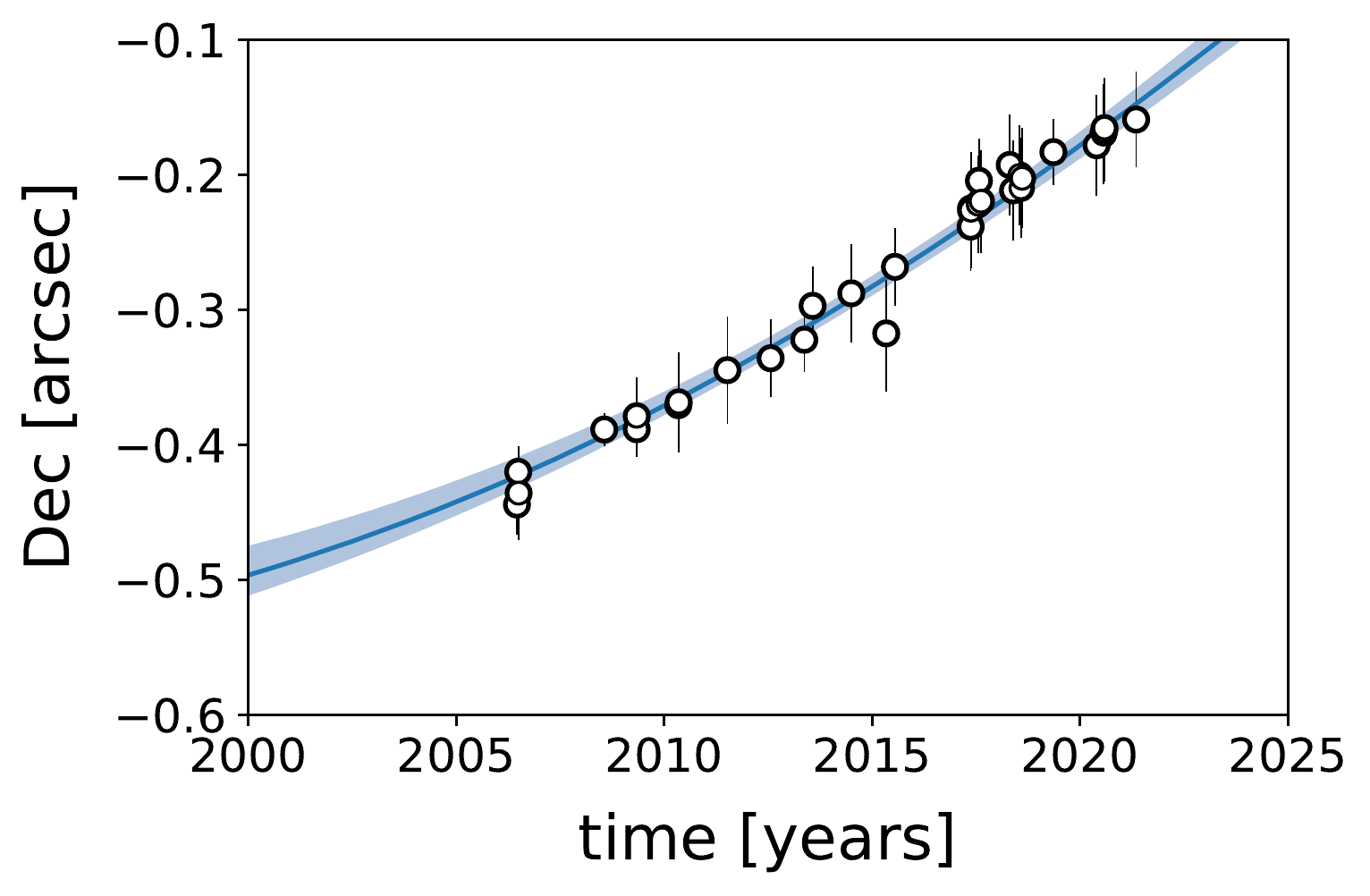}
    \includegraphics[width=5.9cm]{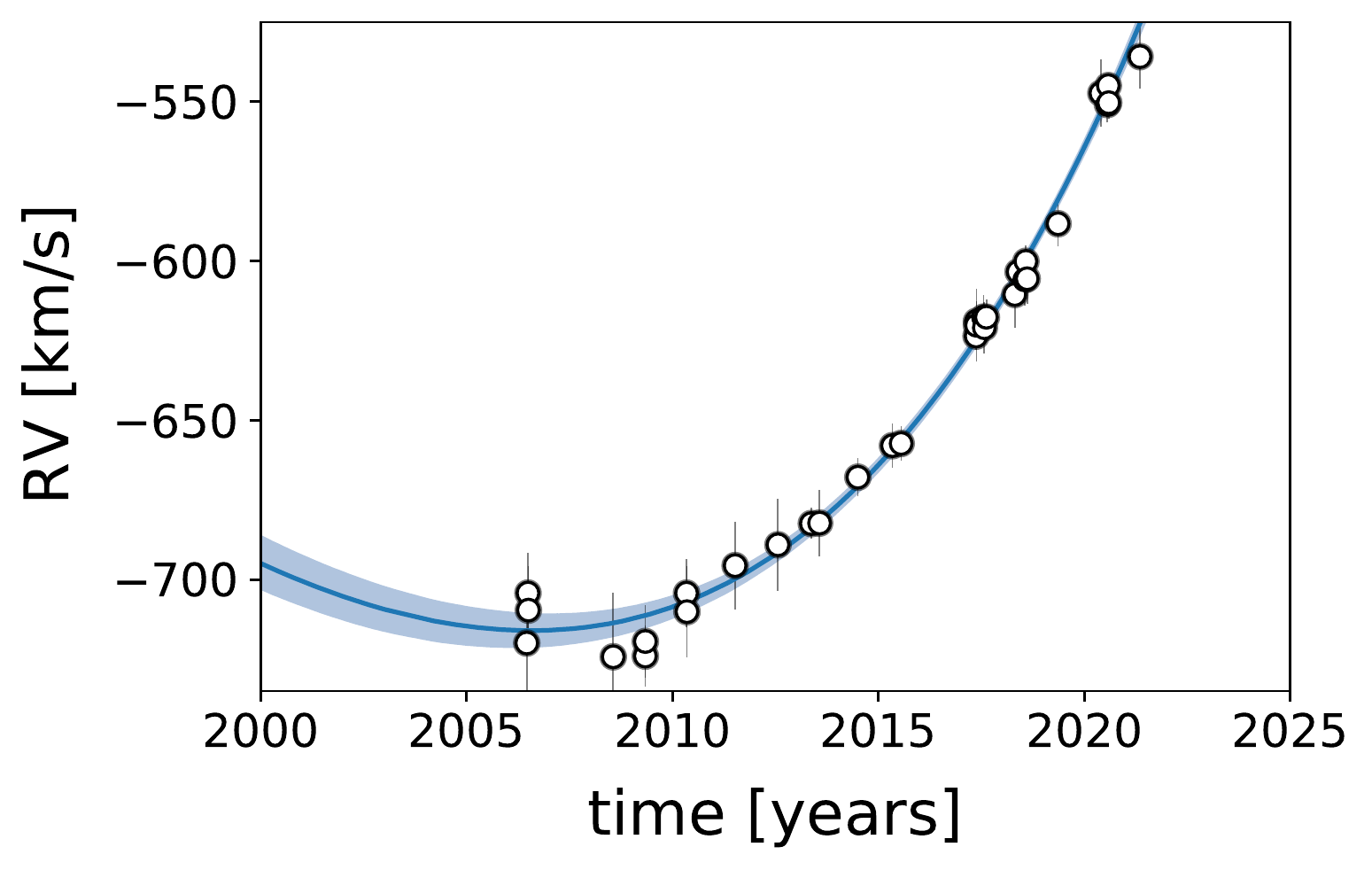}
    \caption{Orbital motion of the leading tip of X7. {\it Top left:} Orbital fit (blue line, with 1-$\sigma$ error envelope) superimposed on the astrometric measurements of Br-$\gamma$ emission (empty circles, obtained with OSIRIS). 
    {\it Top right:} evolution of the radial velocity of the tip of X7, illustrated with a sub-sample of one tip spectrum per year.  The blue-shifted Br-$\gamma$ emission line is fitted with a Gaussian profile (red curve) to extract the tip's radial velocity.
    {\it Bottom:} astrometry (RA and Dec are reported as offsets from Sgr ~A* with positive offsets Westward) and radial velocity fit for the leading tip of X7. Measurements and their uncertainties (in black) are compared to the resulting orbit model (with 1-$\sigma$ error envelopes in light blue). 
   All data points are consistent with the orbit model to within 2-sigma.
    }
    \label{fig:orbplot}
\end{figure*}

We determined that X7's motion on the plane of the sky over a 19 year period is approximately 0.35$^"$, mostly northward and arcing northeastward, with a proper motion velocity of $\sim$135~km/s East and $\sim$690~km/s North. 
The radial velocity measurements show that the tip decelerated from -725 to -550~km/sec over a 15 year time span (Figure~\ref{fig:orbplot}, top-right panel).
We can now combine both the astrometric and radial velocity measurements and fit the orbit of X7's tip. 

We have two sets of astrometric measurements: 1) the gas proper motion and radial velocities measured from OSIRIS data, and 2) the dust proper motion measured from NIRC2 data. 
Both measurements agree within the uncertainties, but we observe a partial offset (more details in Appendix~\ref{app:astro}). Given the extended nature of X7, astrometric measurements are not straightforward, especially across different instruments. Therefore, we opt for the most conservative approach and use OSIRIS measurements only: OSIRIS astrometry data is self-consistent with the radial velocity measurements. 

Our analysis employs a new orbit-fitting software package developed by the GCOI, {\it NStarOrbits} (Martinez et al., in prep.). We perform a Bayesian analysis with a multi-modal nested sampling algorithm, MultiNest \citep{Feroz09} -- the same orbit-fitting methodology described in previous GCOI publications \citep[for example, see][]{Do19GR, Ciurlo20}, but implemented in a much more efficient and modifiable way with \textit{NStarOrbits} (further details on the orbit-fitting procedure are described in Appendix~\ref{app:orbfitnso}).
 In Table~\ref{tab:orbparam}, we present the weighted median and associated 68\% confidence intervals (statistical uncertainties only) for X7's orbital parameters. The orbital fit results are also illustrated in Figure~\ref{fig:orbplot}.
 For the reported orbit fit, we use observable-based priors \citep{ONeil19}, and fix the central potential parameters to the average of the latest published estimates from \cite{GRAVITY19, GRAVITY20} and \cite{Do19GR}. To confirm that this orbit fitting strategy is robust, we also test several other orbit fitting setups.
For example, as described in Appendix~\ref{app:orbfitnso}, we also fit X7 and S0-2 simultaneously. The resulting parameter estimates from the joint fit are consistent with the case we report to within 1-$\sigma$. In Appendix~\ref{app:fitNIRC2}, we show that including the NIRC2 astrometry in addition to the OSIRIS data produces compatible results within the combined uncertainties.

\begin{table}[ht]
\centering
\begin{singlespace}
\begin{tabular}{|c|c|}
\hline
Eccentricity                &  0.34  $\pm$  0.05    \\ 
Period                      &  165  $\pm$  19 years \\ 
Epoch of Periapse Passage   &  2036 $\pm$  2        \\ 
Semi-major Axis             &  4800 $\pm$  1100 AU.  \\ 
Inclination                 &  58   $\pm$  2$^{\circ}$   \\ 
Angle of Ascending Node     &  43   $\pm$  1$^{\circ}$   \\ 
Argument of Periapse        &  -76  $\pm$  9$^{\circ}$   \\ 
\hline
\end{tabular}
\caption{Keplerian orbital parameters of the X7 tip, based on OSIRIS data only. The listed numbers are median values and 68\% confidence intervals. 
Uncertainties are statistical only.}
\label{tab:orbparam}
\end{singlespace}
\end{table}

We find that X7's tip is moving toward us, in front of the plane of the sky containing Sgr~A*, and orbiting Sgr~A* with a period of nearly 200 years.  
It is currently at a distance of $\sim$4000~AU from the SMBH and will reach its closest point ($\sim$3200~AU) just before the year 2040.

\subsection{Association with nearby objects} 
\label{subsec:pmstars}

\begin{figure*}[ht]
    \centering
    \includegraphics[width=18cm]{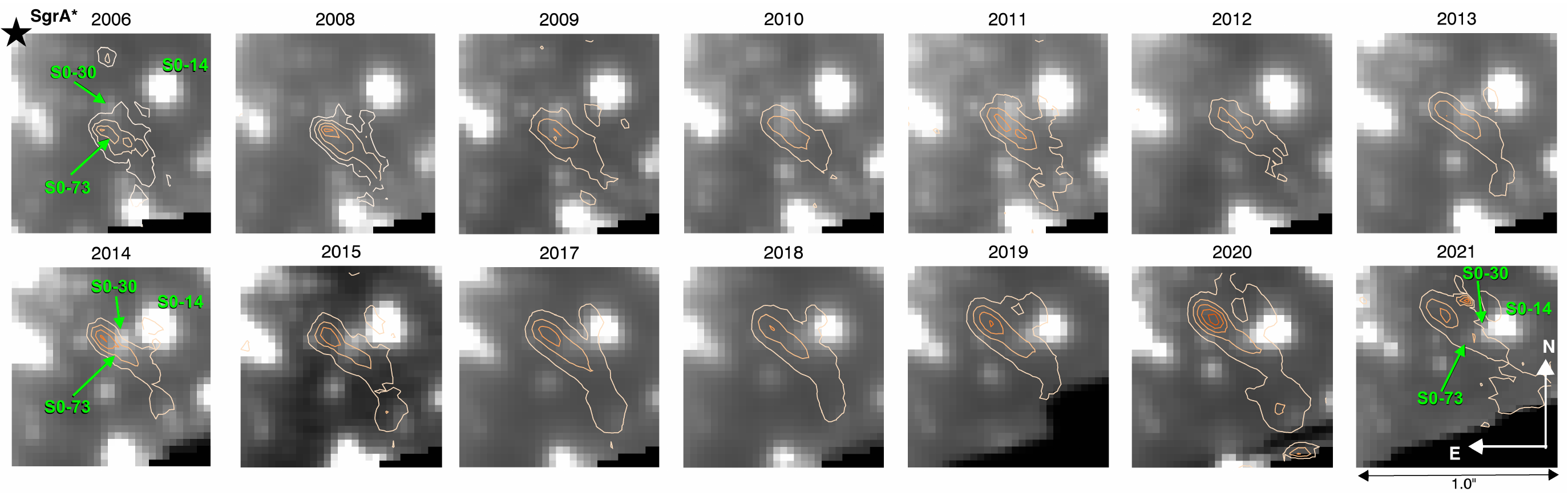}
    \includegraphics[width=8.9cm]{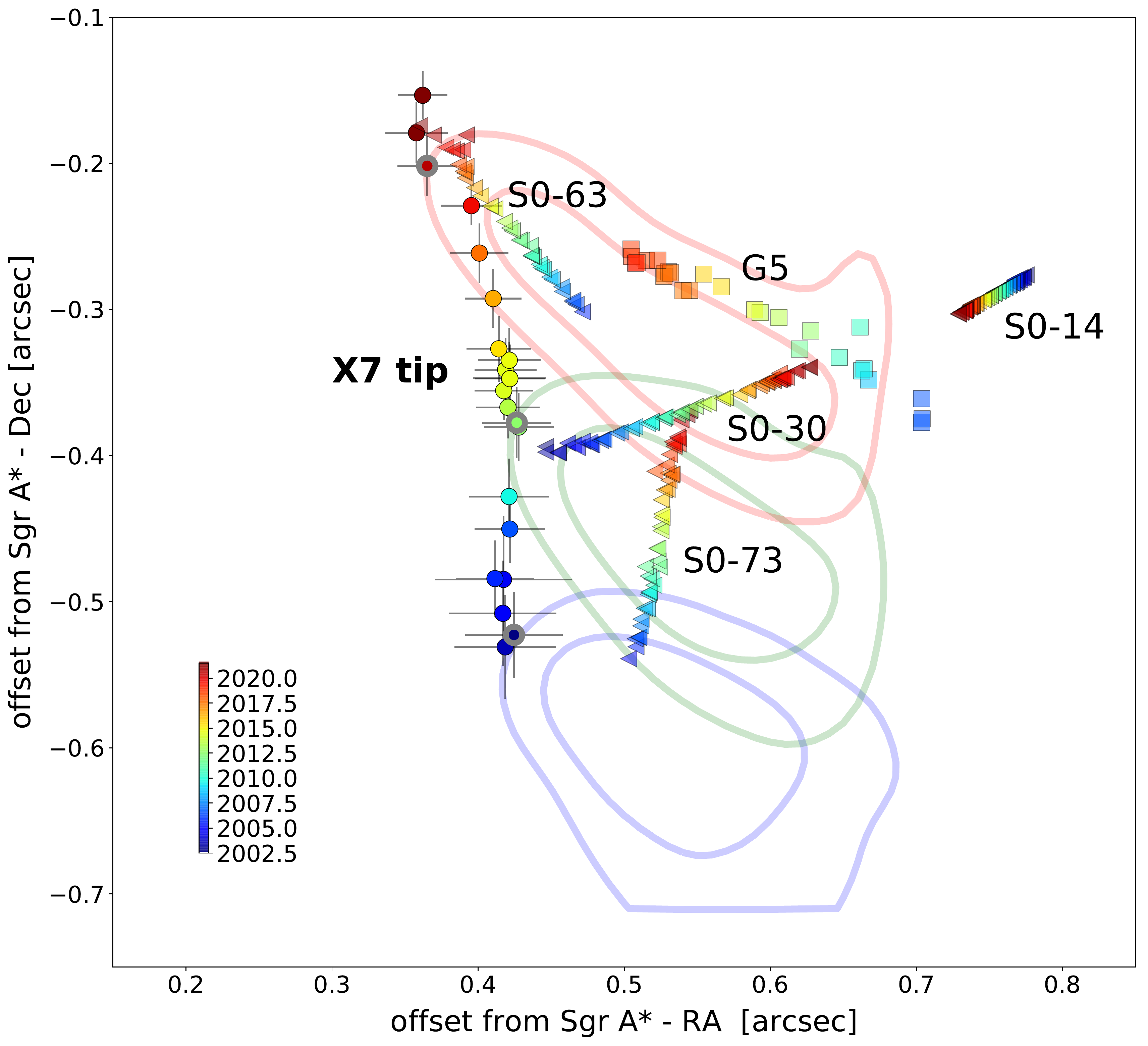}
    \includegraphics[width=7.5cm]{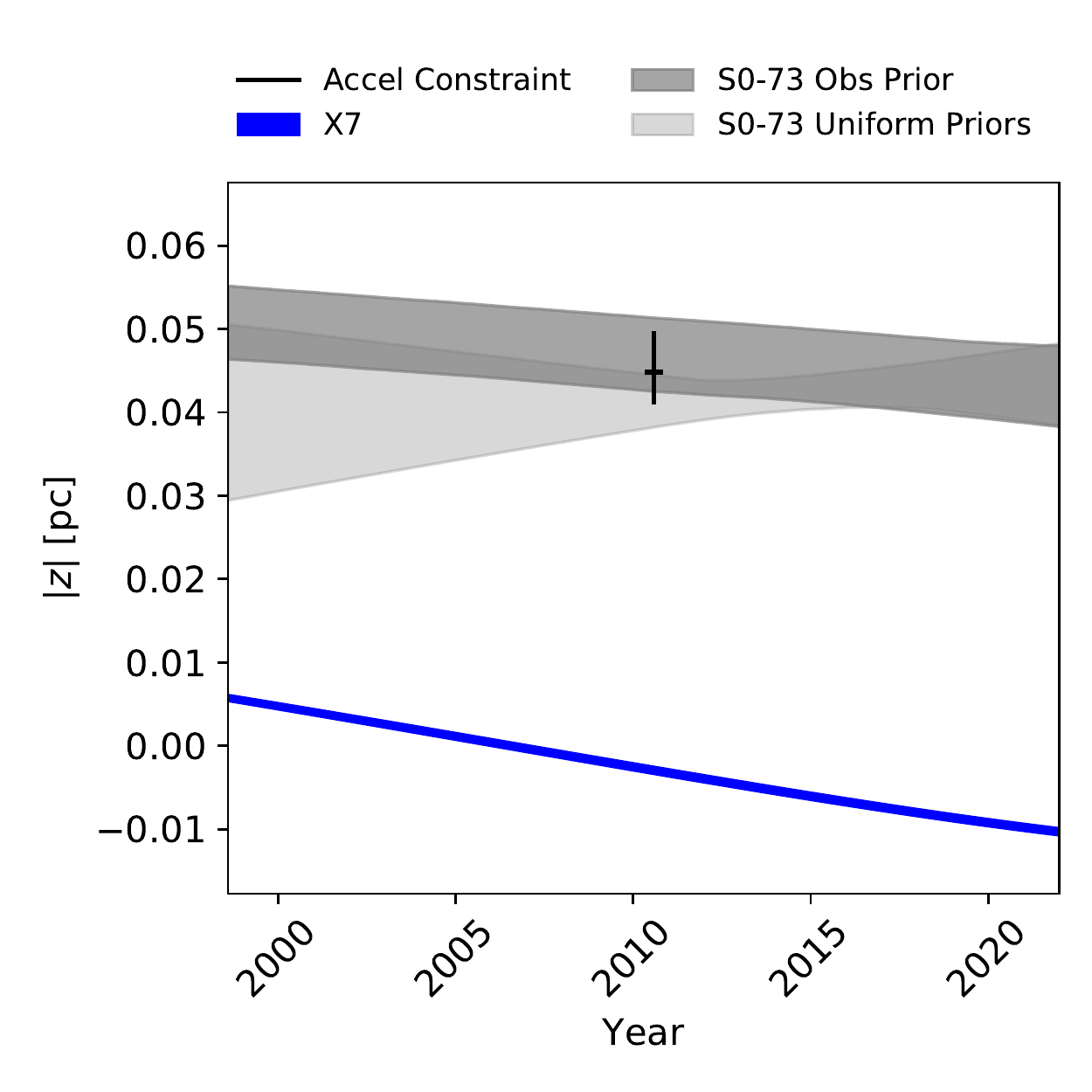}
    \caption{
    X7 is not associated with any of the nearby, detectable stellar sources that have crossed its path.
    The top panel shows contours of Br-$\gamma$ emission over-plotted onto maps of stellar continuum emission for all epochs. In this view, S0-73 and S0-30 show line-of-sight crossings with X7.
    However, the proper motions of both of these stars have different directions than that of X7, making them unlikely sources of the X7 gas filament.
    The bottom left panel shows the  proper motion measurements of the tip of X7, contours of the whole feature (obtained from NIRC2-Lp for 2003, 2012 and 2020), and the proper motions of nearby stellar objects (stars as triangles and the G object G5 as squares \citep{Do19GR, Ciurlo20}). The proper motion measurements of the tip have much larger uncertainties than those of the stars because of the extended nature of X7.  
    The bottom right panel shows the modulo of the z position of X7 and S0-73, according to their orbital models (one sigma uncertainties are shown) together with S0-73 z position inferred from our acceleration measurement. 
    It is evident from the 3D orbits that S0-73 is not coincident with X7.
    }
    \label{fig:pmstars}
\end{figure*}

In Figure~\ref{fig:pmstars}, we compare X7's proper motion and shape evolution to the trajectories of nearby objects, in order to investigate any possible association with X7.

We find that S0-73 (S50 in \citealt{Peissker21}), previously suggested to be associated with X7 \citep{Muzic10, Peissker21}, is not associated with X7. S0-73 does indeed  overlap with X7 along the line of sight in early observations, but appears to move towards the end of the tail and lags behind the northern motion of X7 in recent years. 
X7's motion and ridge inclination are both at a significantly different position angle (to the Northeast) compared to the proper motion of S0-73 (to the North-Northwest).
To confirm that the two sources do not only diverge on the plane of the sky, but also have not been coincident in three-dimensional space, we fit a polynomial to S0-73's astrometry (reduced $\chi ^2$ of 1.17). 
We find an 3-dimensional radial acceleration of 1.35$^{+0.17}_{-0.18}$~AU/yr$^2$ and no significant tangential acceleration (0.487$^{+1.685}_{-1.687}$~arcsec/yr$^2$).
We also fit a three-dimensional orbit to S0-73 astrometry (the faintness of S0-73 prevents us from extracting its spectrum and measuring its radial velocity). 
For the orbital fit, we adopt the same methodology as for X7 (Section~\ref{subsec:orbit}): we fix the black hole parameters and use an observable-based prior (but our choice of prior does not impact our final conclusions, as demonstrated in Figure~\ref{fig:pmstars}).  
We find that the z position of S0-73 inferred from polynomial and the orbital fit are compatible with each other and both constraints exclude S0-73 as being directly connected to, or responsible for, the shape and motion of X7 (Figure~\ref{fig:pmstars}, bottom-right panel).

S0-30 and S0-63 are also somewhat coincident with the ridge of X7 in some epochs, but also are unlikely sources or associates of X7 since they show different directions of motion and are not well aligned for the entire period of observation.

G5 \citep{Ciurlo20} overlaps with X7 in the most recent observations, but the two sources are completely unrelated since: 1) G5 is red-shifted (at +350~km/sec) while X7 is blue-shifted (at -600~km/sec) and 2) the two objects have completely different trajectories.
As we describe in  detail in Appendix~\ref{app:xtps}, the Br-$\gamma$ emission-line images show additional fainter structure, at least partially co-moving with X7, that might be associated with X7 (albeit difficult to trace systematically). This too is completely unrelated to G5, contrary to what has been suggested by \citet{Peissker21}, for the same reason stated above.

The area of sky that covers the central arcsecond near Sgr~A* is extremely crowded, with $\sim$30-50 detectable stars per arcsec$^{2}$.  Therefore X7, an extended feature 0.3'' in length with proper motion greater than 0.4'' across the plane of the sky over a nearly 20 year period during which it has been observed, is very likely to have temporarily coincided on the plane of the sky with several objects at various times. 

\begin{figure}[ht]
    \centering
    \includegraphics[width=8cm]{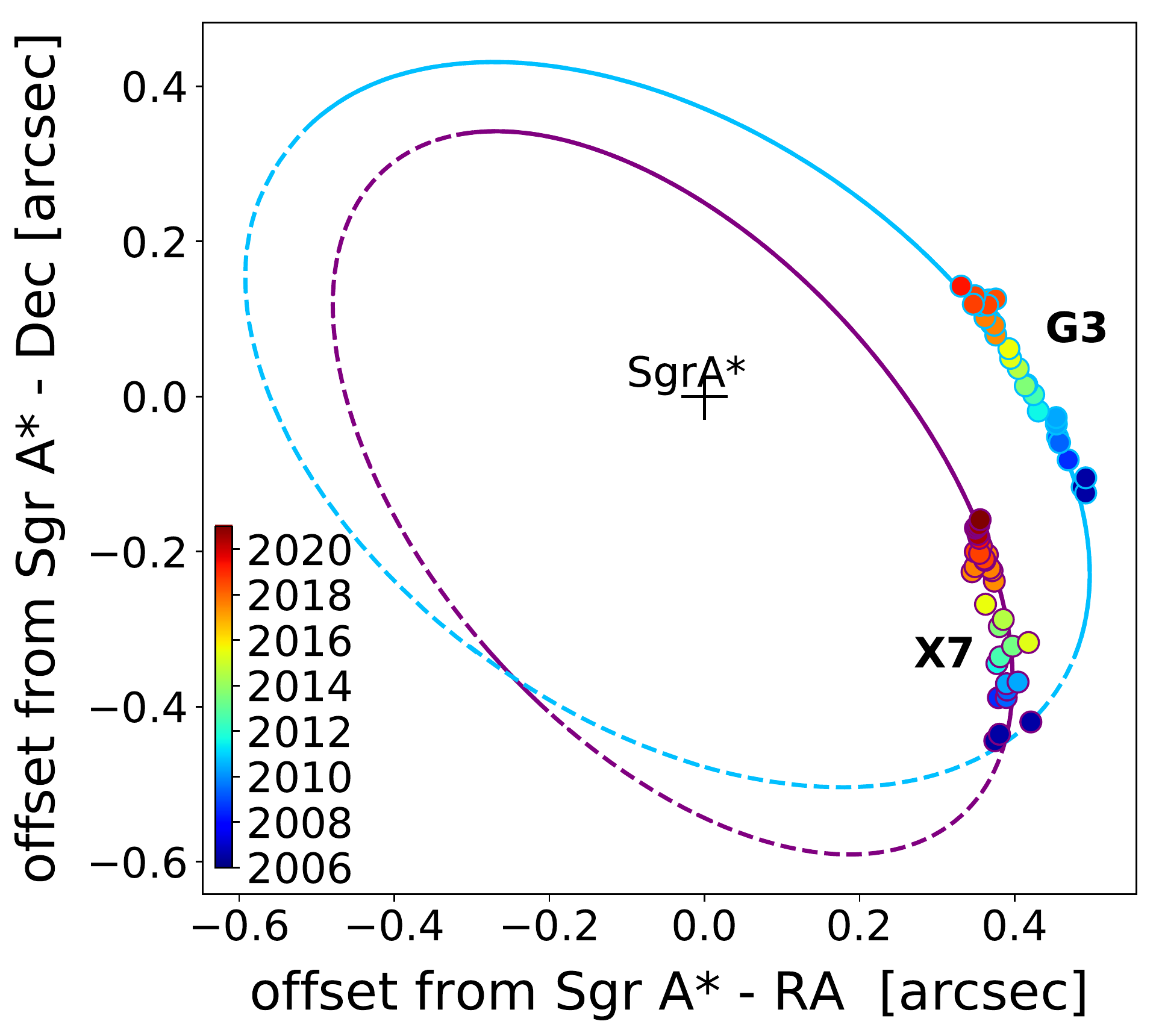}
\caption{
Orbit comparison of X7 and G3. X7 (purple curve) and G3 (blue curve) move in similar ways on the plane of the sky and have similar orbits. We find that X7's period, inclination, angle of ascending node and argument of the periapse (respectively 165 years,  58$^{\circ}$, 43$^{\circ}$and -76$^{\circ}$) are similar to those obtained in \cite{Ciurlo20} for G3 (respectively 156 years, 52$^{\circ}$, 55$^{\circ}$and -99 deg). The eccentricities are modest for X7 (0.34) and low for G3 (0.11). Additionally, both objects are in the same phase of the orbits and are both blueshifted (with a difference in radial velocity of 100~km/s in 2006).
}
    \label{fig:g3}
\end{figure}

However, there is one object that has an orbit and emission characteristics (Lp, Br$\gamma$ and [Fe~III]) that are remarkably similar to that of X7:  the dust-enshrouded object G3. Although G3's proper motion does not coincide with any portion of the ridge of X7 during the 20 years of our observations, its orbit \citep{Ciurlo20} is strikingly similar to that of X7's tip (Figure~\ref{fig:g3}). This correspondence raises the interesting possibility that X7 and G3 are dynamically linked, as we discuss in Section~\ref{subsubsec:collision}. 

Additionally, we find that X7 orbit is not oriented in the same plane and direction of clock-wise stellar disk \citep{Paumard06} nor other stellar features in the region \citep{Fellenberg22}.

\subsection{Brightness variability}
\label{subsec:variab}

We find that, in both Br-$\gamma$ and Lp, the surface brightness remains unchanged during our monitoring period, within an uncertainty of $\sim$35$\%$ for Br-$\gamma$ and $\sim$15$\%$ for Lp, with most of the uncertainty due to AO performance variations. 

This constant surface brightness in our 20 years of monitoring is somewhat surprising given the substantial stretching that X7 has undergone.
One might expect the substantial stretching of X7 to be accompanied by a progressive dimming of its surface brightness, even when one accounts for the secularly changing projection of the tail of X7 on the plane of the sky during its orbital evolution (see Section~\ref{sec:toymod}). 
Possible reasons for the constancy of the surface brightnesses are discussed further in Appendix~\ref{app:brightness}.

\subsection{Mass}
\label{subsec:mass}

The observed total Br-$\gamma$ line flux can be used to derive the density and mass of X7 (we assume that the gas emission is optically thin.).
In 2017, we find a total flux of 3.68$\times$10$^{-16}$~ergs~s$^{-1}$~cm$^{-2}$ which, after correction for extinction \citep{Fritz11}, corresponds to $F_\lambda=3.55\times10^{-15}$ ergs~s$^{-1}$~cm$^{-2}$. 
The corresponding volume emissivity can then be calculated as: 
\begin{equation}
    \epsilon = \frac{F_\lambda \cdot 4\pi R_{0}^2}{V},
\end{equation}
where $R_{0}$ is the distance to the Galactic Center (Section~\ref{subsec:orbit}) and $V$ is the emitting volume.  
In 2017, the length of X7 was $\sim$0.36'', or 2891~AU.  The width of X7 is unresolved in the 2017 OSIRIS data with the 35 mas pixel scale (measured width = 2.6$\pm$0.4~pixels), but we estimate the width of the X7 ridge by appealing to the 2020 observation that used the 20 mas pixels.  There, we find a width of 68~mas for X7 which, corrected by the resolution (the FWHM of the star S0-2, 50~mas), corresponds to an intrinsic width 48~mas, or 390~AU. Assuming a cylindrical emitting volume, we find $\epsilon\simeq$9.26$\times10^{-18}$~ergs~s$^{-1}$~cm$^{-3}$.
To estimate the corresponding density, we use the tabulations of \citet{StoreyHummer95} for emissivity as a function of density and temperature.
For the temperature, we adopt 7000~K, the typical gas temperature for Sgr A West derived by \citet{RobertsGoss93} from their radio recombination line data. 
The emissivity then yields a hydrogen number density of  $\sim$4.3$\times$10$^{4}$~cm$^{-3}$. At this density, the total hydrogen mass of X7 is 2.2$\times$10$^{29}$~gm. Correcting for an assumed 25$\%$ helium abundance by mass, this corresponds to $\sim$50~Earth masses, or $\sim$3~Neptune masses. 
Given the assumptions made this estimates is good to within about 30$\%$--50$\%$, but we underline that the order of magnitude is what is relevant and it indicates a planetary mass rather than a stellar mass.

\section{Modeling the tidal evolution}
\label{sec:toymod}

We can gain insight on how X7 is expected to behave purely under the influence of the gravity of the central black hole through a simple parsimonious model\footnote{This model is similar to the one originally used in \cite{Gillessen12} for G2 under the assumption that G2 is a pure gas cloud, except that our initial conditions have the points distributed along a line with a linear velocity gradient, whereas the points in the Gillessen et al. model are initially distributed in a spherically Gaussian fashion in phase space.} 

In this setup, we model X7 as a set of non-interacting test particles having initial conditions provided by our observations. 
We assume that the ridge of X7 is initially linear, and use its observed length and position angle in 2003 (the earliest epoch available with a good FWHM resolution), along with our determined orbit of X7's tip (Section~\ref{subsec:orbit}), to calculate the subsequent 3-dimensional vector positions of 10 points equally spaced along X7's length.  
The orbit fit for the tip provides the tip's initial 3-dimensional velocity and, in order to assign initial velocities to the 10 points along X7, we assume that the initial velocity gradient is directed along the ridge of X7, and that it increases linearly from tip to tail.
This assumption is consistent with our observation of a radial velocity gradient along the tail of X7 in the earliest epochs. 
Moreover, if X7 was created at some point in time by gas being impulsively ejected in a particular direction, then it is natural to expect that the points furthest from the tip would have the highest velocity.
In this setup, there are only two free parameters for the determination of the position and velocity vectors in our model:  
the initial angle of the ridge with respect to the line of sight, ($\theta_{z}$ in degrees), and the coefficient of the linear increase ($m$ in AU/years/dx, where $dx$ is the initial distance between points 175~AU) of the velocity along X7.

Given these prescriptions, we calculate the state vectors (3-dimensional position and velocity) of 10 non-interacting points along X7 in 2003.
Starting from these state vectors we calculate, for each point along X7, the corresponding orbital parameters \citep{Grould17}.
Given the orbital parameters, we can then predict the position and velocity vectors of each point along X7 at any given moment in time using purely Keplerian orbits.
Our method for assigning the initial state vectors of the points along X7's ridge leads to some of the points potentially following unbound orbits. 
Therefore, we ensure that our orbital determinations can encompass hyperbolic orbits. 

Provided with this model, we can vary our two free parameters ($\theta_{z}$ and $m$) to reproduce several of the observed properties of X7: the constant position angle on the plane of the sky,  the absence of an obvious curvature of the ridge, the lengthening of the ridge with time and the observed evolution of radial velocities along the ridge.
These features are illustrated in Figure~\ref{fig:rvevol}. Here, we characterize the evolution in radial velocities by extracting a spectrum at each of the 10 reference points along X7 in every epoch. 
These measured radial velocities approximately map to the modeled ones. There could be some mismatch between the two because 1) in the model we start with 10 points equally spaced in 3 dimensions whereas our observations are from points equally spaced in the plane of the sky, 2) in the model 10 points equally spaced don't necessarily maintain equal spacing and 3) given the extended and evolving nature of X7 and the absence of resolved internal structure, there is no assurance that any given point each point traces exactly the same gas from epoch to epoch. 

\begin{figure*}[ht]
     \centering
     \includegraphics[width=7.2cm]{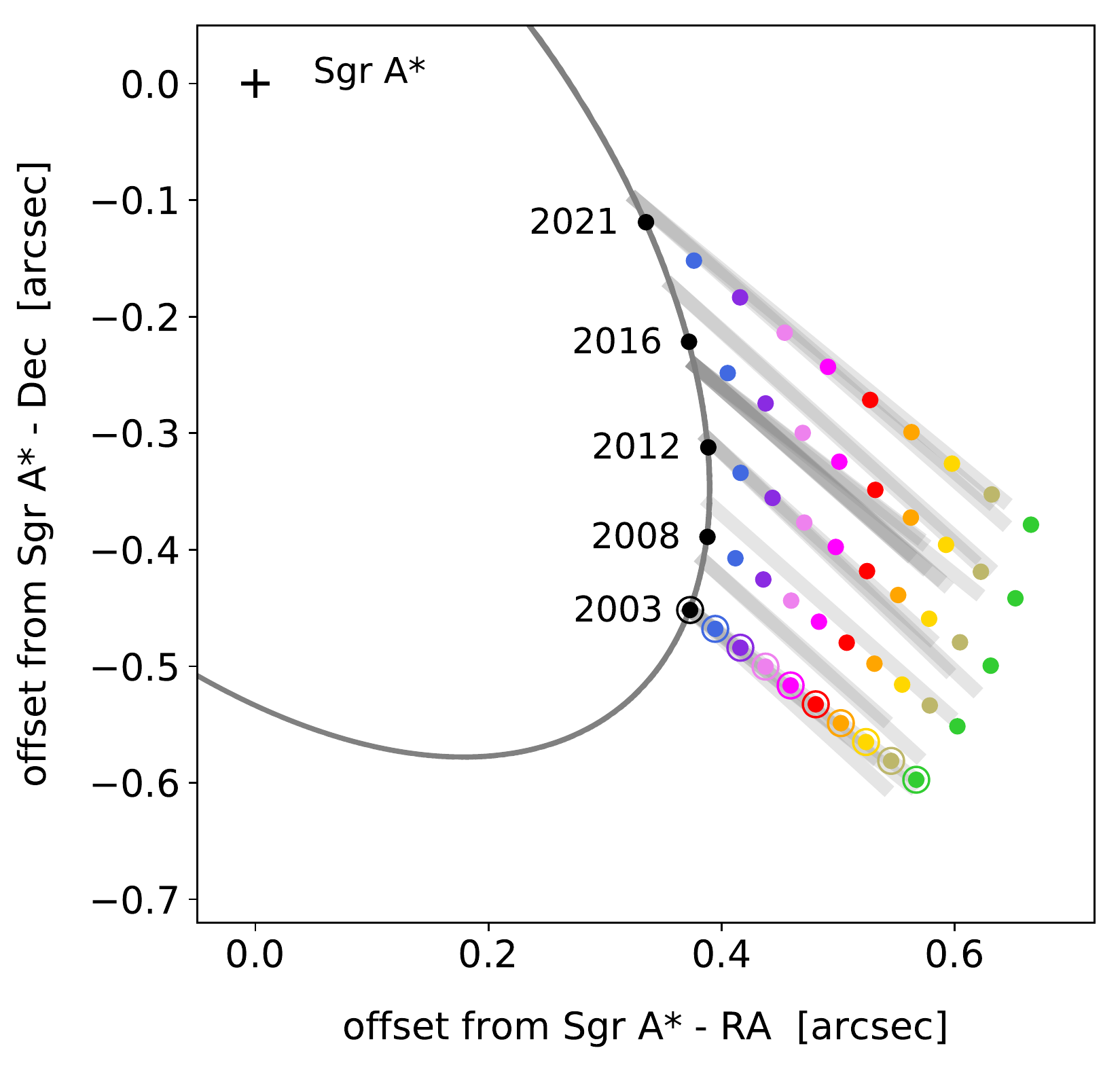}
     \includegraphics[width=10cm]{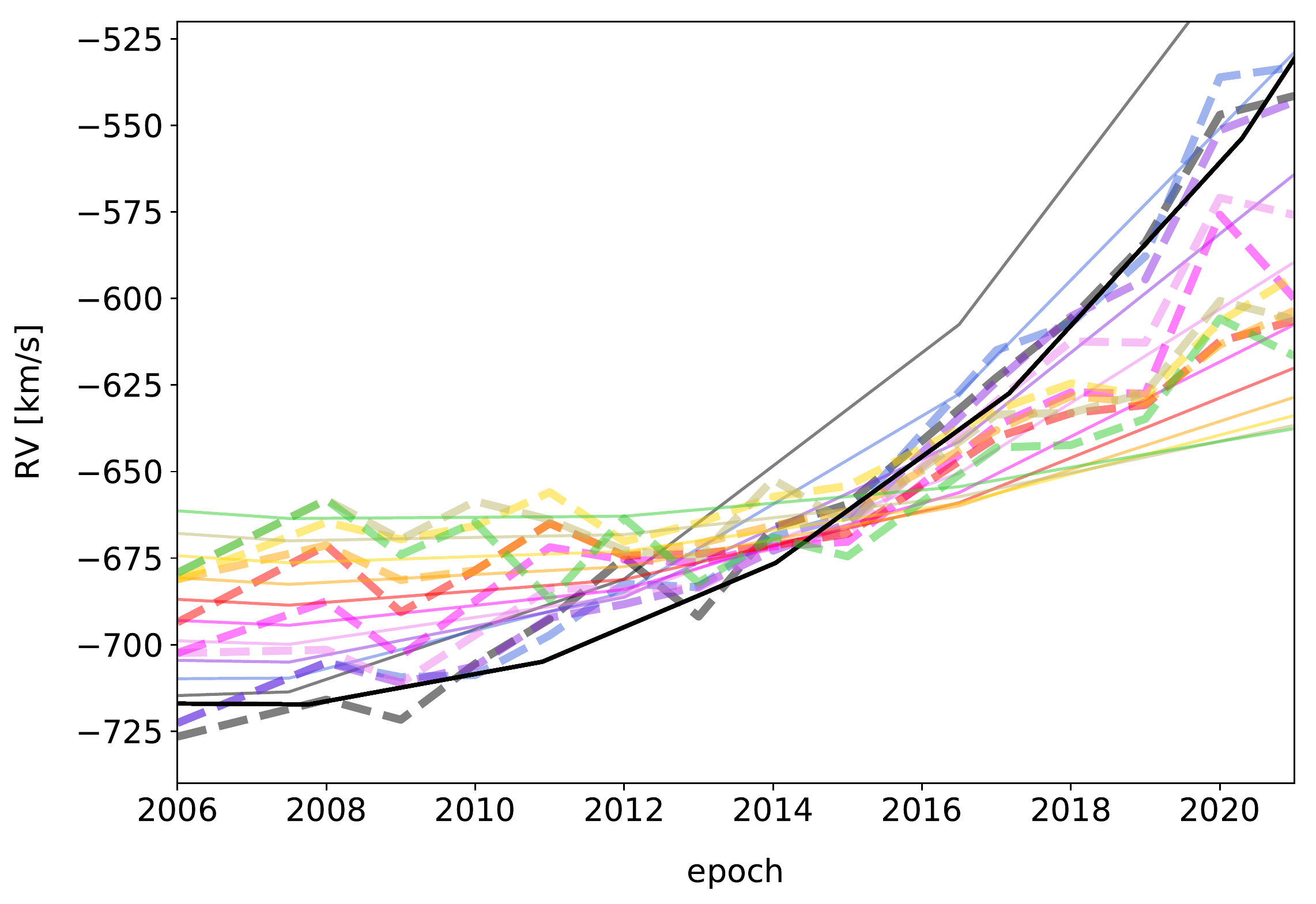} \\
      \caption{Model results best matching the dynamical and morphological evolution of X7. In panels each point along X7 (modeled individually) is identified by a different color (the tip is shown in black). 
      $Left$: the modeled orbital motion of X7 (sampled at 5 epochs during our observing period) reproduce well the observed inclination and lengthening (all measurements are represented, illustrated by gray, thick lines). The gray thin line shows the orbital fit for the tip and the open circles the initial positions fed to the model.
      $Right:$ the radial velocity structure and evolution is well reproduced as well. The solid lines represent the model velocities (the colors match those of the points in the left panel) and the dashed lines depict our measurements of the radial velocities along the ridge of X7. The thick black line represent the fit for the tip.}
     \label{fig:looptoymodel}
 \end{figure*}

We have run the model on a grid of parameter values: angles between 0 and 90 degrees and $m$ between 1 and 10~AU/years/dx. 
The combination of free parameters that most closely reproduces the observed motion of X7, its orientation, its length on the plane of the sky, as well as its radial velocity is $\theta_{z}$=20~degrees and $m$=0.02~AU/year/dx. The best model is shown in Figure \ref{fig:looptoymodel} along with our observations.

Even though this model is simple it illustrates quite well that gravity alone can reproduce many of the observed properties quite well: (1) the position angle of X7 on the plane of the sky is approximately constant, (2) the model predicts a crossover of the radial velocities of the tail and tip, as observed, (3) X7 undergoes tidal stretching by about the observed amount as it approaches its orbital periapse near Sgr~A* and (4) the ridge of X7 remains linear and un-curved. 
This model also predicts that X7 was even more compact in the years prior to our first observation in 2002, as we expect. 

While this simplistic model is based on gravity alone, several other forces can be at play in shaping X7 (magnetic drag, the drag caused by motion through the accretion flow onto Sgr~A*, local winds, etc.). The constraints on these models are discussed in Section~\ref{subsec:orientation}.


\section{Discussion} \label{sec:discussion}

What can be inferred about the nature of X7 depends fundamentally on whether the observed gas and dust that compose it are associated with an object as massive as a star. 
We have shown that X7 does not have a {\it detectable} stellar counterpart (Section~\ref{subsec:pmstars}). 
Its observable manifestation consists entirely of unbound gas and dust in orbit around the SMBH. For the past 20 years, the size of X7 has far exceeded the tidal radius of any conceivable stellar object that would not be easily detected.  The 2002 projected distance of X7 from the SMBH was $\sim 0.7''$ (Fig. \ref{fig:shapevol}). 
At this projected distance, the tidal radius of an object of mass 1~M$_{\odot}$ would be greater than 30~AU ($>$62~AU for a 10~M$_{\odot}$ object). The tidal radius would be even larger at earlier times. Given such required radii, it is unlikely that the gas and dust constituting X7 could consist of material that has been tidally removed from a stellar object that has remained undetectable. Therefore, while we have previously interpreted the G objects as being distended, dust-enshrouded stars having outer radii that can, during their periapse passages near the SMBH, exceed their tidal radii (\citealt{Ciurlo20}, but see \citealt{Gillessen+19} for alternative scenarios), there is no evidence that the same is true for X7.

In the following, we first examine the constraints on the past and future lifetime of X7, and then consider various possibilities for how it might have formed.

\subsection{When could X7 have formed?}
\label{subsec:lifetime}

 If X7 consists only of gas and dust, then it is inevitable that tidal forces will cause it to continue to stretch and ultimately dissipate as it passes through periapse on a timescale of 15 to 20 years. Even if there is a massive, undetected object within X7 that is somehow the source of the gas and dust that we observe as X7, then as argued above, X7 cannot be bound to that massive unseen object.  Consequently, the gas and dust that constitute X7 will dissipate in any case, and the feature that we now observe as X7 will not survive periapse passage, even if it is associated with a putative massive unseen object that does survive.  
 The dissipated material might well be observable for a relatively brief interval after the moment of periapse passage of the tip, because different segments of the tail will reach periapse at progressively later times. 

The orbit that we have derived for the tip of X7 sets a limit on when X7 could have formed. Since X7 cannot survive a passage through its orbital periapse, its age as an unbound gas/dust feature is less than an orbital time, about 200 years.  It is clearly a transient structure.  Furthermore, the fact that X7 has become increasingly elongated, by a factor of $\sim$2 since our first observations in 2002, suggests that it was originally more compact. 
Indeed, extrapolating backward in time from 2003, we do find that the model ridge becomes more compact, but the extrapolation becomes unreliable beyond a few years because our assumption of an initial linear velocity gradient along a linear ridge in 2003 leads to possibly unrealistic non-linearities in the velocity field at earlier times. Thus, at present, we are not able to derive a precise date for the formation of X7.

 \subsection{How could X7 have formed? }
 \label{subsec:formation}
 
 \subsubsection{Collisional Formation}
 \label{subsubsec:collision}
 
The constraints on the age of X7 lead us to consider that X7 could have been created by an impulsive event in the recent past. 
A promising possibility for impulsively producing an unbound stream of gas is direct stellar collisions.  More generally, a collision responsible for the relatively small mass of X7 could have occurred not only between stars \citep[e.g.,][]{DaleDavies,Davies+11,Rose+20}, but also between stars and stellar remnants \citep[e.g.,][]{Rose+22}, or between stars and massive planets or brown dwarfs.
Despite the large stellar density in the central light-year of the Galaxy, direct stellar collisions are relatively rare except in two circumstances:  1) the merger of a binary pair of stars in the presence of the SMBH as a result of the eccentric Kozai-Lidov (EKL) effect \citep{Naoz16,Stephan19} and 2) the collision of a normal star or a compact stellar remnant with a red giant. 

\paragraph{Gas ejection in an EKL-induced stellar merger}
\label{subsubsub:EKL}

The EKL mechanism has been invoked to account for the G objects as products of the induced merger of stars in binary systems. \citep{Phifer13,Witzel+14,Prodan+15,Stephan16, Witzel17,Ciurlo20}.  However, there are major differences between X7 and the G objects that lead us to conclude that X7 is not such a merged stellar system:
 \begin{itemize}
     \item G objects are observed to be much more compact (mostly point-like), in contrast to the extended, roughly uniform-brightness, linear morphology of X7.  The spatial extents of the G objects, when they can be resolved, are also smaller and fainter than the extended ridge of X7.
     \item Extended material that has presumably been tidally removed from G objects is apparently produced at or near their orbital periapse, whereas X7 has had a very extended, linear morphology for the entire time that it has been observed, and all of that long before its projected periapse passage in $\sim$2036. 
     \item  The  material removed from G objects is spread out along their orbits \citep{Witzel17, Gillessen+19}, as expected for a tidal removal process, in contrast to X7, for which the ridge of emission is oriented at a large angle with respect to the orbital trajectory.
 \end{itemize}
We also note that the star-centered model for G objects is not universally accepted. The original notion that they are composed only of gas and dust continues to be investigated \citep{Gillessen12,Gillessen+19}, but the profound structural and dynamical differences between X7 and the G objects mentioned here indicate to us that different phenomena are at play in the two circumstances.

When an EKL-induced direct stellar encounter
takes place in a binary, it happens as the eccentricity approaches unity, and because the 
encounter 
takes place very near periapse, the orbital speeds are maximized \citep[e.g.,][]{Naoz+14}. As a consequence, such collisional encounters are likely to unbind a considerable quantity of material even in the likely case in which the initial
encounter 
is a grazing collision \citep[e.g.,][]{Salas+19}. 
 That leads to two considerations relevant to X7: first, the two objects involved in the collisional encounter (and ultimate merger) do not necessarily follow the same subsequent orbit as the ejected, unbound gas, so a pure gas/dust feature can be produced that is not physically co-located with the star or stars that produced it (although their past orbits would have an intersection point). Second, the gas ejected in such a collision would likely take the form of a quasi-linear stream of unbound material. Note that such a dynamical situation is consistent with the assumptions upon which the simple dynamical model presented in section \ref{sec:toymod} is based. 
 Depending on the binary mass ratio and the stellar sizes, the amount of material ejected in an EKL-induced collisional encounter, and the degree of collimation of the ejecta, can presumably be quite variable, but this has not yet been investigated in detail. 

In light of this EKL-induced collision scenario for producing X7, our finding that G3 has a similar orbit to that of X7 is particularly interesting  (Section~\ref{subsec:pmstars}). 
X7 and G3 can be regarded as a candidate pairing resulting from a recent EKL-induced merger, with the resulting merger product being G3 and the ejecta from the violent collisional encounter being X7.  

Assuming that X7 was ejected from G3, one can ask whether the similarity of their orbital parameters is plausible, given that there is no constraint on the direction of the ejection (there need be no relationship between the orientation of the pre-merger binary orbit and that of the orbit of the binary around the black hole).   
The similarity of the semi-major axes of G3 and X7 can be accommodated with a wide range of ejection velocities. 
Following the equations outlined in \citet{Lu+19}, we find that X7 could have been ejected with a wide range of velocities, with an upper limit of two times the Keplerian velocity of G3 about Sgr~A* and the lower limit being at least the escape velocity from G3. 
This wide range  of allowable ejection velocities yields a similarly wide range of differential orbital orientations between G3 and X7 \citep[defined as the angle between the ejection velocity and G2 velocity][]{Lu+19}, for X7 to have been ejected in the merger process. The details of the parameter space are beyond the scope of this paper, but we note that a similarly wide range of the parameter space seems to exist for yielding an eccentricity for X7 that is slightly higher than that of G3 (i.e., by $\sim0.1$). 

\paragraph{Collisions of red giants with field objects}
\label{subsubsub:redgiants}

For ordinary collisions between isolated field objects, the most likely targets would be red giants because of their large cross-section.  Even a Jupiter-mass object flying through the atmosphere of a red giant could unbind a mass comparable to that of X7, depending on the relative velocity of the collision partners \citep[e.g.,][]{Sahai+03,Salas+19}.
The red giant-compact object collision scenario has been investigated \citep{Bailey+Davies99, Dale+09,Davies+11}, and offers an interesting possibility as a mechanism for producing X7. 
Because of the high stellar density in the Galactic Center, red giants within 0.1 pc of the SMBH can be expected to undergo multiple direct collisions with stars and stellar remnants during their time on the red giant branch \citep{Dale+09}.  
At the typical relative velocities found in the the central 0.1~pc of the GC (several hundred km/s), main-sequence stars and stellar remnants will pass through the atmosphere of the red giant, unbinding some quantity of gas and giving a velocity impulse to the red giant core.  If the collision partner is a main-sequence star (less than a few solar masses, or a white dwarf or neutron star), then the velocity impulse given to the stellar core is relatively small, and most of the atmosphere remains bound to the core \citep{Bailey+Davies99}. 
A more dramatic encounter occurs if the impactor is a $\sim10$ M$_{\odot}$ black hole.  This is less frequent than collisions with main-sequence stars \citep{Dale+09}, even if black hole remnants have been strongly concentrated toward the center by dynamical mass segregation \citep{Morris93, Miralda-Escude+Gould_00, Freitag+06, Antonini+14,Rose+22}. 
In black hole-red giant collisions, almost all of the red giant envelope could be removed and the core would receive a much larger kick \citep{Davies+11}.  

In any collision with a red giant, the amount of gas released depends on the masses of the impactor and the red giant, their relative velocity, the impact parameter, and the evolutionary stage of the giant.  
The relatively small amount of mass that we infer for X7 ($\sim$50 Earth masses) can be attributed to a collision with either kind of partner, but in the black hole case, the collision parameters would be constrained to some combination of a relatively large impact parameter and a relatively high relative velocity. 
Otherwise the resulting mass of the unbound gas would be much larger.
 
Red giant collisions of the sort that might produce X7 would leave the red giant in a distended and dynamically agitated state that would only settle down on a Kelvin-Helmholtz timescale, so such stars could thereafter appear as G objects for a timescale much longer than the observable lifetime of the unbound ejecta. 
The production of an X7-like feature in this manner could therefore be accompanied by the production of a G object. Consequently, we again have a situation in which the similar orbits of X7 and G3 could potentially be understood in terms of a single dynamical encounter. 
 
However, we note that collisions between red giants and main-sequence stars probably happen only about once every 10$^5$~yr in a flat stellar core such as that found at the Galactic center \citep{Rose+20,Rose+22}, not frequently enough for the recent production of X7 in this manner to be very likely.
 
\subsubsection{Other Possible Formation Mechanisms}
\label{subsubsec:other}
 
\paragraph{Infalling gas cloud}\label{subsubsub:junk}

The small semi-major axis (5000~AU) and modest eccentricity ($0.34$) of our calculated orbit imply that X7 could not have been produced in a straightforward manner from a distant gas reservoir.  
The closest reservoir could be the broad northern arm gas stream of Sgr A West, lying to the south of the X7 orbit.  The northern edge of the northern arm is $\sim10^4$ AU from the nearest point of X7's past orbital path (Fig. \ref{fig:shape}), so there is no obvious intersection point.
Furthermore, while the northern arm is strongly blue-shifted at the point where it wraps around Sgr~A* to the south of X7, its radial velocity is far less extreme than that of X7 (up to -300~km/s compared to -700~km/s for X7).

A number of small gas and dust clouds are located within a few arcseconds of Sgr~A* (see Figure~\ref{fig:rgb}).
However, besides the much lower-mass G objects, no gaseous structures are known to be present in the volume occupied by the orbit of X7, although some might be found at larger radii.  There are a few dust features seen in projection near X7, but their velocity structure and morphological evolution are inconsistent with being as close to Sgr~A* as any point in the orbit of X7.  

Collisions of orbiting gas clouds are one way to produce low-angular-momentum parcels of gas that can fall inward in the aftermath of the collision, but such events are more likely to produce dynamically complex systems on eccentric orbits rather than the isolated, relatively compact gas blob on a mildly eccentric orbit that would have been the initial state of X7.
 
It remains possible that X7 is a piece of gaseous "space-junk", detached from a local larger gas structure. Collisions between such structures and the
strong stellar winds from the WR stars in the region might create blobs of gas with low angular momenta, but again, the expected eccentricities of such blobs would likely be large, given the depth to which they would have to fall in the black hole's gravitational potential well.

\paragraph{Colliding winds}
\label{subsubsubsec:collidingwinds}
One possible mechanism that has been suggested for producing the G objects, based on the assumption that they are purely gaseous features, is the formation of dense clumps in colliding stellar winds \citep[e.g.,][]{Burkert+12,Calderon+16}. A similar mechanism might be considered for producing X7. Indeed, \citet{Cuadra06} had earlier argued that colliding winds from massive stars can form cold clumps and filaments as they undergo thermal instability after being shocked and compressed. 
Such clumps could have a filamentary morphology with a velocity gradient along them \citep{Pfuhl15, Plewa+17}, which could possibly describe the initial stage of X7.  However, the requisite physical conditions for clump formation by this process occur only rarely in the GC \citep{Calderon+16}, and when they are produced, they are subject to quick evaporation by thermal conduction in the hot medium of the GC \citep{Calderon+18}. 
Furthermore, a detailed computational study by \citet{Calderon+20b} has shown that the maximum masses of clumps formed in this way are too small to account for the G objects, and that conclusion would hold even more strongly for X7.  
We therefore conclude that the formation of X7 in stellar wind shocks is unlikely.

\subsection{Alternatives to determine X7's tail orientation}
\label{subsec:orientation}

\begin{figure}[ht]
    \centering
    \includegraphics[width=8cm]{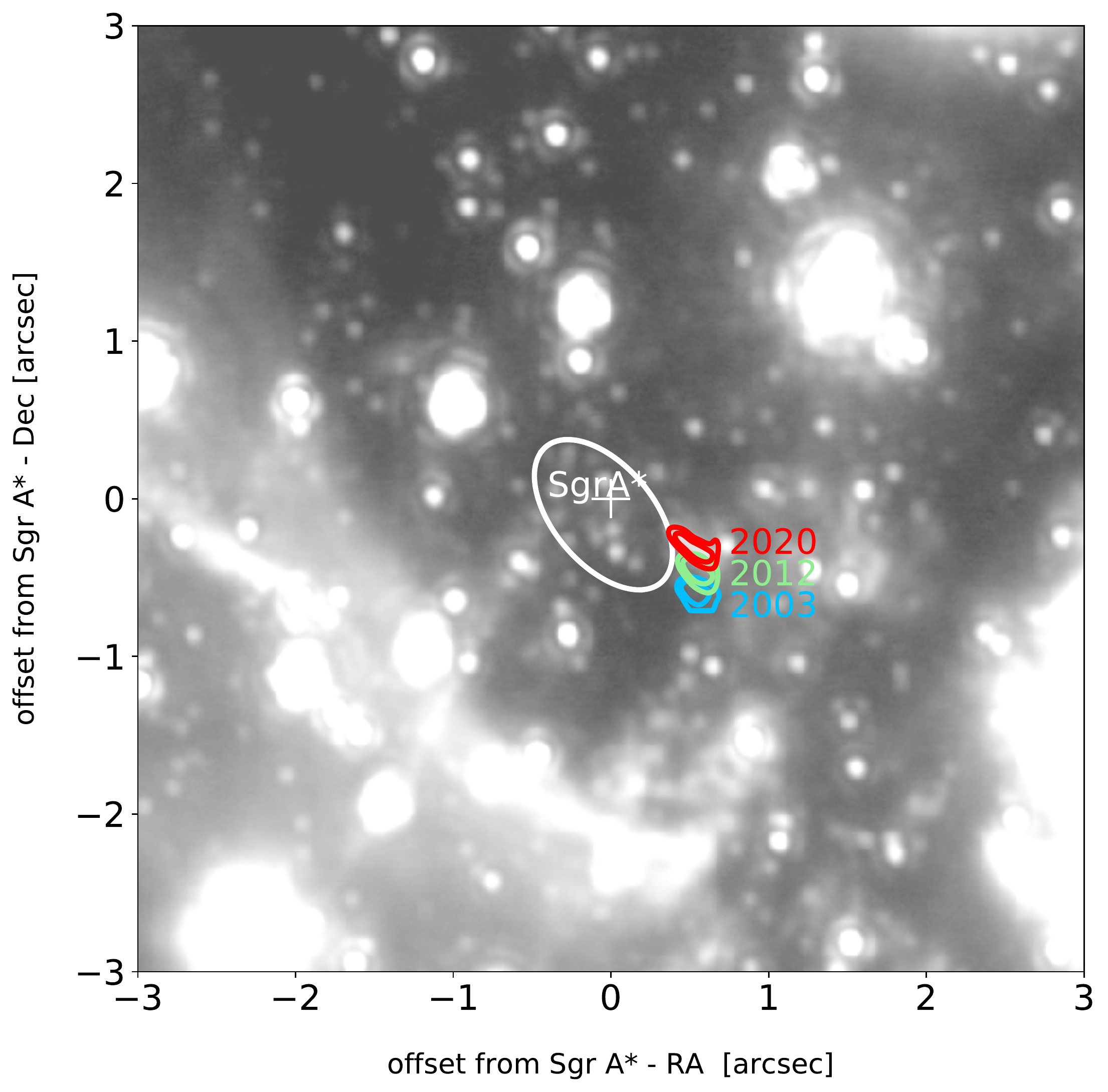}
    \caption{Projected orbit of the X7 tip (white) overlaid on a 2020 gray-scale Lp image.  Contours show the X7 Lp emission at three different epochs, illustrating that X7 is not elongating along its orbital trajectory. X7 orbit is compact and does not overlap with nearby large-scale extended features, such as the mini-spiral, whose Northern Arm is visible at lower-left of the image.}
    \label{fig:shape}
\end{figure}

The pure-gravity model presented in Section~\ref{sec:toymod} reproduces well most of the observed dynamical and morphological characteristics of X7. 
In the following, we discuss several additional phenomena that could conceivably play a role in producing the observed orientation of the tail. 
In particular, all scenarios are constrained by the constant position angle of the tail, as well as the fact that the tail is not aligned with the orbital direction of the tip's motion (Figure~\ref{fig:shape}). 
However, we conclude from the success of our pure-gravity model that the effect of these phenomena are likely to be secondary or negligible.

\subsubsection{A spherical wind from  Sgr~A*}
\begin{figure}[ht]
    \centering
    \includegraphics[width=8cm]{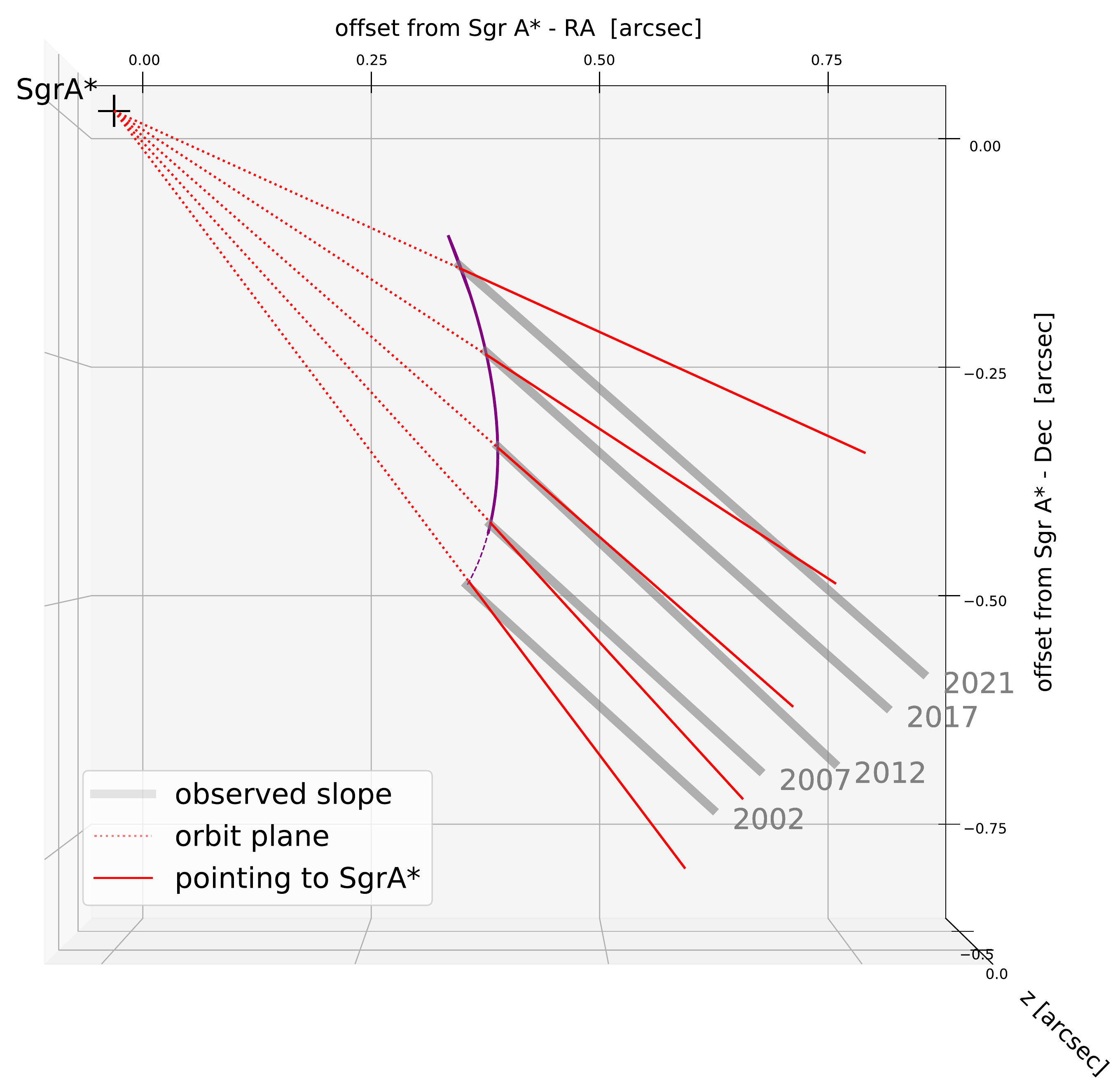}
    \caption{The X7 tail does not point to Sgr~A*. 
    Red lines show the expected tail orientation on the sky if it always points toward Sgr A* and is in the plane of the feature's orbit, while the thick gray lines show the observed constant orientation.} 
    \label{fig:sim}
\end{figure}
X7 in early observations is roughly elongated toward Sgr~A*, while the orbital direction of the tip and the proper motion of X7's overall structure are oriented in a different direction.  
In earlier epochs X7's elongation and flared appearance were deemed consistent with the hypothesis that the morphology of X7 results from a wind arising at or near Sgr~A* \citep{Muzic10}.
However, if that were the case, the tail would always lie in the plane of the orbit and point toward  Sgr~A* in three dimensions.
With our inferred orbit of X7's tip we can project the expected tail inclination onto the plane of the sky on the assumption that the tail always points to Sgr A*.
The result is illustrated in Figure~\ref{fig:sim} along with the observed orientation of X7's tail.
Given X7's orbit, if it were always pointing toward Sgr~A*, we would expect to observe a variation in the tail orientation on the plane of the sky of greater than 30 degrees during the course of our observations.
Since the tail orientation does not change by anywhere near that amount, we exclude a spherical wind from  Sgr~A* as the cause of the orientation of X7's tail.

\subsubsection{Stellar winds}
Another possibility is that X7 is shaped by the combined winds of the Wolf-Rayet stars in the region.
Recently \citet{Calderon+20a} and \citet{Ressler20} proposed two independent versions of such models. 
In both publications the authors take into account the known motions and mass-loss characteristics of the Wolf-Rayet stars to show how their winds evolve and interact, producing 3-dimensional maps of the wind vectors in the central parsec.

We use the orbit of X7's tip to determine its line-of-sight distance relative to Sgr~A*.
Given the 3-dimensional position and shape evolution implied by our model, we can compare X7's orientation to the modelled local wind direction at X7's location.

In the case of \cite{Calderon+20a} the direction of the winds at the location of X7 is uncorrelated with the shape of X7: the winds are blowing toward the East and therefore are oriented almost 45 degrees away from the sky projection of X7's tail (Calderon, private communication).

\citet{Ressler20} additionally take into account the effect of the strength and geometry of the magnetic field on the wind direction.
Their results indicate that, for stronger magnetic fields the direction of the combined winds is toward the southwest (Ressler, personal communication), consistent with X7's shape.
However, according to their model, over roughly a decade during the period of our observations the wind turned significantly (few tens of degrees) north whereas the X7 tail did not.
Therefore, even choosing a magnetic field strength that gives results that best match our observations, we cannot reproduce the  evolution of X7's orientation.

Moreover, for both models, there is still substantial uncertainty in the wind direction, since it depends strongly on the assumed parameters (magnetic field strength, geometry, stellar mass loss rates and velocities).
For example \citet{Ressler20} assume that the magnetic field is perpendicular to the Galactic plane, which is justifiable at much larger scales, but in the neighborhood of X7, the field is likely distorted by winds and by the  Sgr~A* accretion flow.
We conclude that there is no obvious correlation between the orientation of X7 and the local wind direction, but the two very different estimations of the local wind direction obtained in two independent investigations illustrate the considerable uncertainty that presently exists.
Therefore, we cannot at present rule out the possibility that stellar winds contribute to the shaping of X7.

\subsubsection{Magnetic field}

A sufficiently strong magnetic field could also be a direct cause of X7's morphology and orientation, as is the case for some other filamentary structures in the Galactic center region.
\citet{Roche18} showed the magnetic field direction averaged over the line of sight and projected onto the plane of the sky using a relatively high-resolution polarization map of the 12.5~$\mu$m dust emission in much of the Galaxy's inner parsec.
According to this map, X7's tail lies roughly perpendicular to the magnetic field lines. Consequently, it appears unlikely that the magnetic field plays a role in orienting X7.  However, because the polarization measurement results from an integral over all contributions along the line of sight, the possibility that the magnetic field direction at the 3D location of X7 is aligned with X7 cannot be completely excluded.


\section{Conclusions} \label{sec:concl}

Using two decades of imaging and spectroscopic data gathered with the Keck Observatory, we analyze the morphological and dynamical evolution of the extended, linear dust and gas structure, X7, presently located $\sim$0.5'' ($\sim$4000~AU) from the Galactic black hole.
We observe several properties of this unique feature:
\begin{itemize}
    \item X7 exhibits relatively rapid proper motion, comparable to that typical of the S-stars orbiting closely around the Galactic black hole, yet its orientation remains remarkably constant even as it has moved through a substantial portion of its orbit;  
   \item the internal spatio-velocity structure of X7 is changing with time: the tip has decelerated (by approximately 200 km/sec from 2006 to 2021) whereas the radial velocity of the tail remains relatively unchanged;
   \item The 3D motion of X7's leading tip is consistent with a tightly bound orbit around Sgr~A* having a period of only $\sim200$ years;
   \item the shape of X7 has changed with time, morphing from a bow-shock-like structure to a more elongated, linear structure as it approaches the black hole;
    \item X7 has lengthened considerably as it has approached the central black hole, but there is no evidence that this elongation has been accompanied by a corresponding significant change in surface brightness with time;
    \item higher-resolution data obtained in 2020 show that X7 might be fragmenting; 
    \item The proper motion of X7 is quite different from that of all detected stars in its immediate environment, indicating that X7 is unlikely to be associated with any of the known stellar sources that have coincided with it along the line of sight during the period of our observations.

\end{itemize}

From these observations we draw the following conclusions. 
\begin{itemize}
    \item The constant position angle of X7's linear tail rules out shaping by a spherical wind from Sgr~A*. The observed orientation also appears to be inconsistent with the local direction of the collective winds from nearby Wolf-Rayet stars as well as with the projected magnetic field orientation in this region given the current models and observations.
    \item The rapidly decreasing radial velocity of X7's tip is strong evidence of the dominant gravitational influence of the SMBH.  Indeed, we can reproduce the observed properties of X7 (evolution of the radial velocity and its gradient along the ridge, and the elongation and constant position angle of the ridge on the plane of the sky) with a simple test-particle model in which the particles respond {\it only} to the gravitational field of the black hole. Other phenomena, including strong stellar winds, the accretion flow onto the SMBH, and a potentially strong local magnetic field, therefore appear to have, at most, a minor secondary effect on the dynamical evolution of X7.
\end{itemize}

While we are not able to definitively determine the origin of X7, we have explored and emphasized the prospect that X7 was formed by an event such as a stellar merger or a collision with a stellar or substellar object, or even with a stellar remnant. 
In this light we note that the dust-enshrouded stellar object, G3, has an orbit that is remarkably similar to that of X7, eliciting the possibility that X7 could be the ejecta that resulted from the EKL-induced binary merger that created G3.  Further assessment of this intriguing scenario will require both improved orbital determinations and detailed dynamical modelling.
Alternatively, we cannot rule out the possibility that X7 was stripped or shed from one of the larger-scale interstellar structures in the region, although we cannot trace its dynamics in a straightforward way to those of nearby gas structures.

Regardless of its origin, the X7 wisp of gas and dust will continue to undergo even more dramatic evolution in the next 10 or 20 years as it swings through its closest approach to the black hole, becomes even more tidally stretched, gets fragmented by instabilities, and interacts with the accretion flow, potentially triggering enhanced accretion activity.  
Continued monitoring of X7 will allow us to closely witness these extreme changes, ending with the ultimate tidal dissipation of the remnants of this intriguing structure.

\acknowledgments
The authors wish to acknowledge S. M. Ressler, D. Calderon and S. Rose for the helpful insights provided by their theoretical models and A. Huddleston for assisting with photometrical measurements.  
Support for this work was provided by NSF AAG grant AST-1412615, Jim and Lori Keir, the W. M. Keck Observatory Keck Visiting Scholar program, the Gordon and Betty Moore Foundation, the Heising-Simons Foundation, and Howard and Astrid Preston. A. M. G. acknowledges support from her Lauren B. Leichtman and Arthur E. Levine Endowed Astronomy Chair. 
R. S. acknowledges financial support from the State Agency for Research of the Spanish MCIU through the ''Center of Excellence Severo Ochoa'' award for the Istituto de Astrof\'isica de Andaluc\'ia (SEV-2017-0709) and financial support from national project PGC2018-095049-B- C21 (MCIU/AEI/FEDER, UE).
S.N. acknowledges the partial support from NASA ATP 80NSSC20K0505,  NSF through grant No.~2206428,  and thanks Howard and Astrid Preston for their generous support.
The W. M. Keck Observatory is operated as a scientific partnership among the California Institute of Technology, the University of California, and the National Aeronautics and Space Administration. The Observatory was made possible by the generous financial support of the W. M. Keck Foundation. 
The authors wish to recognize and acknowledge the very significant cultural role and reverence that the summit of Maunakea has always had within the indigenous Hawaiian community. We are most fortunate to have the opportunity to conduct observations from this mountain.


\appendix

\section{Extended structure}
\label{app:xtps}

 \begin{figure}[ht]
    \centering
    \includegraphics[width=8cm]{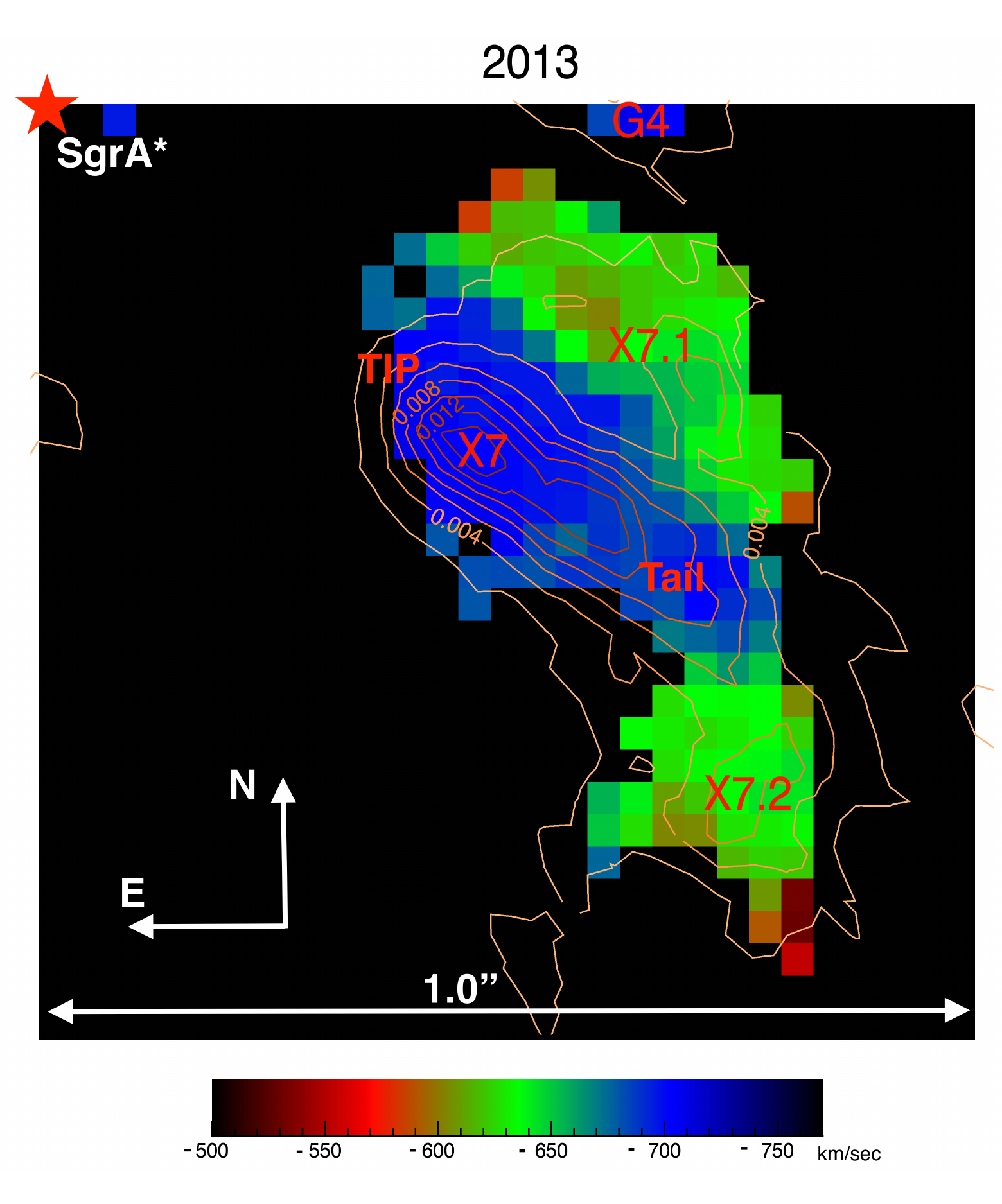}
    \caption{Illustration of X7's complex morphology using OSIRIS Br-$\gamma$ emission as measured in 2013. The color scale illustrates the velocity (km/sec), while the contours trace the intensity. The map is constructed using a relatively narrow range of velocities between -540 and -730 km/sec. The X7 ridge (tip and tail) consists of the highest blue-shifted velocity gas. Additionally, the X7 complex includes an arc of material to the northeast labeled X7.1 and a clump of gas off of the tail to the south, labeled X7.2. All of these blue-shifted Br-$\gamma$-emitting features are co-moving northward. 
    G4 is included in the slice because of its similar blueshifted radial velocity but is unrelated to X7 as its proper motion is in a completely different direction than that of X7 \citep{Ciurlo20}. 
    G5, which is also unrelated to X7, is not included in this slice because of its redshifted radial velocity.
     }
    \label{appfig:morph}
\end{figure}

Figure~\ref{appfig:morph} shows a cut of the OSIRIS data around X7's blueshifted Br-$\gamma$ emission.
This cut includes G4, which is blueshifted at similar radial velocities as its proper motion is in a completely different direction than that of X7 \citep{Ciurlo20}.
The extended, lower-intensity gas emission includes an arc-shaped feature to the Northwest of X7 which \citet{Peissker21} labeled as X7.1 and argued that it has a possible association with the nearby compact G object, G5 \citep{Ciurlo20}.
However, our OSIRIS integral field data indicate that G5 is red-shifted at +350~km/sec and X7.1 is highly blue-shifted around -600~km/sec. Therefore G5 does not appear in the slice and the two are clearly unrelated to each other. 
Moreover, X7.1 \citep{Peissker21} seems to be moving in the same direction as the rest of X7 whereas G5 has a completely different trajectory, even though the two features overlap in the most recent observations (see Figure~\ref{fig:pmstars}, bottom-left panel). 
Neither G5 or X7.1 are well detected in the NIRC2 Lp data.

The Br-$\gamma$ emission line map also reveals a knot-like feature to the South of the X7 tail that we label X7.2. 
X7.2, like X7.1, has a proper motion similar to that of X7 and is also similarly blue-shifted.

Even though X7.1 and X7.2 seem to be moving in the the same general direction as X7,  it is difficult to provide quantitative measurements consistently over time, due to their extended nature and evolving morphology. 
Therefore, these features are likely associated with X7; they might have a common origin with X7 or might be evidence of material that is separating from X7. 
However, we cannot determine their dynamical history in the same way we can for X7 and thus cannot raw definitive conclusions about their relationship to X7.

The early cometary appearance of X7 in 2002--2006 that led \cite{Muzic10} to interpret X7 as a bow-shock might be attributable to the relative placement of these fainter features (or alternatively to the presence of nearby sources in this crowded environment, for example S0-73).

\section{Astrometry}
\label{app:astro}
\begin{figure}[ht]
    \centering
    \includegraphics[width=6cm]{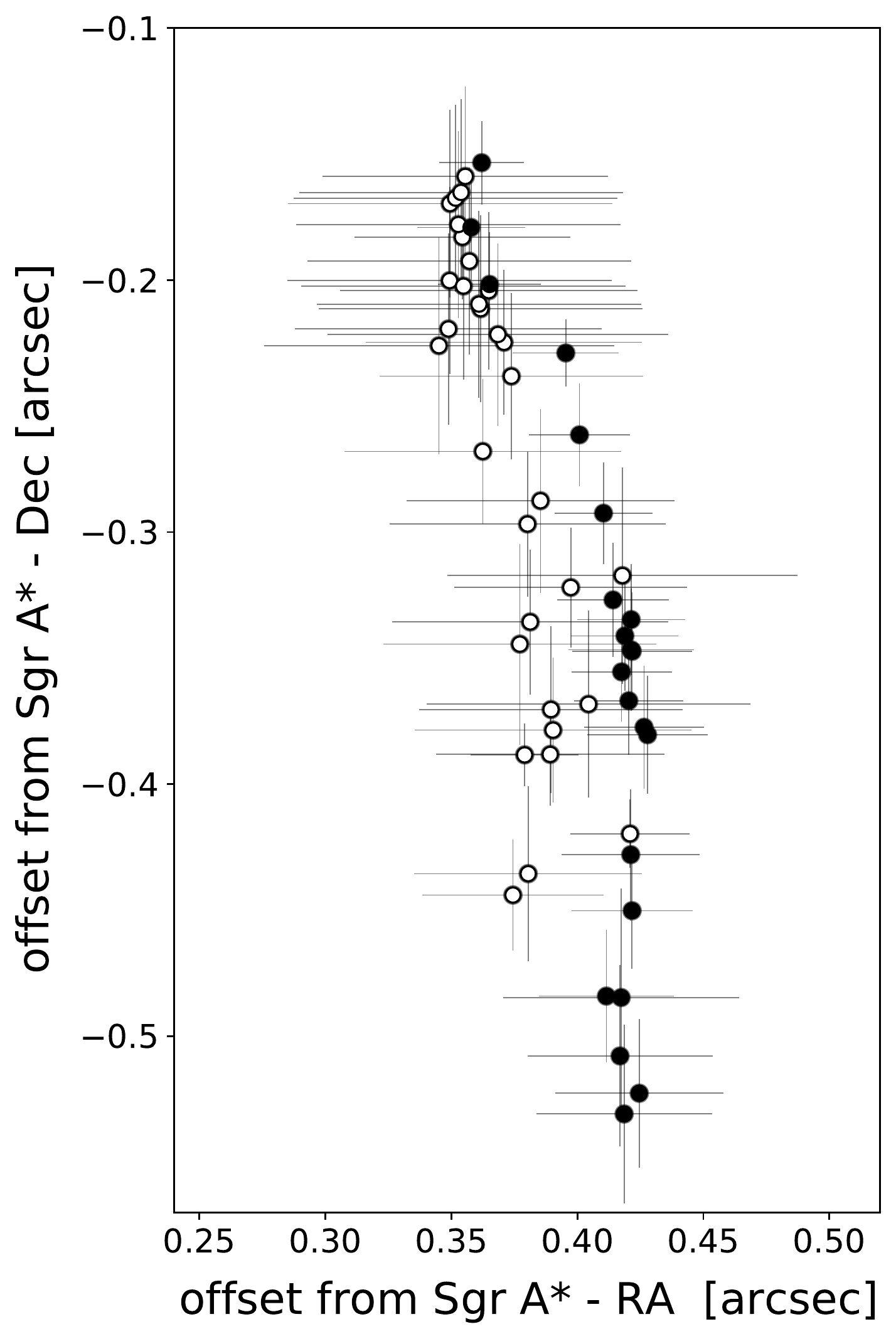}
    \caption{X7 astrometry as measured in OSIRIS (white points) and NIRC2 (black points).}
    \label{figA:astros}
\end{figure}

The two sets of astrometric measurements of the X7 tip agree to within estimated errors (see Figure~\ref{figA:astros}), however, we observe that the NIRC2 points tend to be systematically shifted West by less than 1-sigma uncertainty during roughly half of the monitoring period.
This difference might be due to physical differences between the gas and dust emission of the tip, given that the dust emissivity and the Br-$\gamma$ emissivity have different dependencies on density and temperature and that the dust and gas temperatures can be quite different in this high-ionization environment.  
However, we can't rule out a systematic error in the measurements.
Such error might originate in the very different nature of the two datasets (see for example Figure~\ref{fig:cuts}, bottom right panel). NIRC2 data have a relatively high signal-to-noise ratio and are over-sampled at the pixel scale of 0.010". 
In contrast, OSIRIS data have lower signal-to-noise ratio and they are slightly under-sampled at the 0.035" platescale, with an inherent PSF of $\sim$0.050"~FWHM.
We also note that, because of the extended nature of X7, given one set of measurements made over time with a particular instrument, there is no guarantee that the tip location always follows the same parcel of orbiting material in both datasets.
Given these limitations, in the main text we opt for using only OSIRIS astrometry and radial velocity measurements for orbit fitting, which is self-consistent. We investigate the influence of NIRC2 measurements on the orbit in Appendix~\ref{app:fitNIRC2}.

\section{Dust Temperature}
\label{app:temp}

\begin{figure}[ht]
    \centering
    \includegraphics[width=8cm]{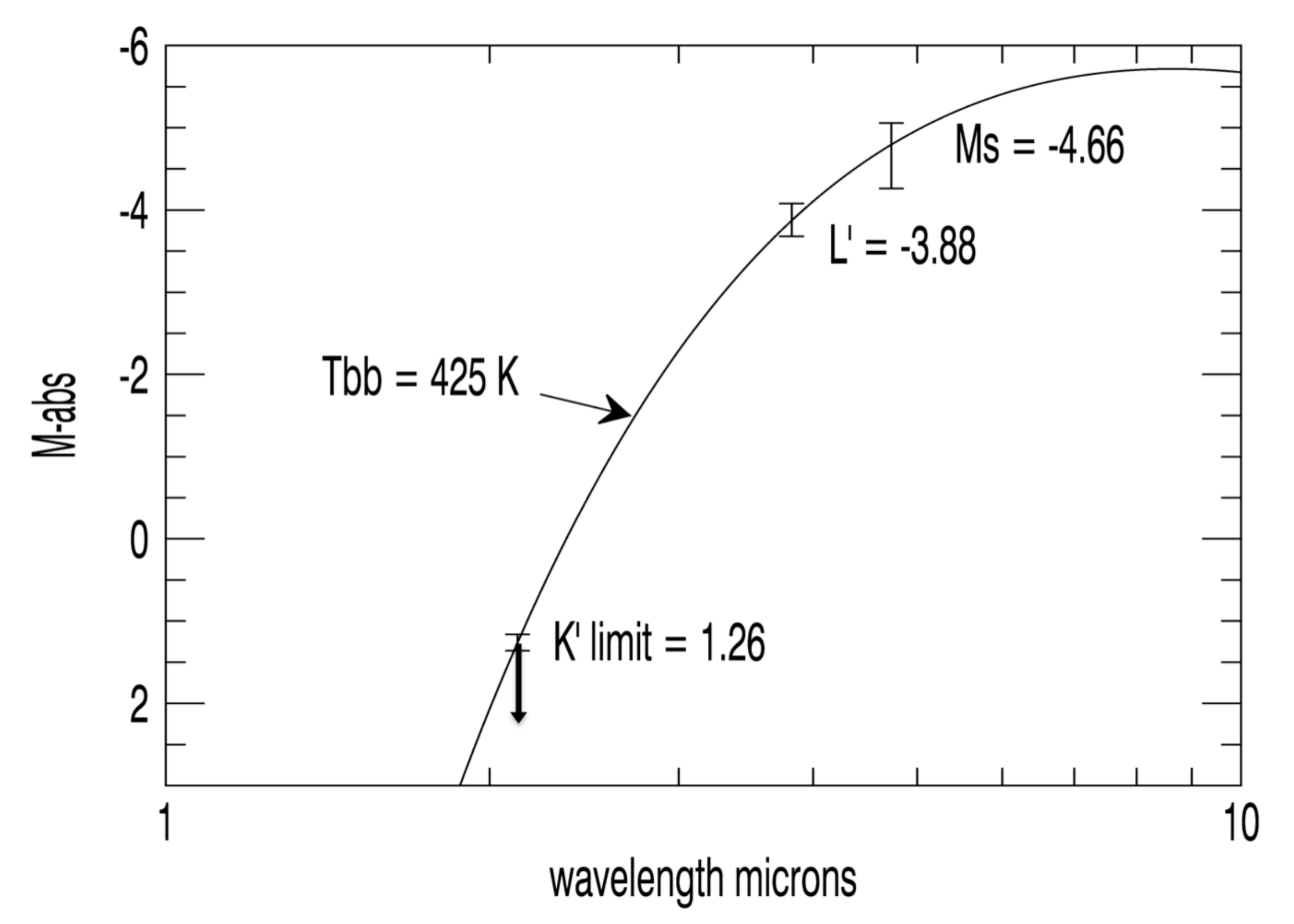}
    \caption{SED of X7's tip. Fluxes measured at Kp (upper limit), Lp, and Ms, plotted with an absolute magnitude scale.  The overlaid curve is a 425~K black body profile, which fits the measurements to within 50~K.}
    \label{fig:SED}
\end{figure}

Using the NIRC2 observations made in three filters (Kp, Lp and Ms at 2.1, 3.8, and 4.5~$\mu$m respectively), we can constrain the spectral energy distribution (SED) using photometric measurements of the tip in these three bands. 
We use differential aperture photometry between X7's tip and the star GCIRS~16C (L-mag = 8.2), assuming the absolute magnitude of GCIRS~16C is the same for both Lp and Ms bands (similar to \citealt{Schodel11}). 
To do so, we used the same circular aperture for both the stellar source and X7's tip: 9~pixel aperture radius for Lp and 11~pixel aperture radius for Ms.
In both cases, the background level was subtracted using the DAOphot \citep{Stetson_1987} algorithm over an annulus of inner radius 30~pixels and outer radius 39~pixels.
The DAOphot algorithm is designed to deal with crowded regions and it eliminates outliers.
We estimate the uncertainties in the photometry from the variances of the measurements when changing background annular size, photometry aperture size and the centering aperture size.
Since X7 is not detected in the Kp band, we use as upper detection limit of $\sim$18~mag in Kp \citep{Do13}. 
We adopt the extinction values A$_{Ks}$= 2.22, A$_{Lp}$=1.07 and A$_{M}$=0.94 \citep{Gillessen12} to estimate the dust color temperature at the tip.
This procedure of obtaining the SED is similar to what was previously done for G2 by \cite{Gillessen12}.  
We then fit a Planck function to the SED, yielding a color temperature of $425\pm50$~K (Figure~\ref{fig:SED}). 

The temperature is comparable but somewhat lower than what was measured for G2 \citep{Gillessen12}, possibly due to X7 not having a stellar core that heats the dust.
The relatively high Br-$\gamma$ surface brightness observed for X7 is comparable to that of the G objects \citep{Gillessen12, Ciurlo20}.

Performing photometric measurements along the ridge of X7, to look for variations or a gradient in dust temperature, is compromised by the occasional passage of background/foreground stellar sources. 
However, the 3-color image in Figure~\ref{fig:rgb} does not show any gradient, therefore we expect the dust temperature to be nearly constant along the extent of the feature. 

\section{Orbital fit with NStarOrbits}
\label{app:orbfitnso}
The orbit model assumes Keplerian motion dominated by a central SMBH point potential. In this model, seven black hole parameters describe the mass, 3-dimensional position, and 3-dimensional linear motion of the central potential in a common reference frame, and six Keplerian orbital elements parameterize the orbit:  eccentricity, orbital period, epoch of closest approach to the SMBH, inclination, angle of the ascending node, and the argument of periapse \citep[e.g.][]{Grould17}.  While other stars closer to the SMBH have been shown to be consistent with a post-Newtonian General Relativistic model \citep{Do19GR, GRAVITY19, GRAVITY20}, the Keplerian approximation is sufficient here to constrain the orbit of X7. 

Because the orbital period of X7 is relatively long compared to the time baseline of observations, X7 alone cannot robustly constrain the black hole parameters. As such, we fix the central potential parameters using information from S0-2 -- the brightest short-period star in the Galactic Center for which more than one full orbit has been observed with both astrometric and radial velocity measurements \citep[for example, see][]{Do19GR, GRAVITY19, GRAVITY20}. We tested several sets of black hole parameters: an average of the latest estimates reported in the literature (\cite{GRAVITY20} and \cite{Do19GR}), as well as the values from each of these references independently. We find that the orbit of X7 is robust to slight changes in the central potential parameters, as the fitted orbit parameters are consistent to within 1-$\sigma$ for all of the above cases. For completeness, we also fitted X7 simultaneously with S0-2, and again find that the results are consistent with all fits described above. The results presented in Table~\ref{tab:orbparam} and Figure~\ref{fig:orbplot} represent the case in which the mass of, and distance to, the black hole are fixed to the average of the literature measurements (4.07$\times$10$^6$~M$_\odot$ and 8.1~kpc, respectively).  

To ensure an unbiased orbital solution, we adopt an observable-based prior -- a prior that has been shown to mitigate biases in estimated parameters when the orbital period is much longer than the time baseline of observations, as is the case for X7 \citep{ONeil19}. Generally, orbits that are fit within the Bayesian framework use priors that are uniform in the orbit model parameters. However, standard uniform priors have been shown to cause biases in the estimated orbit parameters for low-phase-coverage orbits. The observable-based prior is designed to mitigate biases in such cases by enabling all measurements to be equally likely before observations \citep{ONeil19}. In this case, despite the relatively low orbital phase coverage of X7, the radial velocity data provide enough constraining power that the results do not appear biased by low phase coverage. In other words, uniform priors and observable-based priors produce consistent results. While our choice of prior does not impact the results for X7, it does make a substantial difference in the inferred orbit of S0-73 (for which there is no radial velocity data). Since the orbits of X7 and S0-73 are compared in Section~\ref{subsec:pmstars}, we report the values from the observable-based prior case for consistency.

\section{Orbit fit including NIRC2 data}\label{app:fitNIRC2}
\begin{figure}[ht]
    \centering
    \includegraphics[width=13cm]{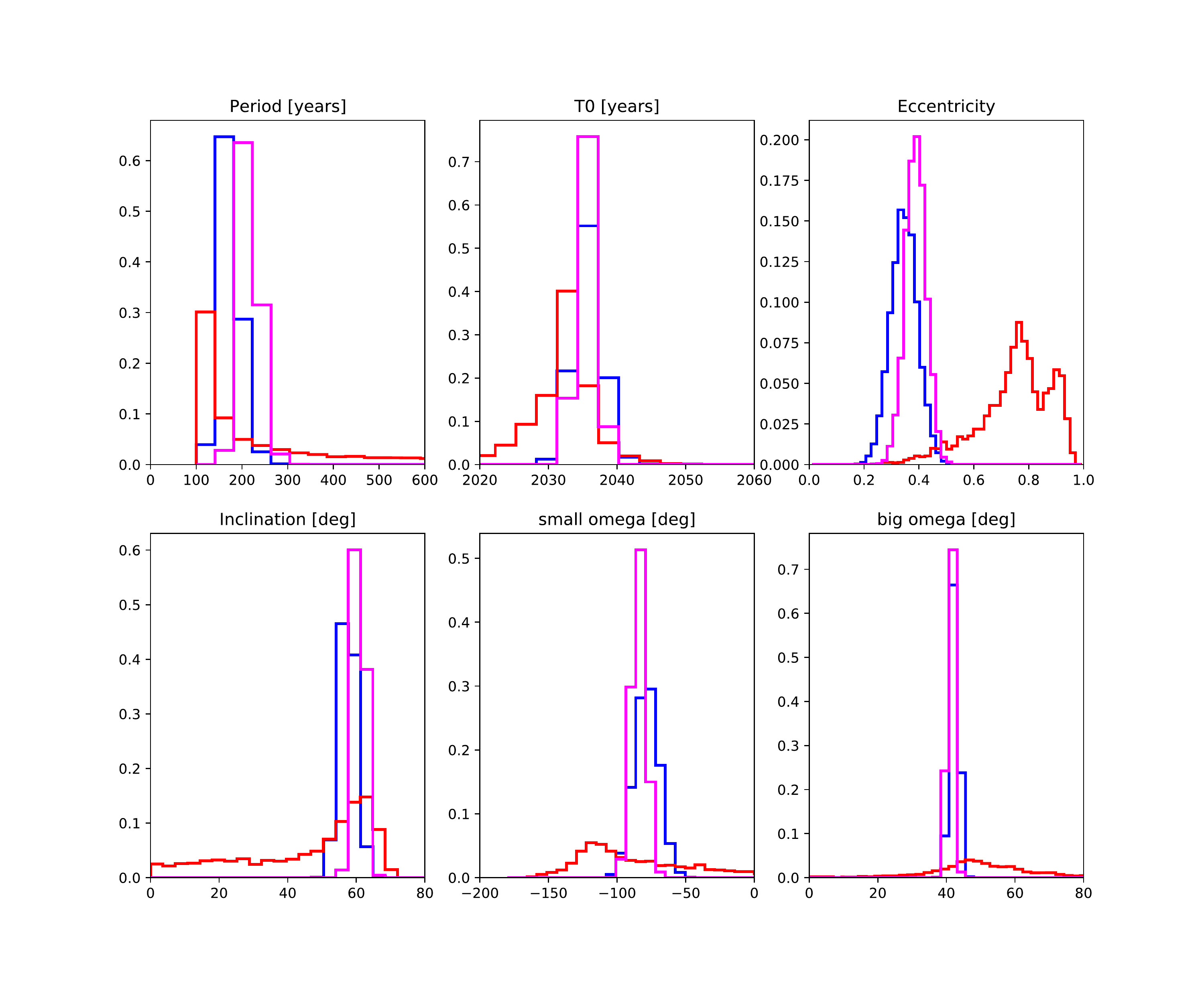}
    \caption{Orbital parameters for different datasets: OSIRIS-only orbit in blue, OSIRIS plus NIRC2 orbit in magenta, NIRC2-only in red. Note that the NIRC2-only orbit does not have any RV information and provides poor constraints on several parameters.}
    \label{figA:n2osrs_orb}
\end{figure}
As shown in Figure~\ref{figA:astros}, and described in Appendix \ref{app:astro}, NIRC2 and OSIRIS astrometric measurement are compatible but they have a residual offset that could be systematic or physical. 
In the main body of the paper we opt for a conservative approach and only use OSIRIS measurements. 
In Figure~\ref{figA:n2osrs_orb}, we report the orbital parameter obtained including NIRC astrometric points as well as a pure-astrometry, NIRC2-only orbit fit for comparison. 
Adding NIRC2 astrometry leads to better constrained orbital parameters but might introduce biased results because of a residual offset between the datasets.
Fitting the orbit using NIRC2 astrometry only we find less well constrained but mostly compatible results.

\section{Expected brightness evolution}
\label{app:brightness}

We observed no significant temporal change in the brightness of X7's tip in either the gas or dust emission (Section~\ref{subsec:mass}).
There are several reasons to expect the surface brightness of X7 to evolve as it orbits closer to Sgr~A*. First, our model indicates that the projection of the X7 tail along the line of sight has been  decreasing with time, which progressively increases the column density through the tail, and thereby the surface brightness.  On the other hand, the stretching by the tidal force from Sgr~A* reduces the linear mass density along the ridge.  Consequently, these are competing effects that could together reduce the brightness evolution below our threshold for detecting any significant changes. Tidal compression toward the central axis of the ridge is also operating, but because the X7 ridge is unresolved in its narrow dimension, such compression would not affect the optically thin Lp flux density that we measure. 

In contrast, the volume emissivity of the Br-$\gamma$ line depends on the square of the gas density, so tidal compression in the narrow dimension of the ridge can counteract tidal stretching to some extent in determining the surface brightness of Br-$\gamma$ along the X7 ridge. The relative importance of tidal stretching and tidal compression is a function of orbital phase, orbital eccentricity, black hole mass, the ambient gas pressure, and the distance to the black hole.

The other physical considerations affecting the evolution of the Lp and Br-$\gamma$ surface brightness of X7 involve the environment through which X7 is orbiting. Given the orbit that we have determined (Section~\ref{subsec:orbit}), we find that the 3D distance of X7 from the black hole is comparable to or even inside the Bondi radius, a region seen clearly in X-rays \citep{Baganoff+03, WangQD+13}, where the accretion flow encounters a shock that heats the inflowing gas to temperatures exceeding $10^7$ K. The hot gas surrounding X7 and the X-rays emitted there are likely to raise the dust temperature, which would counteract to some extent the decline of the surface brightness of the Lp dust emission resulting from tidal stretching.  In addition, the increased pressure of the external medium as X7 gets closer to the black hole, owing to both the higher temperature within and inside of the accretion shock and the higher densities implied by the converging accretion flow, act to compress the ridge of X7, adding to the tidal compression and therefore to the density-dependent emissivity in the Br-$\gamma$ line. A quantitative investigation of these environmental effects, as well as the net effects of tidal forces, is outside the scope of this paper, but continued monitoring of both the line and continuum surface brightness of X7 should provide important constraints on the myriad physical processes at play.

The comments in this subsection have assumed that there is no source of dust and gas continuously contributing to the mass of X7.  We note, however, that if such a source were present (and as discussed in Section~\ref{subsec:pmstars}, we have not identified any luminous source that could serve as a candidate), then its contributed material could add to the overall column density within X7, and thereby be yet another factor contributing to the surface brightness.

\bibliography{biblio_X7}

\begin{thebibliography}{}
\expandafter\ifx\csname natexlab\endcsname\relax\def\natexlab#1{#1}\fi
\providecommand{\url}[1]{\href{#1}{#1}}

\bibitem[{{Antonini}(2014)}]{Antonini+14}
{Antonini}, F. 2014, \apj, 794, 106

\bibitem[{{Baganoff} {et~al.}(2003){Baganoff}, {Maeda}, {Morris}, {Bautz},
  {Brandt}, {Cui}, {Doty}, {Feigelson}, {Garmire}, {Pravdo}, {Ricker}, \&
  {Townsley}}]{Baganoff+03}
{Baganoff}, F.~K., {Maeda}, Y., {Morris}, M., {et~al.} 2003, \apj, 591, 891

\bibitem[{{Bailey} \& {Davies}(1999)}]{Bailey+Davies99}
{Bailey}, V.~C., \& {Davies}, M.~B. 1999, \mnras, 308, 257

\bibitem[{{Boehle} {et~al.}(2016){Boehle}, {Ghez}, {Sch{\"o}del}, {Meyer},
  {Yelda}, {Albers}, {Martinez}, {Becklin}, {Do}, {Lu}, {Matthews}, {Morris},
  {Sitarski}, \& {Witzel}}]{Boehle16}
{Boehle}, A., {Ghez}, A.~M., {Sch{\"o}del}, R., {et~al.} 2016, \apj, 830, 17

\bibitem[{Bond {et~al.}(2020)Bond, Cetre, Lilley, Wizinowich, Mawet, Chun,
  Wetherell, Jacobson, Lockhart, Warmbier, Ragland, Álvarez, Guyon, Goebel,
  Delorme, Jovanovic, Hall, Wallace, Taheri, Plantet, \&
  Chambouleyron}]{Bond20}
Bond, C.~Z., Cetre, S., Lilley, S., {et~al.} 2020, Journal of Astronomical
  Telescopes, Instruments, and Systems, 6, 1 .
\newblock \url{https://doi.org/10.1117/1.JATIS.6.3.039003}

\bibitem[{{Burkert} {et~al.}(2012){Burkert}, {Schartmann}, {Alig}, {Gillessen},
  {Genzel}, {Fritz}, \& {Eisenhauer}}]{Burkert+12}
{Burkert}, A., {Schartmann}, M., {Alig}, C., {et~al.} 2012, \apj, 750, 58

\bibitem[{{Calder{\'o}n} {et~al.}(2016){Calder{\'o}n}, {Ballone}, {Cuadra},
  {Schartmann}, {Burkert}, \& {Gillessen}}]{Calderon+16}
{Calder{\'o}n}, D., {Ballone}, A., {Cuadra}, J., {et~al.} 2016, \mnras, 455,
  4388

\bibitem[{{Calder{\'o}n} {et~al.}(2018){Calder{\'o}n}, {Cuadra}, {Schartmann},
  {Burkert}, {Plewa}, {Eisenhauer}, \& {Habibi}}]{Calderon+18}
{Calder{\'o}n}, D., {Cuadra}, J., {Schartmann}, M., {et~al.} 2018, \mnras, 478,
  3494

\bibitem[{{Calder{\'o}n} {et~al.}(2020{\natexlab{a}}){Calder{\'o}n}, {Cuadra},
  {Schartmann}, {Burkert}, {Prieto}, \& {Russell}}]{Calderon+20b}
---. 2020{\natexlab{a}}, \mnras, 493, 447

\bibitem[{{Calder{\'o}n} {et~al.}(2020{\natexlab{b}}){Calder{\'o}n}, {Cuadra},
  {Schartmann}, {Burkert}, \& {Russell}}]{Calderon+20a}
{Calder{\'o}n}, D., {Cuadra}, J., {Schartmann}, M., {Burkert}, A., \&
  {Russell}, C. M.~P. 2020{\natexlab{b}}, \apjl, 888, L2

\bibitem[{{Campbell} {et~al.}(2021){Campbell}, {Ciurlo}, {Morris}, {Do}, \&
  {Ghez}}]{Campbell21}
{Campbell}, R.~D., {Ciurlo}, A., {Morris}, M.~R., {Do}, T., \& {Ghez}, A.~M.
  2021, in Astronomical Society of the Pacific Conference Series, Vol. 528, New
  Horizons in Galactic Center Astronomy and Beyond, ed. M.~{Tsuboi} \&
  T.~{Oka}, 235

\bibitem[{{Chen} {et~al.}(2018){Chen}, {Do}, {Witzel}, {Ghez}, {Sch{\"o}del},
  {Gallego}, {Sitarski}, {Lu}, {Becklin}, {Dehghanfar}, {Gautam}, {Hees},
  {Jia}, {Matthews}, \& {Morris}}]{Chen18}
{Chen}, Z., {Do}, T., {Witzel}, G., {et~al.} 2018, in American Astronomical
  Society Meeting Abstracts, Vol. 231, American Astronomical Society Meeting
  Abstracts \#231, 237.06

\bibitem[{{Chu} {et~al.}(2018){Chu}, {Do}, {Hees}, {Ghez}, {Naoz}, {Witzel},
  {Sakai}, {Chappell}, {Gautam}, {Lu}, \& {Matthews}}]{Chu18}
{Chu}, D.~S., {Do}, T., {Hees}, A., {et~al.} 2018, \apj, 854, 12

\bibitem[{{Ciurlo} {et~al.}(2020){Ciurlo}, {Campbell}, {Morris}, {Do}, {Ghez},
  {Hees}, {Sitarski}, {Kosmo O'Neil}, {Chu}, {Martinez}, {Naoz}, \&
  {Stephan}}]{Ciurlo20}
{Ciurlo}, A., {Campbell}, R.~D., {Morris}, M.~R., {et~al.} 2020, \nat, 577, 337

\bibitem[{{Cl{\'e}net} {et~al.}(2004){Cl{\'e}net}, {Rouan}, {Gendron},
  {Lacombe}, {Lagrange}, {Mouillet}, {Magnard}, {Rousset}, {Fusco}, {Montri},
  {Genzel}, {Sch{\"o}del}, {Ott}, {Eckart}, {Marco}, \&
  {Tacconi-Garman}}]{Clenet04}
{Cl{\'e}net}, Y., {Rouan}, D., {Gendron}, E., {et~al.} 2004, \aap, 417, L15

\bibitem[{{Cuadra} {et~al.}(2006){Cuadra}, {Nayakshin}, {Springel}, \& {Di
  Matteo}}]{Cuadra06}
{Cuadra}, J., {Nayakshin}, S., {Springel}, V., \& {Di Matteo}, T. 2006, \mnras,
  366, 358

\bibitem[{{Dale} \& {Davies}(2006)}]{DaleDavies}
{Dale}, J.~E., \& {Davies}, M.~B. 2006, \mnras, 366, 1424

\bibitem[{{Dale} {et~al.}(2009){Dale}, {Davies}, {Church}, \&
  {Freitag}}]{Dale+09}
{Dale}, J.~E., {Davies}, M.~B., {Church}, R.~P., \& {Freitag}, M. 2009, \mnras,
  393, 1016

\bibitem[{{Davies} {et~al.}(2011){Davies}, {Church}, {Malmberg}, {Nzoke},
  {Dale}, \& {Freitag}}]{Davies+11}
{Davies}, M.~B., {Church}, R.~P., {Malmberg}, D., {et~al.} 2011, in
  Astronomical Society of the Pacific Conference Series, Vol. 439, The Galactic
  Center: a Window to the Nuclear Environment of Disk Galaxies, ed. M.~R.
  {Morris}, Q.~D. {Wang}, \& F.~{Yuan}, 212

\bibitem[{{Do} {et~al.}(2013){Do}, {Lu}, {Ghez}, {Morris}, {Yelda}, {Martinez},
  {Wright}, \& {Matthews}}]{Do13}
{Do}, T., {Lu}, J.~R., {Ghez}, A.~M., {et~al.} 2013, \apj, 764, 154

\bibitem[{{Do} {et~al.}(2019){Do}, {Hees}, {Ghez}, {Martinez}, {Chu}, {Jia},
  {Sakai}, {Lu}, {Gautam}, {O'Neil}, {Becklin}, {Morris}, {Matthews},
  {Nishiyama}, {Campbell}, {Chappell}, {Chen}, {Ciurlo}, {Dehghanfar},
  {Gallego-Cano}, {Kerzendorf}, {Lyke}, {Naoz}, {Saida}, {Sch{\"o}del},
  {Takahashi}, {Takamori}, {Witzel}, \& {Wizinowich}}]{Do19GR}
{Do}, T., {Hees}, A., {Ghez}, A., {et~al.} 2019, Science, 365, 664

\bibitem[{{Feroz} {et~al.}(2009){Feroz}, {Hobson}, \& {Bridges}}]{Feroz09}
{Feroz}, F., {Hobson}, M.~P., \& {Bridges}, M. 2009, \mnras, 398, 1601

\bibitem[{{Freitag} {et~al.}(2006){Freitag}, {Amaro-Seoane}, \&
  {Kalogera}}]{Freitag+06}
{Freitag}, M., {Amaro-Seoane}, P., \& {Kalogera}, V. 2006, \apj, 649, 91

\bibitem[{{Fritz} {et~al.}(2011){Fritz}, {Gillessen}, {Dodds-Eden}, {Lutz},
  {Genzel}, {Raab}, {Ott}, {Pfuhl}, {Eisenhauer}, \& {Yusef-Zadeh}}]{Fritz11}
{Fritz}, T.~K., {Gillessen}, S., {Dodds-Eden}, K., {et~al.} 2011, \apj, 737, 73

\bibitem[{{Genzel} {et~al.}(2010){Genzel}, {Eisenhauer}, \&
  {Gillessen}}]{Genzel10}
{Genzel}, R., {Eisenhauer}, F., \& {Gillessen}, S. 2010, Reviews of Modern
  Physics, 82, 3121

\bibitem[{{Genzel} {et~al.}(2003){Genzel}, {Sch{\"o}del}, {Ott}, {Eisenhauer},
  {Hofmann}, {Lehnert}, {Eckart}, {Alexander}, {Sternberg}, {Lenzen},
  {Cl{\'e}net}, {Lacombe}, {Rouan}, {Renzini}, \& {Tacconi-Garman}}]{Genzel03}
{Genzel}, R., {Sch{\"o}del}, R., {Ott}, T., {et~al.} 2003, \apj, 594, 812

\bibitem[{{Ghez} {et~al.}(2005{\natexlab{a}}){Ghez}, {Salim}, {Hornstein},
  {Tanner}, {Lu}, {Morris}, {Becklin}, \& {Duch{\^e}ne}}]{Ghez05}
{Ghez}, A.~M., {Salim}, S., {Hornstein}, S.~D., {et~al.} 2005{\natexlab{a}},
  \apj, 620, 744

\bibitem[{{Ghez} {et~al.}(2004){Ghez}, {Wright}, {Matthews}, {Thompson}, {Le
  Mignant}, {Tanner}, {Hornstein}, {Morris}, {Becklin}, \& {Soifer}}]{Ghez04}
{Ghez}, A.~M., {Wright}, S.~A., {Matthews}, K., {et~al.} 2004, \apjl, 601, L159

\bibitem[{{Ghez} {et~al.}(2005{\natexlab{b}}){Ghez}, {Hornstein}, {Lu},
  {Bouchez}, {Le Mignant}, {van Dam}, {Wizinowich}, {Matthews}, {Morris},
  {Becklin}, {Campbell}, {Chin}, {Hartman}, {Johansson}, {Lafon}, {Stomski}, \&
  {Summers}}]{Ghez05b}
{Ghez}, A.~M., {Hornstein}, S.~D., {Lu}, J.~R., {et~al.} 2005{\natexlab{b}},
  \apj, 635, 1087

\bibitem[{{Ghez} {et~al.}(2008){Ghez}, {Salim}, {Weinberg}, {Lu}, {Do}, {Dunn},
  {Matthews}, {Morris}, {Yelda}, {Becklin}, {Kremenek}, {Milosavljevic}, \&
  {Naiman}}]{Ghez08}
{Ghez}, A.~M., {Salim}, S., {Weinberg}, N.~N., {et~al.} 2008, \apj, 689, 1044

\bibitem[{{Gillessen} {et~al.}(2012){Gillessen}, {Genzel}, {Fritz}, {Quataert},
  {Alig}, {Burkert}, {Cuadra}, {Eisenhauer}, {Pfuhl}, {Dodds-Eden}, {Gammie},
  \& {Ott}}]{Gillessen12}
{Gillessen}, S., {Genzel}, R., {Fritz}, T.~K., {et~al.} 2012, \nat, 481, 51

\bibitem[{{Gillessen} {et~al.}(2019){Gillessen}, {Plewa}, {Widmann}, {von
  Fellenberg}, {Schartmann}, {Habibi}, {Jimenez Rosales}, {Baub{\"o}ck},
  {Dexter}, {Gao}, {Waisberg}, {Eisenhauer}, {Pfuhl}, {Ott}, {Burkert}, {de
  Zeeuw}, \& {Genzel}}]{Gillessen+19}
{Gillessen}, S., {Plewa}, P.~M., {Widmann}, F., {et~al.} 2019, \apj, 871, 126

\bibitem[{{Gravity Collaboration} {et~al.}(2019){Gravity Collaboration},
  {Abuter}, {Amorim}, {Baub{\"o}ck}, {Berger}, {Bonnet}, {Brandner},
  {Cl{\'e}net}, {Coud{\'e} Du Foresto}, {de Zeeuw}, {Dexter}, {Duvert},
  {Eckart}, {Eisenhauer}, {F{\"o}rster Schreiber}, {Garcia}, {Gao}, {Gendron},
  {Genzel}, {Gerhard}, {Gillessen}, {Habibi}, {Haubois}, {Henning}, {Hippler},
  {Horrobin}, {Jim{\'e}nez-Rosales}, {Jocou}, {Kervella}, {Lacour},
  {Lapeyr{\`e}re}, {Le Bouquin}, {L{\'e}na}, {Ott}, {Paumard}, {Perraut},
  {Perrin}, {Pfuhl}, {Rabien}, {Rodriguez Coira}, {Rousset}, {Scheithauer},
  {Sternberg}, {Straub}, {Straubmeier}, {Sturm}, {Tacconi}, {Vincent}, {von
  Fellenberg}, {Waisberg}, {Widmann}, {Wieprecht}, {Wiezorrek}, {Woillez}, \&
  {Yazici}}]{GRAVITY19}
{Gravity Collaboration}, {Abuter}, R., {Amorim}, A., {et~al.} 2019, \aap, 625,
  L10

\bibitem[{{Gravity Collaboration} {et~al.}(2020){Gravity Collaboration},
  {Abuter}, {Amorim}, {Baub{\"o}ck}, {Berger}, {Bonnet}, {Brandner}, {Cardoso},
  {Cl{\'e}net}, {de Zeeuw}, {Dexter}, {Eckart}, {Eisenhauer}, {F{\"o}rster
  Schreiber}, {Garcia}, {Gao}, {Gendron}, {Genzel}, {Gillessen}, {Habibi},
  {Haubois}, {Henning}, {Hippler}, {Horrobin}, {Jim{\'e}nez-Rosales}, {Jochum},
  {Jocou}, {Kaufer}, {Kervella}, {Lacour}, {Lapeyr{\`e}re}, {Le Bouquin},
  {L{\'e}na}, {Nowak}, {Ott}, {Paumard}, {Perraut}, {Perrin}, {Pfuhl},
  {Rodr{\'\i}guez-Coira}, {Shangguan}, {Scheithauer}, {Stadler}, {Straub},
  {Straubmeier}, {Sturm}, {Tacconi}, {Vincent}, {von Fellenberg}, {Waisberg},
  {Widmann}, {Wieprecht}, {Wiezorrek}, {Woillez}, {Yazici}, \&
  {Zins}}]{GRAVITY20}
---. 2020, \aap, 636, L5

\bibitem[{{Grould} {et~al.}(2017){Grould}, {Vincent}, {Paumard}, \&
  {Perrin}}]{Grould17}
{Grould}, M., {Vincent}, F.~H., {Paumard}, T., \& {Perrin}, G. 2017, \aap, 608,
  A60

\bibitem[{{Jia} {et~al.}(2019){Jia}, {Lu}, {Sakai}, {Gautam}, {Do}, {Hosek},
  {Service}, {Ghez}, {Gallego-Cano}, {Sch{\"o}del}, {Hees}, {Morris},
  {Becklin}, \& {Matthews}}]{Jia19}
{Jia}, S., {Lu}, J.~R., {Sakai}, S., {et~al.} 2019, \apj, 873, 9

\bibitem[{{Larkin} {et~al.}(2006){Larkin}, {Barczys}, {Krabbe}, {Adkins},
  {Aliado}, {Amico}, {Brims}, {Campbell}, {Canfield}, {Gasaway}, {Honey},
  {Iserlohe}, {Johnson}, {Kress}, {LaFreniere}, {Lyke}, {Magnone}, {Magnone},
  {McElwain}, {Moon}, {Quirrenbach}, {Skulason}, {Song}, {Spencer}, {Weiss}, \&
  {Wright}}]{Larkin06}
{Larkin}, J., {Barczys}, M., {Krabbe}, A., {et~al.} 2006, in \procspie, Vol.
  6269, Society of Photo-Optical Instrumentation Engineers (SPIE) Conference
  Series, 62691A

\bibitem[{{Lo} \& {Claussen}(1983)}]{Lo83}
{Lo}, K.~Y., \& {Claussen}, M.~J. 1983, \nat, 306, 647

\bibitem[{{Lu} \& {Naoz}(2019)}]{Lu+19}
{Lu}, C.~X., \& {Naoz}, S. 2019, \mnras, 484, 1506

\bibitem[{{Lu} {et~al.}(2013){Lu}, {Do}, {Ghez}, {Morris}, {Yelda}, \&
  {Matthews}}]{Lu13}
{Lu}, J.~R., {Do}, T., {Ghez}, A.~M., {et~al.} 2013, \apj, 764, 155

\bibitem[{{Miralda-Escud{\'e}} \& {Gould}(2000)}]{Miralda-Escude+Gould_00}
{Miralda-Escud{\'e}}, J., \& {Gould}, A. 2000, \apj, 545, 847

\bibitem[{{Morris}(1993)}]{Morris93}
{Morris}, M. 1993, \apj, 408, 496

\bibitem[{{Morris} \& {Serabyn}(1996)}]{Morris96}
{Morris}, M., \& {Serabyn}, E. 1996, \araa, 34, 645

\bibitem[{{Morris} \& {Yusef-Zadeh}(1987)}]{MYZ87}
{Morris}, M., \& {Yusef-Zadeh}, F. 1987, in American Institute of Physics
  Conference Series, Vol. 155, The Galactic Center, ed. D.~C. {Backer},
  127--132

\bibitem[{{Mu{\v z}i{\'c}} {et~al.}(2007){Mu{\v z}i{\'c}}, {Eckart},
  {Sch{\"o}del}, {Meyer}, \& {Zensus}}]{Muzic+07}
{Mu{\v z}i{\'c}}, K., {Eckart}, A., {Sch{\"o}del}, R., {Meyer}, L., \&
  {Zensus}, A. 2007, \aap, 469, 993

\bibitem[{{Mu{\v{z}}i{\'c}} {et~al.}(2010){Mu{\v{z}}i{\'c}}, {Eckart},
  {Sch{\"o}del}, {Buchholz}, {Zamaninasab}, \& {Witzel}}]{Muzic10}
{Mu{\v{z}}i{\'c}}, K., {Eckart}, A., {Sch{\"o}del}, R., {et~al.} 2010, \aap,
  521, A13

\bibitem[{{Naoz}(2016)}]{Naoz16}
{Naoz}, S. 2016, \araa, 54, 441

\bibitem[{{Naoz} \& {Fabrycky}(2014)}]{Naoz+14}
{Naoz}, S., \& {Fabrycky}, D.~C. 2014, \apj, 793, 137

\bibitem[{{O'Neil} {et~al.}(2019){O'Neil}, {Martinez}, {Hees}, {Ghez}, {Do},
  {Witzel}, {Konopacky}, {Becklin}, {Chu}, {Lu}, {Matthews}, \&
  {Sakai}}]{ONeil19}
{O'Neil}, K.~K., {Martinez}, G.~D., {Hees}, A., {et~al.} 2019, \aj, 158, 4

\bibitem[{{Paumard} {et~al.}(2006){Paumard}, {Genzel}, {Martins}, {Nayakshin},
  {Beloborodov}, {Levin}, {Trippe}, {Eisenhauer}, {Ott}, {Gillessen}, {Abuter},
  {Cuadra}, {Alexander}, \& {Sternberg}}]{Paumard06}
{Paumard}, T., {Genzel}, R., {Martins}, F., {et~al.} 2006, \apj, 643, 1011

\bibitem[{{Pei{\ss}ker} {et~al.}(2021){Pei{\ss}ker}, {Ali}, {Zaja{\v{c}}ek},
  {Eckart}, {Hosseini}, {Karas}, {Cl{\'e}net}, {Sabha}, {Labadie}, \&
  {Subroweit}}]{Peissker21}
{Pei{\ss}ker}, F., {Ali}, B., {Zaja{\v{c}}ek}, M., {et~al.} 2021, \apj, 909, 62

\bibitem[{{Pfuhl} {et~al.}(2015){Pfuhl}, {Gillessen}, {Eisenhauer}, {Genzel},
  {Plewa}, {Ott}, {Ballone}, {Schartmann}, {Burkert}, {Fritz}, {Sari},
  {Steinberg}, \& {Madigan}}]{Pfuhl15}
{Pfuhl}, O., {Gillessen}, S., {Eisenhauer}, F., {et~al.} 2015, \apj, 798, 111

\bibitem[{{Phifer} {et~al.}(2013){Phifer}, {Do}, {Meyer}, {Ghez}, {Witzel},
  {Yelda}, {Boehle}, {Lu}, {Morris}, {Becklin}, \& {Matthews}}]{Phifer13}
{Phifer}, K., {Do}, T., {Meyer}, L., {et~al.} 2013, \apjl, 773, L13

\bibitem[{{Plewa} {et~al.}(2017){Plewa}, {Gillessen}, {Pfuhl}, {Eisenhauer},
  {Genzel}, {Burkert}, {Dexter}, {Habibi}, {George}, {Ott}, {Waisberg}, \& {von
  Fellenberg}}]{Plewa+17}
{Plewa}, P.~M., {Gillessen}, S., {Pfuhl}, O., {et~al.} 2017, \apj, 840, 50

\bibitem[{{Prodan} {et~al.}(2015){Prodan}, {Antonini}, \& {Perets}}]{Prodan+15}
{Prodan}, S., {Antonini}, F., \& {Perets}, H.~B. 2015, \apj, 799, 118

\bibitem[{{Ressler} {et~al.}(2020){Ressler}, {Quataert}, \&
  {Stone}}]{Ressler20}
{Ressler}, S.~M., {Quataert}, E., \& {Stone}, J.~M. 2020, \mnras, 492, 3272

\bibitem[{{Roberts} \& {Goss}(1993)}]{RobertsGoss93}
{Roberts}, D.~A., \& {Goss}, W.~M. 1993, \apjs, 86, 133

\bibitem[{{Roche} {et~al.}(2018){Roche}, {Lopez-Rodriguez}, {Telesco},
  {Sch{\"o}del}, \& {Packham}}]{Roche18}
{Roche}, P.~F., {Lopez-Rodriguez}, E., {Telesco}, C.~M., {Sch{\"o}del}, R., \&
  {Packham}, C. 2018, \mnras, 476, 235

\bibitem[{{Rose} {et~al.}(2020){Rose}, {Naoz}, {Gautam}, {Ghez}, {Do}, {Chu},
  \& {Becklin}}]{Rose+20}
{Rose}, S.~C., {Naoz}, S., {Gautam}, A.~K., {et~al.} 2020, \apj, 904, 113

\bibitem[{{Rose} {et~al.}(2022){Rose}, {Naoz}, {Sari}, \& {Linial}}]{Rose+22}
{Rose}, S.~C., {Naoz}, S., {Sari}, R., \& {Linial}, I. 2022, \apjl, 929, L22

\bibitem[{{Sahai} {et~al.}(2003){Sahai}, {Morris}, {Knapp}, {Young}, \&
  {Barnbaum}}]{Sahai+03}
{Sahai}, R., {Morris}, M., {Knapp}, G.~R., {Young}, K., \& {Barnbaum}, C. 2003,
  \nat, 426, 261

\bibitem[{{Sakai} {et~al.}(2019){Sakai}, {Lu}, {Ghez}, {Jia}, {Do}, {Witzel},
  {Gautam}, {Hees}, {Becklin}, {Matthews}, \& {Hosek}}]{Sakai19}
{Sakai}, S., {Lu}, J.~R., {Ghez}, A., {et~al.} 2019, \apj, 873, 65

\bibitem[{{Salas} {et~al.}(2019){Salas}, {Naoz}, {Morris}, \&
  {Stephan}}]{Salas+19}
{Salas}, J.~M., {Naoz}, S., {Morris}, M.~R., \& {Stephan}, A.~P. 2019, \mnras,
  487, 3029

\bibitem[{{Sch{\"o}del} {et~al.}(2011){Sch{\"o}del}, {Morris}, {Muzic},
  {Alberdi}, {Meyer}, {Eckart}, \& {Gezari}}]{Schodel11}
{Sch{\"o}del}, R., {Morris}, M.~R., {Muzic}, K., {et~al.} 2011, \aap, 532, A83

\bibitem[{{Stephan} {et~al.}(2016){Stephan}, {Naoz}, {Ghez}, {Witzel},
  {Sitarski}, {Do}, \& {Kocsis}}]{Stephan16}
{Stephan}, A.~P., {Naoz}, S., {Ghez}, A.~M., {et~al.} 2016, \mnras, 460, 3494

\bibitem[{{Stephan} {et~al.}(2019){Stephan}, {Naoz}, {Ghez}, {Morris},
  {Ciurlo}, {Do}, {Breivik}, {Coughlin}, \& {Rodriguez}}]{Stephan19}
---. 2019, \apj, 878, 58

\bibitem[{Stetson(1987)}]{Stetson_1987}
Stetson, P.~B. 1987, Publications of the Astronomical Society of the Pacific,
  99, 191.
\newblock \url{https://doi.org/10.1086/131977}

\bibitem[{{Storey} \& {Hummer}(1995)}]{StoreyHummer95}
{Storey}, P.~J., \& {Hummer}, D.~G. 1995, \mnras, 272, 41

\bibitem[{{von Fellenberg} {et~al.}(2022){von Fellenberg}, {Gillessen},
  {Stadler}, {Baub{\"o}ck}, {Genzel}, {de Zeeuw}, {Pfuhl}, {Amaro Seoane},
  {Drescher}, {Eisenhauer}, {Habibi}, {Ott}, {Widmann}, \&
  {Young}}]{Fellenberg22}
{von Fellenberg}, S.~D., {Gillessen}, S., {Stadler}, J., {et~al.} 2022, \apjl,
  932, L6

\bibitem[{{Wang} {et~al.}(2013){Wang}, {Nowak}, {Markoff}, {Baganoff},
  {Nayakshin}, {Yuan}, {Cuadra}, {Davis}, {Dexter}, {Fabian}, {Grosso},
  {Haggard}, {Houck}, {Ji}, {Li}, {Neilsen}, {Porquet}, {Ripple}, \&
  {Shcherbakov}}]{WangQD+13}
{Wang}, Q.~D., {Nowak}, M.~A., {Markoff}, S.~B., {et~al.} 2013, Science, 341,
  981

\bibitem[{{Witzel} {et~al.}(2014){Witzel}, {Ghez}, {Morris}, {Sitarski},
  {Boehle}, {Naoz}, {Campbell}, {Becklin}, {Canalizo}, {Chappell}, {Do}, {Lu},
  {Matthews}, {Meyer}, {Stockton}, {Wizinowich}, \& {Yelda}}]{Witzel+14}
{Witzel}, G., {Ghez}, A.~M., {Morris}, M.~R., {et~al.} 2014, \apjl, 796, L8

\bibitem[{{Witzel} {et~al.}(2017){Witzel}, {Sitarski}, {Ghez}, {Morris},
  {Hees}, {Do}, {Lu}, {Naoz}, {Boehle}, {Martinez}, {Chappell}, {Sch{\"o}del},
  {Meyer}, {Yelda}, {Becklin}, \& {Matthews}}]{Witzel17}
{Witzel}, G., {Sitarski}, B.~N., {Ghez}, A.~M., {et~al.} 2017, \apj, 847, 80

\bibitem[{{Wizinowich} {et~al.}(2006){Wizinowich}, {Le Mignant}, {Bouchez},
  {Campbell}, {Chin}, {Contos}, {van Dam}, {Hartman}, {Johansson}, {Lafon},
  {Lewis}, {Stomski}, {Summers}, {Brown}, {Danforth}, {Max}, \&
  {Pennington}}]{Wizinowich06}
{Wizinowich}, P.~L., {Le Mignant}, D., {Bouchez}, A.~H., {et~al.} 2006, \pasp,
  118, 297

\bibitem[{{Yusef-Zadeh} {et~al.}(1990){Yusef-Zadeh}, {Morris}, \&
  {Ekers}}]{Yusef-Zadeh90}
{Yusef-Zadeh}, F., {Morris}, M., \& {Ekers}, R.~D. 1990, \nat, 348, 45

\end{thebibliography}
\bibliographystyle{aasjournal}

\end{document}